\documentclass[prd,aps,nofootinbib,floatfix,11pt]{revtex4}

\usepackage{amsmath,graphicx,epsfig,amssymb,dsfont,mathtools,}
\usepackage[usenames]{color}
\usepackage{ulem} 
\usepackage{bigstrut}
\usepackage{slashed}
\usepackage{multirow}

\allowdisplaybreaks

\begin{document}
\title{Weak Decays of Doubly Heavy Baryons: the $1/2\to 1/2$ case  }
\author{Wei Wang$^1$~\footnote{Email:wei.wang@sjtu.edu.cn}, Fu-Sheng Yu$^2$~\footnote{Email:yufsh@lzu.edu.cn}, and Zhen-Xing Zhao$^1$~\footnote{Email:star\_0027@sjtu.edu.cn}}
\affiliation{$^1$ INPAC, Shanghai Key Laboratory for Particle Physics and Cosmology, School of Physics and Astronomy, Shanghai Jiao Tong University, Shanghai  200240,   China\\
$^2$   School of Nuclear Science and Technology, Lanzhou University Lanzhou 730000, People's Republic of China }

\begin{abstract} 
Very recently, the LHCb collaboration has observed  in the   final state $\Lambda_c^+ K^-\pi^+\pi^+$ a resonant structure  that is identified as the doubly-charmed baryon $\Xi_{cc}^{++}$.  Inspired by this observation,  we investigate  the weak decays of doubly heavy baryons
$\Xi_{cc}^{++}$, $\Xi_{cc}^{+}$, $\Omega_{cc}^{+}$, $\Xi_{bc}^{(\prime)+}$,
$\Xi_{bc}^{(\prime)0}$, $\Omega_{bc}^{(\prime)0}$, $\Xi_{bb}^{0}$, $\Xi_{bb}^{-}$ and $\Omega_{bb}^{-}$ and focus on the decays into  spin $1/2$ baryons  in this paper. At the
quark level these decay processes  are induced by the  $c\to d/s$ or $b\to u/c$ transitions, and the   two spectator quarks can be  viewed as a scalar or axial vector diquark. We first derive  the hadronic form factors for these transitions in the light-front approach and then apply them to predict the partial widths for the semi-leptonic and non-leptonic decays  of doubly heavy baryons. We find that a number of  decay channels are sizable and can be examined in future measurements at experimental facilities like LHC, Belle II and CEPC. 
\end{abstract}

\maketitle

\section{Introduction}

The constituent quark model predicts the existence of multiplets of baryon and meson states~\cite{GellMann:1964nj}. 
Including the heavy charm and bottom quark,   baryons made of three quarks are in a big family of hadron spectroscopy.  When considering $u,d,s$ and $c$, the baryon ground states, those with no orbital or radial excitations, consist of  a 20-plet  with $J^P=1/2^+$ and a 20-plet with $J^P=3/2^+$. For the five flavors, $u,d,s,c,b$, the ground states are then  composed  of a 40-plet with $J^P=1/2^+$  and 35-plet with $J^P=3/2^+$.  All these ground states with zero or one heavy  quark have been well established on experimental side~\cite{Olive:2016xmw}.

The search for doubly heavy baryons is  a long-standing    problem in the last decade. The only     evidence in the past  from the experimental side was found for $\Xi_{cc}^+$ by the SELEX collaboration  \cite{Mattson:2002vu,Ocherashvili:2004hi}.  However this evidence  has not been confirmed by any other experiments \cite{Kato:2013ynr,Aaij:2013voa,Aubert:2006qw,Ratti:2003ez}. 
Very recently, the LHCb collaboration  has observed the doubly charmed baryon $\Xi_{cc}^{++}$ with  the mass   given as~\cite{1707.01621}
\begin{eqnarray}
m_{\Xi_{cc}^{++}} = (3621.40\pm 0.72\pm 0.27\pm 0.14) {\rm MeV}. \label{eq:LHCb_measurement}
\end{eqnarray}
It is anticipated that this observation will  have a great impact on the hadron spectroscopy and with no doubt  it  will trigger much more  interests in this research field. On the other hand, 
after the observation in the $\Xi_{cc}^{++}\to \Lambda_c^+ K^-\pi^+\pi^+$ decay mode, we also believe that   experimental investigations  should be conducted  in a number of other decay channels.   Thus from this viewpoint   theoretical  studies on   weak decays of doubly heavy baryons, not only $\Xi_{cc}^{++}$,  will be  of great importance and are highly demanded. Some   attempts have been made  in Refs.~\cite{Guo:1998yj,SanchisLozano:1994vh,Faessler:2001mr,Egolf:2002nk,Ebert:2004ck,Albertus:2006wb,Hernandez:2007qv,Flynn:2007qt,Albertus:2009ww,Faessler:2009xn,Li:2017ndo,Yu:2017zst},  but a comprehensive study is not available in the recent literature, and the aim of this work and the forthcoming ones   is to fill this gap. To do so, we will calculate the transition form factors and use these results to study the weak decays of bottom quark and charm quark.

The quantum numbers of the doubly heavy baryons are given in Table~\ref{tab:JPC}.  
Among various doubly charmed baryons, three of them can decay only through weak interactions, a $\Xi_{cc}$ isodoublet $ccu, ccd$, and an $\Omega_{cc}$ isosinglet $ccs$.  There are three  doubly bottom baryons similarly. For the bottom-charm baryons, the ones with two different heavy flavors, there are two sets of SU(3) triplets, $\Xi_{bc},\Omega_{bc}$ and $\Xi_{bc}',\Omega_{bc}'$. These two triplets have different total  spin for the   heavy quark system $bc$, but in reality they will probably mix with each other. Only the lighter ones can weak decay  with sizable branching fractions.  The  mixing scheme between the two triplets is unknown yet, and we will  consider both types   in this work.   All  these baryons that can weak decay have spin $1/2$.  The ones with spin $3/2$ can radiatively decay into the lowest-lying ones if the mass splitting is not large enough, or decay into the lowest-lying ones with the emission of a light  pion when they are heavy enough.

\begin{table*}[!htb]
\footnotesize
\caption{Quantum numbers and quark content for the ground state of doubly heavy baryons.  The $S_{h}^{\pi}$ denotes the spin of the heavy quark system. The light quark $q$ corresponds to $u,d$ quark.   }\label{tab:JPC}
\begin{center}
\begin{tabular}{cccc|cccccc} \hline \hline
Baryon      & Quark Content  &  $S_h^\pi$  &$J^P$   & Baryon & Quark Content &   $S_h^\pi$  &$J^P$   \\ \hline  
$\Xi_{cc}$ & $\{cc\}q$  & $1^+$ & $1/2^+$ &   $\Xi_{bb}$ & $\{bb\}q$  & $1^+$ & $1/2^+$ & \\  
$\Xi_{cc}^*$ & $\{cc\}q$  & $1^+$ & $3/2^+$ &   $\Xi_{bb}^*$ & $\{bb\}q$  & $1^+$ & $3/2^+$ & \\ \hline  
$\Omega_{cc}$ & $\{cc\}s$  & $1^+$ & $1/2^+$ &   $\Omega_{bb}$ & $\{bb\}s$  & $1^+$ & $1/2^+$ & \\  
$\Omega_{cc}^*$ & $\{cc\}s$  & $1^+$ & $3/2^+$ &   $\Omega_{bb}^*$ & $\{bb\}s$  & $1^+$ & $3/2^+$ &  \\ \hline  
$\Xi_{bc}'$ & $\{bc\}q$  & $0^+$ & $1/2^+$ &   $\Omega_{bc}'$ & $\{bc\}s$  & $0^+$ & $1/2^+$ & \\  
$\Xi_{bc}$ & $\{bc\}q$  & $1^+$ & $1/2^+$ &   $\Omega_{bc}$ & $\{bc\}s$  & $1^+$ & $1/2^+$ & \\  
$\Xi_{bc}^*$ & $\{bc\}q$  & $1^+$ & $3/2^+$ &   $\Omega_{bc}^*$ & $\{bc\}s$  & $1^+$ & $3/2^+$ & 
 \\ \hline \hline
\end{tabular}
\end{center}
\end{table*}

\begin{figure}
\includegraphics[width=0.5\columnwidth]{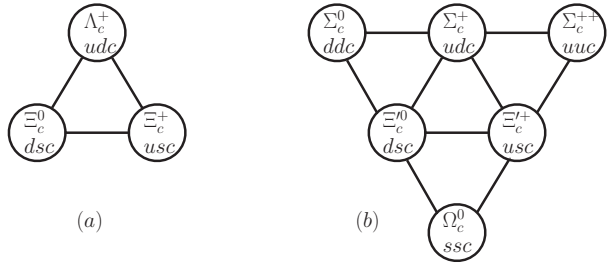} 
\caption{Anti-triplets (panel a) and sextets (panel b) of charmed baryons with one charm quark and two light quarks.  It is similar for the baryons with a bottom quark. The total spin of these baryons is $1/2$, while   another  sextets have spin $3/2$.  }
\label{fig:one_heavy}
\end{figure}

The decay final state of the $\Xi_{cc}$ and $\Omega_{cc}$ contains the baryons with one charm quark. These baryons form an anti-triplets  and sextets   of charmed baryons, as shown in Fig.~\ref{fig:one_heavy}. This is also similar for baryons with one bottom quark.  The total spin of the  baryons  in Fig.~\ref{fig:one_heavy}  is $1/2$, while   another  sextet  has the  spin $3/2$.   In this work,  we shall focus on the $1/2\to 1/2$ transition, and leave the   $1/2\to 3/2$ transition  in a forthcoming publication.

To be more explicit, we will investigate  the following  decay modes of doubly-heavy baryons. 
\begin{itemize}
	\item $cc$ sector
	\begin{align*}
		\Xi_{cc}^{++}(ccu) & \to\Lambda_{c}^{+}(dcu)/\Sigma_{c}^{+}(dcu)/\Xi_{c}^{+}(scu)/\Xi_{c}^{\prime+}(scu),\\
		\Xi_{cc}^{+}(ccd) & \to\Sigma_{c}^{0}(dcd)/\Xi_{c}^{0}(scd)/\Xi_{c}^{\prime0}(scd),\\
		\Omega_{cc}^{+}(ccs) & \to\Xi_{c}^{0}(dcs)/\Xi_{c}^{\prime0}(dcs)/\Omega_{c}^{0}(scs),
	\end{align*}
	\item $bb$ sector
\begin{align*}
	\Xi_{bb}^{0}(bbu) & \to\Sigma_{b}^{+}(ubu)/\Xi_{bc}^{+}(cbu)/\Xi_{bc}^{\prime+}(cbu),\\
	\Xi_{bb}^{-}(bbd) & \to\Lambda_{b}^{0}(ubd)/\Sigma_{b}^{0}(ubd)/\Xi_{bc}^{0}(cbd)/\Xi_{bc}^{\prime0}(cbd),\\
	\Omega_{bb}^{-}(bbs) & \to\Xi_{b}^{0}(ubs)/\Xi_{b}^{\prime0}(ubs)/\Omega_{bc}^{0}(cbs)/\Omega_{bc}^{\prime0}(cbs),
\end{align*}
	\item $bc$ sector with the $c$ quark decay
\begin{align*}
	\Xi_{bc}^{+}(cbu)/\Xi_{bc}^{\prime+}(cbu) & \to\Lambda_{b}^{0}(dbu)/\Sigma_{b}^{0}(dbu)/\Xi_{b}^{0}(sbu)/\Xi_{b}^{\prime0}(sbu),\\
	\Xi_{bc}^{0}(cbd)/\Xi_{bc}^{\prime0}(cbd) & \to\Sigma_{b}^{-}(dbd)/\Xi_{b}^{-}(sbd)/\Xi_{b}^{\prime-}(sbd),\\
	\Omega_{bc}^{0}(cbs)/\Omega_{bc}^{\prime0}(cbs) & \to\Xi_{b}^{-}(dbs)/\Xi_{b}^{\prime-}(dbs)/\Omega_{b}^{-}(sbs),
\end{align*}
	\item $bc$ sector with the $b$ quark decay
\begin{align*}
	\Xi_{bc}^{+}(bcu)/\Xi_{bc}^{\prime+}(bcu) & \to\Sigma_{c}^{++}(ucu)/\Xi_{cc}^{++}(ccu),\\
	\Xi_{bc}^{0}(bcd)/\Xi_{bc}^{\prime0}(bcd) & \to\Lambda_{c}^{+}(ucd)/\Sigma_{c}^{+}(ucd)/\Xi_{cc}^{+}(ccd),\\
	\Omega_{bc}^{0}(bcs)/\Omega_{bc}^{\prime0}(bcs) & \to\Xi_{c}^{+}(ucs)/\Xi_{c}^{\prime+}(ucs)/\Omega_{cc}^{+}(ccs).
\end{align*}
\end{itemize}
In the above, the quark components have been explicitly shown in the brackets,
in which the first quarks denote the quarks participating in the  weak decays.

To deal with the strong interaction  in the transition,  we will  adopt the light front approach and calculate the decay form factors.  This approach has been widely applied to various mesonic  transitions~\cite{Jaus:1999zv,Jaus:1989au,Jaus:1991cy,Cheng:1996if,Cheng:2003sm,Cheng:2004yj,Ke:2009ed,Ke:2009mn,Cheng:2009ms,Lu:2007sg,Wang:2007sxa,Wang:2008xt,Wang:2008ci,Wang:2009mi,Chen:2009qk,Li:2010bb,Verma:2011yw,Shi:2016gqt}, and  some analyses of baryonic transitions in this approach can be found in Refs.~\cite{Ke:2007tg,Wei:2009np,Ke:2012wa}. 
We will  use diquark picture and only consider the ground states of baryons, thereby the two spectator quarks are considered to be a scalar diquark
with $J^{P}=0^{+}$ or an axial vector diquark with $J^{P}=1^{+}$.  Both types of diquarks will contribute and their contributions are calculated respectively.

The rest of this paper is organized as follows.  In Sec.~\ref{sec:spectroscopy_lifetime}, we will give a  very   brief overview of the spectroscopy and lifetimes of the doubly heavy baryons. Sec.~\ref{sec:formfactorsLFQM} is devoted to the calculation  of transition form factors  in the covariant light-front quark model.  In Sec.~\ref{sec:semileptonic} and ~\ref{sec:nonleptonic}, we apply our results to calculate  the partial widths for semileptonic $B\to B'\ell\bar\nu_\ell$ decays, and the nonleptonic decays, respectively.  A brief summary and some discussions on the future improvements are given in the last section. 
	
\section{Spectroscopy and lifetimes }
\label{sec:spectroscopy_lifetime}

The doubly heavy baryon systems with the quark contents
$Q_{1}Q_{2}q$ with $Q_{1,2}=b,c$ and $q=u,d,s$ have been studied extensively using various
theoretical methods, such as quark models \cite{Ebert:1996ec,Ebert:2002ig,Roberts:2007ni,Karliner:2014gca}, the bag
model \cite{He:2004px}, QCD sum rules \cite{Zhang:2008rt}, heavy quark effective theory~\cite{Korner:1994nh,Chen:2017sbg}, Lattice QCD
simulation~\cite{LlanesEstrada:2011kc,Wang:2011ae,Flynn:2011gf,Meinel:2012qz,Aliev:2012tt,Aliev:2014lxa,Padmanath:2013zfa}, etc.   For the $\Xi_{cc}$, most predictions are in the range $3.5$ to $3.7$ GeV. 
	For instance Refs.~\cite{Karliner:2014gca} and~\cite{Brown:2014ena} give respectively $m_{\Xi_{cc}}= 3.627$ GeV and  $m_{\Xi_{cc}}= 3.610$ GeV which are very close to the  LHCb measurement in Eq.~\eqref{eq:LHCb_measurement}. Thus in this calculation we will use the results from Refs.~\cite{Karliner:2014gca} and~\cite{Brown:2014ena} if available.
	 These results are collected in Tab.~\ref{Tab:para_doubly_heavy}.

In  Tab.~\ref{Tab:para_doubly_heavy},  we have neglected   the isospin splittings, that is, we have used $m_{\Xi_{cc}^{++}}=m_{\Xi_{cc}^{+}}$, $m_{\Xi_{bc}^{+}}=m_{\Xi_{bc}^{0}}$ and  $m_{\Xi_{bb}^{0}}=m_{\Xi_{bb}^{-}}$.  The isospin splittings for   doubly heavy baryons have been   studied  in Ref.~\cite{Brodsky:2011zs} with the results: 
\begin{eqnarray}
 m_{\Xi_{cc}^{++}}-  m_{\Xi_{cc}^{+}} &=& (1.5\pm2.7){\rm MeV}, \nonumber\\
 m_{\Xi_{bb}^{-}}-  m_{\Xi_{bb}^{0}} &=& (6.3\pm1.7){\rm MeV},\nonumber\\
 m_{\Xi_{bc}^{+}}-  m_{\Xi_{bc}^{0}} &=& (-1.5\pm0.9){\rm MeV}. 
\end{eqnarray}
As one can see from the above equation, the isospin splittings are at most  a few MeV.   We find that their impact on the form factors  and decay  widths of semi-leptonic   and non-leptonic decays is   small especially compared to hadronic uncertainties. This, however, will  be improved   when the  experimental data for the masses are available.

For the baryons with a strange quark, one expects that they should be higher than the corresponding states with a $u$ or $d$ quark.   Theoretical results from Ref.~\cite{Brown:2014ena} did respect this expectation for the $ccq$ and $bbq$ baryons, however, the predicted mass for $\Omega_{bc}$ is only 55 MeV higher than that for $\Xi_{bc}$.   It should be warned that using these results might introduce some theoretical uncertainties to form factors and decay widths.

The lifetime of the baryons is determined by the inclusive decays. Thus one can in principle  use the optical theorem to obtain the total width (lifetime) of the heavy hadron by calculating the absorptive part of the forward-scattering amplitude.   	
The lifetimes of the baryons have been studied in Refs.~\cite{Anikeev:2001rk,Kiselev:2001fw,Karliner:2014gca,Guberina:1999mx,Kiselev:1998sy,Chang:2007xa,Berezhnoy:2016wix}, where the results differ significantly. For the instance, the lifetime of the $\Xi_{cc}^{++}$ baryon is predicted in the range $200$ fs to 700 fs. This large ambiguity will introduce dramatic uncertainties to the decay branching fractions, and we intend to improve the precision of the lifetime in the future.  In this work, for the lifetime of $\Xi_{cc}$ we will use the results from Ref.~\cite{Yu:2017zst}, while other lifetimes are taken from Ref.~\cite{Kiselev:2001fw,Karliner:2014gca}.

\begin{table}[!htb]
\caption{Masses (in units of GeV) and lifetimes (in units of fs) of doubly
heavy baryons. We have used the experimental data for the mass of $\Xi_{cc}^{++}$~\cite{1707.01621} and theoretical  results from Ref.~\cite{Brown:2014ena,Yu:2017zst,Karliner:2014gca,Kiselev:2001fw}.}
\label{Tab:para_doubly_heavy} %
\begin{tabular}{c|c|c|c|c|c|c|c|c|c}
\hline \hline
baryons  & $\Xi_{cc}^{++}$  & $\Xi_{cc}^{+}$  & $\Omega_{cc}^{+}$  & $\Xi_{bc}^{+}$  & $\Xi_{bc}^{0}$  & $\Omega_{bc}^{0}$  & $\Xi_{bb}^{0}$  & $\Xi_{bb}^{-}$  & $\Omega_{bb}^{-}$ \tabularnewline
\hline 
masses  & $3.621$ \cite{1707.01621}  & $3.621$ \cite{1707.01621}  & $3.738$ \cite{Brown:2014ena}  & $6.943$ \cite{Brown:2014ena}  & $6.943$ \cite{Brown:2014ena}  & $6.998$ \cite{Brown:2014ena}  & $10.143$\cite{Brown:2014ena}  & $10.143$ \cite{Brown:2014ena}  & $10.273$\cite{Brown:2014ena}\tabularnewline
\hline 
lifetimes  & 300~\cite{Yu:2017zst}  & 100~\cite{Yu:2017zst}  & $270$ \cite{Kiselev:2001fw}  & $244$ \cite{Karliner:2014gca}  & $93$ \cite{Karliner:2014gca}  & $220$ \cite{Kiselev:2001fw}  & $370$ \cite{Karliner:2014gca}  & $370$ \cite{Karliner:2014gca}  & $800$\cite{Kiselev:2001fw}\tabularnewline
\hline \hline
\end{tabular}
\end{table}

In heavy quark limit, the  interaction between the heavy quark and gluon is independent of the heavy quark spin. Thus the  spin of heavy quarks   is conserved and can be used for classification of hadrons.  For the lowest lying $bbq$ and $ccq$ ($q=u,d,s$) system with $L=0$, the $bb$ and $cc$ must have spin 1 due to the symmetry between the two heavy quarks. For the $bcq$ baryon, the $bc$ system can have spin 0,  corresponding to  $1/2$ baryons,  or spin 1, corresponding to the $1/2$ or $3/2$ baryons as shown in Tab.~\ref{tab:JPC}.  
The physical hadrons,  mass eigenstates,  might be mixtures of the spin eigenstates of heavy quark subsystem: 
\begin{eqnarray}
 \left(   \begin{array}{c}
     \Xi^{(1)}_{bc}\\ 
     \Xi^{(2)}_{bc} \\ 
  \end{array}\right)
 =  \left(   \begin{array}{cc}
    \cos\theta_{\Xi} &  \sin\theta_{\Xi}\\ 
     -\sin\theta_{\Xi}&  \cos\theta_{\Xi} \\ 
  \end{array}\right)
  \left(   \begin{array}{c}
       \Xi_{bc}\\ 
       \Xi^{\prime}_{bc} \\ 
  \end{array}\right),\\
 \left(   \begin{array}{c}
     \Omega^{(1)}_{bc}\\ 
     \Omega^{(2)}_{bc} \\ 
  \end{array}\right)
 =  \left(   \begin{array}{cc}
    \cos\theta_{\Omega} &  \sin\theta_{\Omega}\\ 
     -\sin\theta_{\Omega}&  \cos\theta_{\Omega} \\ 
  \end{array}\right)
  \left(   \begin{array}{c}
       \Omega_{bc}\\ 
       \Omega^{\prime}_{bc} \\ 
  \end{array}\right). 
\end{eqnarray}
We expect the mixing effects are at the order $\Lambda_{QCD}/m_Q$, and in this case $m_Q$ is very probably  the charm quark mass.   But currently we are  unable to determine the mixing angle in a reliable way, and thus in the following we will calculate the decays for both $\Xi_{bc}$ and $\Xi_{bc}'$. It is necessary to point out that once the mixing scheme is determined, only one of the two sets of baryons have sizable branching fractions for weak decays. The ones with higher mass will radiatively decay into the lower one.

\section{Transition form factors in the light-front approach}
\label{sec:formfactorsLFQM}

\subsection{Light-front approach for baryons}

Doubly heavy baryons are made  of two heavy quarks and a light quark.  For a very solid analysis from QCD of  their weak decays, one has to take into account all three quarks, which  is very complicated and far beyond our capability now. 
Starting from the initial doubly heavy baryons, it might be better to treat the two heavy quarks in the initial state as a diquark system. However, if one adopts the heavy-QQ-diquark picture for the doubly heavy baryons, the diquark system must be smashed in the weak transition of doubly heavy system to singly heavy system.  An observation is, in the  decay transition,  one of the two heavy quarks decays, while the other heavy quark and the light quark will act as spectators.   Thus as an approximation it might be plausible to treat the two spectators as a  system.  Here it should be stressed that this system is not tightly bounded as a usual diquark system. Only for brevity, we use the symbol $di$ to  denote the heavy-light system.

In the light-front approach, it is convenient to use the light-front
decomposition of the momentum $p=(p^{-},p^{+},p_{\perp})$, with $p^{\pm}=p^{0}\pm p^{3}$
and $p_{\perp}=(p^{1},p^{2})$ and thus $p\cdot p=p^{-}p^{+}-p_{\perp}^{2}$.
A baryon with total momentum $P$, spin $S=\frac{1}{2}$ and a scalar/axial vector
diquark can be expanded as
\begin{align}
|B(P,S,S_{z})\rangle & =\int\{d^{3}p_{1}\}\{d^{3}p_{2}\}2(2\pi)^{3}\delta^{3}(\tilde{P}-\tilde{p}_{1}-\tilde{p}_{2})\nonumber \\
& \quad\times\sum_{\lambda_{1},\lambda_{2}}\Psi^{SS_{z}}(\tilde{p}_{1},\tilde{p}_{2},\lambda_{1},\lambda_{2})|q_{1}(p_{1},\lambda_{1})[di](p_{2},\lambda_{2})\rangle,\label{eq:stateVector}
\end{align}
where $p_{1}$ and $p_{2}$ are  the momenta of the quark $q_{1}$
and the diquark $[di]$, respectively. The convention is chosen as:
\begin{eqnarray}
\tilde{p}=(p^{+},p_{\perp}),\;\; 
\{d^{3}p\}\equiv\frac{dp^{+}d^{2}p_{\perp}}{2(2\pi)^{3}},\quad\delta^{3}(\tilde{p})=\delta(p^{+})\delta^{2}(p_{\perp}).
\end{eqnarray}

The baryon mass is denoted as  $M$, and the quark $q_{1}$ mass and
the diquark $[di]$ mass are  $m_{1}$ and $m_{2}$ respectively.
The minus  component of the momenta can be determined by their
corresponding on-shell condition, 
\begin{eqnarray}
p^{-}=\frac{p_{\perp}^{2}+m^{2}}{p^{+}}. 
\end{eqnarray}

One can introduce the  momentum fraction $x_{1,2}$ of $q_{1}$ and $[di]$ through
\begin{align}
p_{1}^{+} & =x_{1}P^{+},\quad p_{2}^{+}=x_{2}P^{+},\quad x_{1}+x_{2}=1.\label{eq:momenFrac}
\end{align}
It is often convenient to use  $x\equiv x_{2}$ and hence $x_{1}=1-x$.
Denote $\bar{P}\equiv p_{1}+p_{2}$ and $M_{0}^{2}\equiv\bar{P}^{2}$.
In $\bar{P}$ rest frame, $e_{1,2}$ corresponds to the energy of $q_{1}$
and $[di]$, respectively. The  3-momentum of $[di]$ is $\vec{k}=(k_{\perp},k_{z})$.
Then $M_{0}$ can be expressed as a function of the internal variables
$x$ and $k_{\perp}$:
\begin{align}
M_{0}^{2} & =\frac{k_{\perp}^{2}+m_{1}^{2}}{x_{1}}+\frac{k_{\perp}^{2}+m_{2}^{2}}{x_{2}}.
\end{align}
Using $e_{1}+e_{2}=M_{0}$ and the on-shell conditions of $q_{1}$
and $[di]$, one  can obtain:
\begin{align}
e_{i} & =\frac{x_{i}M_{0}}{2}+\frac{m_{i}^{2}+k_{\perp}^{2}}{2x_{i}M_{0}},\\
k_{z} & =\frac{xM_{0}}{2}-\frac{m_{2}^{2}+k_{\perp}^{2}}{2xM_{0}}.
\end{align}
Here $e_{i}$ and $k_{z}$ have also been expressed in terms  of the internal variables $x$ and $k_{\perp}$.

The momentum-space wave function $\Psi^{SS_{z}}$ is expressed as
\cite{Cheng:2003sm}
\begin{align}
\Psi^{SS_{z}}(\tilde{p}_{1},\tilde{p}_{2},\lambda_{1},\lambda_{2}) & =\sum_{s_{1},s_{2}}\langle\lambda_{1}|{\cal R}_{M}^{\dagger}(x_{1},-k_{\perp},m_{1})|s_{1}\rangle\langle\lambda_{2}|{\cal R}_{M}^{\dagger}(x_{2},k_{\perp},m_{2})|s_{2}\rangle \nonumber \\
& \quad\quad\times\langle\frac{1}{2}s_{1};s_{[di]}s_{2}|\frac{1}{2}S_{z}\rangle\varphi(x,k_{\perp}),
\end{align}
where $\varphi(x,k_{\perp})$ is the light-front wave function which
describes the momentum distribution of the constituents in the bound
state; $\langle\frac{1}{2}s_{1};s_{[di]}s_{2}|\frac{1}{2}S_{z}\rangle$
is the Clebsch-Gordan coefficient with $s_{[di]}=s_{2}=0$ for the
scalar diquark and $s_{[di]}=1,\,s_{2}=0,\pm1$ for the axial vector
diquark. $\langle\lambda_{1}|{\cal R}_{M}^{\dagger}(x_{1},-k_{\perp},m_{1})|s_{1}\rangle$
is the Melosh transformation matrix element which transforms the conventional
spin states in the instant form into the light-front helicity eigenstates.
It can be shown that \cite{Cheng:2003sm}:
\begin{align}\label{eq:Melosh}
& \sum_{s_{1},s_{2}}\langle\lambda_{1}|{\cal R}_{M}^{\dagger}(1-x,-k_{\perp},m_{1})|s_{1}\rangle\langle\lambda_{2}|{\cal R}_{M}^{\dagger}(x,k_{\perp},m_{2})|s_{2}\rangle\langle\frac{1}{2}s_{1};s_{[di]}s_{2}|\frac{1}{2}S_{z}\rangle\nonumber \\
& =\frac{1}{\sqrt{2(p_{1}\cdot\bar{P}+m_{1}M_{0})}}\bar{u}(p_{1},\lambda_{1})\Gamma u(\bar{P},S_{z}),
\end{align}
with $\Gamma=1$ for the scalar diquark and $\Gamma=-\frac{1}{\sqrt{3}}\gamma_{5}\slashed\epsilon^{*}(p_{2},\lambda_{2})$ for the axial vector diquark~\cite{Ke:2012wa}.

The light-front wave function is given as
\begin{equation}
\varphi(x,k_{\perp})=A\phi(x,k_{\perp}),
\end{equation}
where $A=1$ for the scalar diquark and $A=\sqrt{\frac{3(M_{0}m_{1}+p_{1}\cdot\bar{P})}{3M_{0}m_{1}+p_{1}\cdot\bar{P}+(2p_{1}\cdot p_{2}p_{2}\cdot\bar{P})/m_{2}^{2}}}$
for the axial vector diquark. 
The heavy baryon state is normalized as
\begin{equation}
\langle B(P^{\prime},S^{\prime},S_{z}^{\prime})|B(P,S,S_{z})\rangle=2(2\pi)^{3}P^{+}\delta^{3}(\tilde{P}^{\prime}-\tilde{P})\delta_{S^{\prime}S}\delta_{S_{z}^{\prime}S_{z}},
\end{equation}
which implies that the light-front wave function $\phi(x,k_{\perp})$
should satisfy the following constraint
\begin{equation}
\int\frac{dxd^{2}k_{\perp}}{2(2\pi)^{3}}|\phi(x,k_{\perp})|^{2}=1.
\end{equation}
In the  practical calculations, a Gaussian form function is widely used,
\begin{equation}
\phi(x,k_{\perp})=N\sqrt{\frac{\partial k_{z}}{\partial x_{2}}}\exp\left(\frac{-\vec{k}^{2}}{2\beta^{2}}\right),\label{Gaussian_type}
\end{equation}
with
\begin{equation}
N=4\left(\frac{\pi}{\beta^{2}}\right)^{3/4},\quad\frac{\partial k_{z}}{\partial x_{2}}=\frac{e_{1}e_{2}}{x_{1}x_{2}M_{0}},
\end{equation}
where the parameter $\beta$ characterizes the momentum distributions between the constituents, and is usually obtained by fitting the data. 

\subsection{Transitions with scalar diquarks}

\begin{figure}[!]
	\includegraphics[width=0.5\columnwidth]{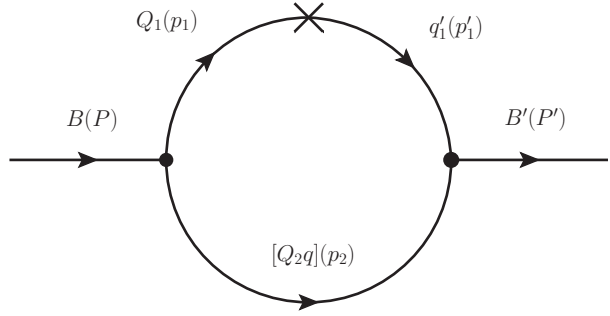} 
	\caption{Feynman diagrams for baryon-baryon transitions in the diquark picture. $P^{(\prime)}$ is the momentum of the incoming (outgoing) baryon, $p_{1}^{(\prime)}$ is the quark momentum, $p_{2}$ is the diquark momentum and the cross mark denotes the corresponding vertex of weak interaction.}
	\label{fig:decay}
\end{figure}

The baryon-baryon weak transitng matrix elements are expressed in terms of form factors as  
\begin{align}
\langle B^{\prime}(P^{\prime},S_{z}^{\prime})|(V-A)_{\mu}|B(P,S_{z})&\rangle  =
\bar{u}(P^{\prime},S_{z}^{\prime})\left[\gamma_{\mu}f_{1}(q^{2})+i\sigma_{\mu\nu}\frac{q^{\nu}}{M}f_{2}(q^{2})+\frac{q_{\mu}}{M}f_{3}(q^{2})\right]u(P,S_{z}),\nonumber \\
& - \bar{u}(P^{\prime},S_{z}^{\prime})\left[\gamma_{\mu}g_{1}(q^{2})+i\sigma_{\mu\nu}\frac{q^{\nu}}{M}g_{2}(q^{2})+\frac{q_{\mu}}{M}g_{3}(q^{2})\right]\gamma_{5}u(P,S_{z}),\label{eq:weakMatrix}
\end{align}
where $(V-A)_{\mu}$ is the weak current, $q=P-P^{\prime}$, and $M$ denotes the mass of the parent baryon
$B$. With the baryon state in the light-front approach in (\ref{eq:stateVector}), the above matrix elements are
\begin{align}
\langle B^{\prime}(P^{\prime},S_{z}^{\prime})|(V-A)_{\mu}|B(P,S_{z})\rangle & =\int\{d^{3}p_{2}\}\frac{\phi^{\prime*}(x^{\prime},k_{\perp}^{\prime})\phi(x,k_{\perp})}{2\sqrt{p_{1}^{+}p_{1}^{\prime+}(p_{1}\cdot\bar{P}+m_{1}M_{0})(p_{1}^{\prime}\cdot\bar{P}^{\prime}+m_{1}^{\prime}M_{0}^{\prime})}}\nonumber \\
& \quad\times\bar{u}(\bar{P}^{\prime},S_{z}^{\prime})\bar{\Gamma}^{\prime}(\slashed p_{1}^{\prime}+m_{1}^{\prime})\gamma_{\mu}(1-\gamma_{5})(\slashed p_{1}+m_{1})\Gamma u(\bar{P},S_{z}),\label{eq:weakMatrix2}
\end{align}
where 
\begin{equation}
\bar{\Gamma}^{\prime}=\gamma_{0}\Gamma^{\dagger}\gamma_{0}=\Gamma=1,\label{eq:GammaBarPrimed}
\end{equation}
for the transitions with scalar diqaruks,
$m_{1}$, $m_{1}^{\prime}$ and $m_{2}$ are the masses of initial
quark, final quark and diquark with momenta $p_{1}$, $p_{1}^{\prime}$
and $p_{2}$ respectively, $P$ and $P^{\prime}$ are the momenta of
initial and final baryons respectively, $\bar{P}$ and $\bar{P}^{\prime}$
are defined as $\bar{P}=p_{1}+p_{2}$ and $\bar{P}^{\prime}=p_{1}^{\prime}+p_{2}$
respectively. The Feynman diagram of the baryon-baryon transitions in the diquark picture is shown in Fig. \ref{fig:decay}.
Since the momentum distribution wavefunction $\phi^{\prime}$
is real, (\ref{eq:weakMatrix2}) becomes
\begin{align}
\langle B^{\prime}(P^{\prime},S_{z}^{\prime})|(V-A)_{\mu}|B(P,S_{z})\rangle & =\int\frac{dxd^{2}k_{\perp}}{2(2\pi)^{3}}\frac{\phi^{\prime}(x^{\prime},k_{\perp}^{\prime})\phi(x,k_{\perp})}{2\sqrt{x_{1}x_{1}^{\prime}(p_{1}\cdot\bar{P}+m_{1}M_{0})(p_{1}^{\prime}\cdot\bar{P}^{\prime}+m_{1}^{\prime}M_{0}^{\prime})}}\nonumber \\
& \quad\times\bar{u}(\bar{P}^{\prime},S_{z}^{\prime})(\slashed p_{1}^{\prime}+m_{1}^{\prime})\gamma_{\mu}(1-\gamma_{5})(\slashed p_{1}+m_{1})u(\bar{P},S_{z}),\label{eq:simWeakMatrix}
\end{align}
With the relations of 
\begin{align}
\langle B^{\prime}(P^{\prime},S_{z}^{\prime})|V^{+}|B(P,S_{z})\rangle & =2\sqrt{P^{+}P^{\prime+}}\left[f_{1}(q^{2})\delta_{S_{z}{}^{\prime}S_{z}}+\frac{f_{2}(q^{2})}{M}(\vec{\sigma}\cdot\vec{q}_{\perp}\sigma^{3})_{S_{z}^{\prime}S_{z}}\right],\nonumber \\
\langle B^{\prime}(P^{\prime},S_{z}^{\prime})|A^{+}|B(P,S_{z})\rangle & =2\sqrt{P^{+}P^{\prime+}}\left[g_{1}(q^{2})(\sigma^{3})_{S_{z}{}^{\prime}S_{z}}+\frac{g_{2}(q^{2})}{M}(\vec{\sigma}\cdot\vec{q}_{\perp})_{S_{z}^{\prime}S_{z}}\right],\label{eq:extractFF}
\end{align}
one can obtain  form factors as \cite{Ke:2007tg}
\begin{align}
f_{1}(q^{2})= & \frac{1}{8P^{+}P^{\prime+}}\int\frac{dxd^{2}k_{\perp}}{2(2\pi)^{3}}\frac{\phi^{\prime}(x^{\prime},k_{\perp}^{\prime})\phi(x,k_{\perp})}{2\sqrt{x_{1}x_{1}^{\prime}(p_{1}\cdot\bar{P}+m_{1}M_{0})(p_{1}^{\prime}\cdot\bar{P}^{\prime}+m_{1}^{\prime}M_{0}^{\prime})}}\nonumber \\
& \times{\rm Tr}[(\bar{\slashed P}+M_{0})\gamma^{+}(\bar{\slashed P^{\prime}}+M_{0}^{\prime})(\slashed p_{1}^{\prime}+m_{1}^{\prime})\gamma^{+}(\slashed p_{1}+m_{1})]\nonumber \\
g_{1}(q^{2})= & \frac{1}{8P^{+}P^{\prime+}}\int\frac{dxd^{2}k_{\perp}}{2(2\pi)^{3}}\frac{\phi^{\prime}(x^{\prime},k_{\perp}^{\prime})\phi(x,k_{\perp})}{2\sqrt{x_{1}x_{1}^{\prime}(p_{1}\cdot\bar{P}+m_{1}M_{0})(p_{1}^{\prime}\cdot\bar{P}^{\prime}+m_{1}^{\prime}M_{0}^{\prime})}}\nonumber \\
& \times{\rm Tr}[(\bar{\slashed P}+M_{0})\gamma^{+}\gamma_{5}(\bar{\slashed P^{\prime}}+M_{0}^{\prime})(\slashed p_{1}^{\prime}+m_{1}^{\prime})\gamma^{+}\gamma_{5}(\slashed p_{1}+m_{1})]\nonumber \\
\frac{f_{2}(q^{2})}{M}= & -\sum_{j=1}^{2}\frac{iq_{\perp}^{j}}{8P^{+}P^{\prime+}q_{\perp}^{2}}\int\frac{dxd^{2}k_{\perp}}{2(2\pi)^{3}}\frac{\phi^{\prime}(x^{\prime},k_{\perp}^{\prime})\phi(x,k_{\perp})}{2\sqrt{x_{1}x_{1}^{\prime}(p_{1}\cdot\bar{P}+m_{1}M_{0})(p_{1}^{\prime}\cdot\bar{P}^{\prime}+m_{1}^{\prime}M_{0}^{\prime})}}\nonumber \\
& \times{\rm Tr}[(\bar{\slashed P}+M_{0})\sigma^{j+}(\bar{\slashed P^{\prime}}+M_{0}^{\prime})(\slashed p_{1}^{\prime}+m_{1}^{\prime})\gamma^{+}(\slashed p_{1}+m_{1})]\nonumber \\
\frac{g_{2}(q^{2})}{M}= &\sum_{j=1}^{2} \frac{iq_{\perp}^{j}}{8P^{+}P^{\prime+}q_{\perp}^{2}}\int\frac{dxd^{2}k_{\perp}}{2(2\pi)^{3}}\frac{\phi^{\prime}(x^{\prime},k_{\perp}^{\prime})\phi(x,k_{\perp})}{2\sqrt{x_{1}x_{1}^{\prime}(p_{1}\cdot\bar{P}+m_{1}M_{0})(p_{1}^{\prime}\cdot\bar{P}^{\prime}+m_{1}^{\prime}M_{0}^{\prime})}}\nonumber \\
& \times{\rm Tr}[(\bar{\slashed P}+M_{0})\sigma^{j+}\gamma_{5}(\bar{\slashed P^{\prime}}+M_{0}^{\prime})(\slashed p_{1}^{\prime}+m_{1}^{\prime})\gamma^{+}\gamma_{5}(\slashed p_{1}+m_{1})].
\end{align}
The final expressions for form factors
are given:
\begin{align}
f_{1}(q^{2})= & \int\frac{dxd^{2}k_{\perp}}{2(2\pi)^{3}}\frac{\phi^{\prime}(x^{\prime},k_{\perp}^{\prime})\phi(x,k_{\perp})[k_{\perp}\cdot k_{\perp}^{\prime}+(x_{1}M_{0}+m_{1})(x_{1}^{\prime}M_{0}^{\prime}+m_{1}^{\prime})]}{\sqrt{\left[(m_{1}+x_{1}M_{0})^{2}+k_{\perp}^{2}\right]\left[(m_{1}^{\prime}+x_{1}^{\prime}M_{0}^{\prime})^{2}+k_{\perp}^{\prime2}\right]}},\nonumber\\
g_{1}(q^{2})= & \int\frac{dxd^{2}k_{\perp}}{2(2\pi)^{3}}\frac{\phi^{\prime}(x^{\prime},k_{\perp}^{\prime})\phi(x,k_{\perp})[-k_{\perp}\cdot k_{\perp}^{\prime}+(x_{1}M_{0}+m_{1})(x_{1}^{\prime}M_{0}^{\prime}+m_{1}^{\prime})]}{\sqrt{\left[(m_{1}+x_{1}M_{0})^{2}+k_{\perp}^{2}\right]\left[(m_{1}^{\prime}+x_{1}^{\prime}M_{0}^{\prime})^{2}+k_{\perp}^{\prime2}\right]}},\nonumber\\
\frac{f_{2}(q^{2})}{M}= & \frac{1}{q_{\perp}^{2}}\int\frac{dxd^{2}k_{\perp}}{2(2\pi)^{3}}\frac{\phi^{\prime}(x^{\prime},k_{\perp}^{\prime})\phi(x,k_{\perp})[-(m_{1}+x_{1}M_{0})k_{\perp}^{\prime}\cdot q_{\perp}+(m_{1}^{\prime}+x_{1}^{\prime}M_{0}^{\prime})k_{\perp}\cdot q_{\perp}]}{\sqrt{\left[(m_{1}+x_{1}M_{0})^{2}+k_{\perp}^{2}\right]\left[(m_{1}^{\prime}+x_{1}^{\prime}M_{0}^{\prime})^{2}+k_{\perp}^{\prime2}\right]}},\nonumber\\
\frac{g_{2}(q^{2})}{M}= & \frac{1}{q_{\perp}^{2}}\int\frac{dxd^{2}k_{\perp}}{2(2\pi)^{3}}\frac{\phi^{\prime}(x^{\prime},k_{\perp}^{\prime})\phi(x,k_{\perp})[-(m_{1}+x_{1}M_{0})k_{\perp}^{\prime}\cdot q_{\perp}-(m_{1}^{\prime}+x_{1}^{\prime}M_{0}^{\prime})k_{\perp}\cdot q_{\perp}]}{\sqrt{\left[(m_{1}+x_{1}M_{0})^{2}+k_{\perp}^{2}\right]\left[(m_{1}^{\prime}+x_{1}^{\prime}M_{0}^{\prime})^{2}+k_{\perp}^{\prime2}\right]}},\label{eq:ff_scalar}
\end{align}
where $x^{\prime}=x$, $x_{1}^{\prime}=x_{1}=1-x$ and $k_{\perp}^{\prime}=k_{\perp}+x_{2}q_{\perp}$
since we choose the coordinate system which satisfies $q^{+}=0$. 

\subsection{Transitions with axial vector diquarks}
With $\Gamma=-\frac{1}{\sqrt{3}}\gamma_{5}\slashed\epsilon^{*}(p_{2},\lambda_{2})$  in (\ref{eq:Melosh}) for the transitions with axial vector diquarks, the form factors can be obtained similarly \cite{Ke:2012wa}, given as
\begin{align}
f_{1}(q^{2})= & \frac{1}{8P^{+}P^{\prime+}}\int\frac{dx_{2}d^{2}k_{\perp}}{2(2\pi)^{3}}\frac{\varphi^{\prime}(x^{\prime},k_{\perp}^{\prime})\varphi(x,k_{\perp})}{6\sqrt{x_{1}x_{1}^{\prime}(p_{1}\cdot\bar{P}+m_{1}M_{0})(p_{1}^{\prime}\cdot\bar{P}^{\prime}+m_{1}^{\prime}M_{0}^{\prime})}},\nonumber \\
& \times{\rm Tr}[(\bar{\slashed P}+M_{0})\gamma^{+}(\bar{\slashed P^{\prime}}+M_{0}^{\prime})\gamma_{5}\gamma_{\alpha}(\slashed p_{1}^{\prime}+m_{1}^{\prime})\gamma^{+}(\slashed p_{1}+m_{1})\gamma_{5}\gamma_{\beta}](\frac{p_{2}^{\alpha}p_{2}^{\beta}}{m_{2}^{2}}-g^{\alpha\beta}),\nonumber \\
g_{1}(q^{2})= & \frac{1}{8P^{+}P^{\prime+}}\int\frac{dx_{2}d^{2}k_{\perp}}{2(2\pi)^{3}}\frac{\varphi^{\prime}(x^{\prime},k_{\perp}^{\prime})\varphi(x,k_{\perp})}{6\sqrt{x_{1}x_{1}^{\prime}(p_{1}\cdot\bar{P}+m_{1}M_{0})(p_{1}^{\prime}\cdot\bar{P}^{\prime}+m_{1}^{\prime}M_{0}^{\prime})}}\nonumber \\
& \times{\rm Tr}[(\bar{\slashed P}+M_{0})\gamma^{+}\gamma_{5}(\bar{\slashed P^{\prime}}+M_{0}^{\prime})\gamma_{5}\gamma_{\alpha}(\slashed p_{1}^{\prime}+m_{1}^{\prime})\gamma^{+}\gamma_{5}(\slashed p_{1}+m_{1})\gamma_{5}\gamma_{\beta}](\frac{p_{2}^{\alpha}p_{2}^{\beta}}{m_{2}^{2}}-g^{\alpha\beta}),\nonumber \\
\frac{f_{2}(q^{2})}{M}= & -\frac{1}{8P^{+}P^{\prime+}}\frac{iq_{\perp}^{i}}{q_{\perp}^{2}}\int\frac{dx_{2}d^{2}k_{\perp}}{2(2\pi)^{3}}\frac{\varphi^{\prime}(x^{\prime},k_{\perp}^{\prime})\varphi(x,k_{\perp})}{6\sqrt{x_{1}x_{1}^{\prime}(p_{1}\cdot\bar{P}+m_{1}M_{0})(p_{1}^{\prime}\cdot\bar{P}^{\prime}+m_{1}^{\prime}M_{0}^{\prime})}}\nonumber \\
& \times{\rm Tr}[(\bar{\slashed P}+M_{0})\sigma^{i+}(\bar{\slashed P^{\prime}}+M_{0}^{\prime})\gamma_{5}\gamma_{\alpha}(\slashed p_{1}^{\prime}+m_{1}^{\prime})\gamma^{+}(\slashed p_{1}+m_{1})\gamma_{5}\gamma_{\beta}](\frac{p_{2}^{\alpha}p_{2}^{\beta}}{m_{2}^{2}}-g^{\alpha\beta}),\nonumber \\
\frac{g_{2}(q^{2})}{M}= & \frac{1}{8P^{+}P^{\prime+}}\frac{iq_{\perp}^{i}}{q_{\perp}^{2}}\int\frac{dx_{2}d^{2}k_{\perp}}{2(2\pi)^{3}}\frac{\varphi^{\prime}(x^{\prime},k_{\perp}^{\prime})\varphi(x,k_{\perp})}{6\sqrt{x_{1}x_{1}^{\prime}(p_{1}\cdot\bar{P}+m_{1}M_{0})(p_{1}^{\prime}\cdot\bar{P}^{\prime}+m_{1}^{\prime}M_{0}^{\prime})}}\nonumber \\
& \times{\rm Tr}[(\bar{\slashed P}+M_{0})\sigma^{i+}\gamma_{5}(\bar{\slashed P^{\prime}}+M_{0}^{\prime})\gamma_{5}\gamma_{\alpha}(\slashed p_{1}^{\prime}+m_{1}^{\prime})\gamma^{+}\gamma_{5}(\slashed p_{1}+m_{1})\gamma_{5}\gamma_{\beta}](\frac{p_{2}^{\alpha}p_{2}^{\beta}}{m_{2}^{2}}-g^{\alpha\beta}).\label{eq:ff_axial}
\end{align}

\subsection{Mixing of transition form factors}

\begin{table}[!]
	\caption{Mixing coefficients of the transition matrix elements for the doubly charmed baryon decays. Taking the  $\Xi_{cc}^{++}\to \Lambda_{c}^{+}$ as an example, the physical transition matrix elements can be evaluated as follows.  $\langle\Lambda_{c}^{+}|(V-A)_{\mu}|\Xi_{cc}^{++}\rangle = c_{S} \langle d[cu]_{S}|(V-A)_{\mu}|c[cu]_{S}\rangle + c_{A} \langle d[cu]_{A}|(V-A)_{\mu}|c[cu]_{A}\rangle$ with $c_{S}=\sqrt{6}/4$ and $c_{A}=\sqrt{6}/4$.}
	\label{Tab:mixing_cc}
\begin{tabular}{c|c|c}
	\hline \hline
	& $\langle q_{1}[cq]_{S}|(V-A)_{\mu}|c[cq]_{S}\rangle$ & $\langle q_{1}[cq]_{A}|(V-A)_{\mu}|c[cq]_{A}\rangle$\tabularnewline
	\hline 
	$\langle\Lambda_{c}^{+}|(V-A)_{\mu}|\Xi_{cc}^{++}\rangle$ & $\frac{\sqrt{6}}{4}$ & $\frac{\sqrt{6}}{4}$\tabularnewline
	\hline 
	$\langle\Sigma_{c}^{+}|(V-A)_{\mu}|\Xi_{cc}^{++}\rangle$ & $-\frac{3\sqrt{2}}{4}$ & $\frac{\sqrt{2}}{4}$\tabularnewline
	\hline 
	$\langle\Xi_{c}^{+}|(V-A)_{\mu}|\Xi_{cc}^{++}\rangle$ & $\frac{\sqrt{6}}{4}$ & $\frac{\sqrt{6}}{4}$\tabularnewline
	\hline 
	$\langle\Xi_{c}^{\prime+}|(V-A)_{\mu}|\Xi_{cc}^{++}\rangle$ & $-\frac{3\sqrt{2}}{4}$ & $\frac{\sqrt{2}}{4}$\tabularnewline
	\hline 
	$\langle\Sigma_{c}^{0}|(V-A)_{\mu}|\Xi_{cc}^{+}\rangle$ & $-\frac{3}{2}$ & $\frac{1}{2}$\tabularnewline
	\hline 
	$\langle\Xi_{c}^{0}|(V-A)_{\mu}|\Xi_{cc}^{+}\rangle$ & $\frac{\sqrt{6}}{4}$ & $\frac{\sqrt{6}}{4}$\tabularnewline
	\hline 
	$\langle\Xi_{c}^{\prime0}|(V-A)_{\mu}|\Xi_{cc}^{+}\rangle$ & $-\frac{3\sqrt{2}}{4}$ & $\frac{\sqrt{2}}{4}$\tabularnewline
	\hline 
	$\langle\Xi_{c}^{0}|(V-A)_{\mu}|\Omega_{cc}^{+}\rangle$ & $-\frac{\sqrt{6}}{4}$ & $-\frac{\sqrt{6}}{4}$\tabularnewline
	\hline 
	$\langle\Xi_{c}^{\prime0}|(V-A)_{\mu}|\Omega_{cc}^{+}\rangle$ & $-\frac{3\sqrt{2}}{4}$ & $\frac{\sqrt{2}}{4}$\tabularnewline
	\hline 
	$\langle\Omega_{c}^{0}|(V-A)_{\mu}|\Omega_{cc}^{+}\rangle$ & $-\frac{3}{2}$ & $\frac{1}{2}$\tabularnewline
	\hline \hline
\end{tabular}
\end{table}

\begin{table}
	\caption{Same as Table \ref{Tab:mixing_cc} but for the doubly bottom baryon decays.}
	\label{Tab:mixing_bb}
\begin{tabular}{c|c|c}
	\hline \hline 
	& $\langle q_{1}[bq]_{S}|(V-A)_{\mu}|b[bq]_{S}\rangle$ & $\langle q_{1}[bq]_{A}|(V-A)_{\mu}|b[bq]_{A}\rangle$\tabularnewline
	\hline 
	$\langle\Sigma_{b}^{+}|(V-A)_{\mu}|\Xi_{bb}^{0}\rangle$ & $-\frac{3}{2}$ & $\frac{1}{2}$\tabularnewline
	\hline 
	$\langle\Xi_{bc}^{+}|(V-A)_{\mu}|\Xi_{bb}^{0}\rangle$ & $\frac{3\sqrt{2}}{4}$ & $\frac{\sqrt{2}}{4}$\tabularnewline
	\hline 
	$\langle\Xi_{bc}^{\prime+}|(V-A)_{\mu}|\Xi_{bb}^{0}\rangle$ & $-\frac{\sqrt{6}}{4}$ & $\frac{\sqrt{6}}{4}$\tabularnewline
	\hline 
	$\langle\Lambda_{b}^{0}|(V-A)_{\mu}|\Xi_{bb}^{-}\rangle$ & $-\frac{\sqrt{6}}{4}$ & $-\frac{\sqrt{6}}{4}$\tabularnewline
	\hline 
	$\langle\Sigma_{b}^{0}|(V-A)_{\mu}|\Xi_{bb}^{-}\rangle$ & $-\frac{3\sqrt{2}}{4}$ & $\frac{\sqrt{2}}{4}$\tabularnewline
	\hline 
	$\langle\Xi_{bc}^{0}|(V-A)_{\mu}|\Xi_{bb}^{-}\rangle$ & $\frac{3\sqrt{2}}{4}$ & $\frac{\sqrt{2}}{4}$\tabularnewline
	\hline 
	$\langle\Xi_{bc}^{\prime0}|(V-A)_{\mu}|\Xi_{bb}^{-}\rangle$ & $-\frac{\sqrt{6}}{4}$ & $\frac{\sqrt{6}}{4}$\tabularnewline
	\hline 
	$\langle\Xi_{b}^{0}|(V-A)_{\mu}|\Omega_{bb}^{-}\rangle$ & $-\frac{\sqrt{6}}{4}$ & $-\frac{\sqrt{6}}{4}$\tabularnewline
	\hline 
	$\langle\Xi_{b}^{\prime0}|(V-A)_{\mu}|\Omega_{bb}^{-}\rangle$ & $-\frac{3\sqrt{2}}{4}$ & $\frac{\sqrt{2}}{4}$\tabularnewline
	\hline 
	$\langle\Omega_{bc}^{0}|(V-A)_{\mu}|\Omega_{bb}^{-}\rangle$ & $\frac{3\sqrt{2}}{4}$ & $\frac{\sqrt{2}}{4}$\tabularnewline
	\hline 
	$\langle\Omega_{bc}^{\prime0}|(V-A)_{\mu}|\Omega_{bb}^{-}\rangle$ & $-\frac{\sqrt{6}}{4}$ & $\frac{\sqrt{6}}{4}$\tabularnewline
	\hline \hline 
\end{tabular}
\end{table}

\begin{table}
	\caption{Same as Table \ref{Tab:mixing_cc} but for the bottom-charm baryons with charm decays.}
	\label{Tab:mixing_bc_c}
\begin{tabular}{c|c|c}
	\hline \hline 
	& $\langle q_{1}[bq]_{S}|(V-A)_{\mu}|c[bq]_{S}\rangle$ & $\langle q_{1}[bq]_{A}|(V-A)_{\mu}|c[bq]_{A}\rangle$\tabularnewline
	\hline 
	$\langle\Lambda_{b}^{0}|(V-A)_{\mu}|\Xi_{bc}^{+}\rangle$ & $\frac{\sqrt{3}}{4}$ & $\frac{\sqrt{3}}{4}$\tabularnewline
	\hline 
	$\langle\Sigma_{b}^{0}|(V-A)_{\mu}|\Xi_{bc}^{+}\rangle$ & $-\frac{3}{4}$ & $\frac{1}{4}$\tabularnewline
	\hline 
	$\langle\Xi_{b}^{0}|(V-A)_{\mu}|\Xi_{bc}^{+}\rangle$ & $\frac{\sqrt{3}}{4}$ & $\frac{\sqrt{3}}{4}$\tabularnewline
	\hline 
	$\langle\Xi_{b}^{\prime0}|(V-A)_{\mu}|\Xi_{bc}^{+}\rangle$ & $-\frac{3}{4}$ & $\frac{1}{4}$\tabularnewline
	\hline 
	$\langle\Lambda_{b}^{0}|(V-A)_{\mu}|\Xi_{bc}^{\prime+}\rangle$ & $-\frac{1}{4}$ & $\frac{3}{4}$\tabularnewline
	\hline 
	$\langle\Sigma_{b}^{0}|(V-A)_{\mu}|\Xi_{bc}^{\prime+}\rangle$ & $\frac{\sqrt{3}}{4}$ & $\frac{\sqrt{3}}{4}$\tabularnewline
	\hline 
	$\langle\Xi_{b}^{0}|(V-A)_{\mu}|\Xi_{bc}^{\prime+}\rangle$ & $-\frac{1}{4}$ & $\frac{3}{4}$\tabularnewline
	\hline 
	$\langle\Xi_{b}^{\prime0}|(V-A)_{\mu}|\Xi_{bc}^{\prime+}\rangle$ & $\frac{\sqrt{3}}{4}$ & $\frac{\sqrt{3}}{4}$\tabularnewline
	\hline 
	$\langle\Sigma_{b}^{-}|(V-A)_{\mu}|\Xi_{bc}^{0}\rangle$ & $-\frac{3\sqrt{2}}{4}$ & $\frac{\sqrt{2}}{4}$\tabularnewline
	\hline 
	$\langle\Xi_{b}^{-}|(V-A)_{\mu}|\Xi_{bc}^{0}\rangle$ & $\frac{\sqrt{3}}{4}$ & $\frac{\sqrt{3}}{4}$\tabularnewline
	\hline 
	$\langle\Xi_{b}^{\prime-}|(V-A)_{\mu}|\Xi_{bc}^{0}\rangle$ & $-\frac{3}{4}$ & $\frac{1}{4}$\tabularnewline
	\hline 
	$\langle\Sigma_{b}^{-}|(V-A)_{\mu}|\Xi_{bc}^{\prime0}\rangle$ & $\frac{\sqrt{6}}{4}$ & $\frac{\sqrt{6}}{4}$\tabularnewline
	\hline 
	$\langle\Xi_{b}^{-}|(V-A)_{\mu}|\Xi_{bc}^{\prime0}\rangle$ & $-\frac{1}{4}$ & $\frac{3}{4}$\tabularnewline
	\hline 
	$\langle\Xi_{b}^{\prime-}|(V-A)_{\mu}|\Xi_{bc}^{\prime0}\rangle$ & $\frac{\sqrt{3}}{4}$ & $\frac{\sqrt{3}}{4}$\tabularnewline
	\hline 
	$\langle\Xi_{b}^{-}|(V-A)_{\mu}|\Omega_{bc}^{0}\rangle$ & $-\frac{\sqrt{3}}{4}$ & $-\frac{\sqrt{3}}{4}$\tabularnewline
	\hline 
	$\langle\Xi_{b}^{\prime-}|(V-A)_{\mu}|\Omega_{bc}^{0}\rangle$ & $-\frac{3}{4}$ & $\frac{1}{4}$\tabularnewline
	\hline 
	$\langle\Omega_{b}^{-}|(V-A)_{\mu}|\Omega_{bc}^{0}\rangle$ & $-\frac{3\sqrt{2}}{4}$ & $\frac{\sqrt{2}}{4}$\tabularnewline
	\hline 
	$\langle\Xi_{b}^{-}|(V-A)_{\mu}|\Omega_{bc}^{\prime0}\rangle$ & $\frac{1}{4}$ & $-\frac{3}{4}$\tabularnewline
	\hline 
	$\langle\Xi_{b}^{\prime-}|(V-A)_{\mu}|\Omega_{bc}^{\prime0}\rangle$ & $\frac{\sqrt{3}}{4}$ & $\frac{\sqrt{3}}{4}$\tabularnewline
	\hline 
	$\langle\Omega_{b}^{-}|(V-A)_{\mu}|\Omega_{bc}^{\prime0}\rangle$ & $\frac{\sqrt{6}}{4}$ & $\frac{\sqrt{6}}{4}$\tabularnewline
	\hline \hline
\end{tabular}
\end{table}

\begin{table}
	\caption{Same as Table \ref{Tab:mixing_cc} but for the bottom-charm baryons with $b$ decays.}
	\label{Tab:mixing_bc_b}
\begin{tabular}{c|c|c}
	\hline \hline 
	& $\langle q_{1}[cq]_{S}|(V-A)_{\mu}|b[cq]_{S}\rangle$ & $\langle q_{1}[cq]_{A}|(V-A)_{\mu}|b[cq]_{A}\rangle$\tabularnewline
	\hline 
	$\langle\Sigma_{c}^{++}|(V-A)_{\mu}|\Xi_{bc}^{+}\rangle$ & $-\frac{3\sqrt{2}}{4}$ & $\frac{\sqrt{2}}{4}$\tabularnewline
	\hline 
	$\langle\Xi_{cc}^{++}|(V-A)_{\mu}|\Xi_{bc}^{+}\rangle$ & $\frac{3\sqrt{2}}{4}$ & $\frac{\sqrt{2}}{4}$\tabularnewline
	\hline 
	$\langle\Sigma_{c}^{++}|(V-A)_{\mu}|\Xi_{bc}^{\prime+}\rangle$ & $-\frac{\sqrt{6}}{4}$ & $-\frac{\sqrt{6}}{4}$\tabularnewline
	\hline 
	$\langle\Xi_{cc}^{++}|(V-A)_{\mu}|\Xi_{bc}^{\prime+}\rangle$ & $\frac{\sqrt{6}}{4}$ & $-\frac{\sqrt{6}}{4}$\tabularnewline
	\hline 
	$\langle\Lambda_{c}^{+}|(V-A)_{\mu}|\Xi_{bc}^{0}\rangle$ & $-\frac{\sqrt{3}}{4}$ & $-\frac{\sqrt{3}}{4}$\tabularnewline
	\hline 
	$\langle\Sigma_{c}^{+}|(V-A)_{\mu}|\Xi_{bc}^{0}\rangle$ & $-\frac{3}{4}$ & $\frac{1}{4}$\tabularnewline
	\hline 
	$\langle\Xi_{cc}^{+}|(V-A)_{\mu}|\Xi_{bc}^{0}\rangle$ & $\frac{3\sqrt{2}}{4}$ & $\frac{\sqrt{2}}{4}$\tabularnewline
	\hline 
	$\langle\Lambda_{c}^{+}|(V-A)_{\mu}|\Xi_{bc}^{\prime0}\rangle$ & $-\frac{1}{4}$ & $\frac{3}{4}$\tabularnewline
	\hline 
	$\langle\Sigma_{c}^{+}|(V-A)_{\mu}|\Xi_{bc}^{\prime0}\rangle$ & $-\frac{\sqrt{3}}{4}$ & $-\frac{\sqrt{3}}{4}$\tabularnewline
	\hline 
	$\langle\Xi_{cc}^{+}|(V-A)_{\mu}|\Xi_{bc}^{\prime0}\rangle$ & $\frac{\sqrt{6}}{4}$ & $-\frac{\sqrt{6}}{4}$\tabularnewline
	\hline 
	$\langle\Xi_{c}^{+}|(V-A)_{\mu}|\Omega_{bc}^{0}\rangle$ & $-\frac{\sqrt{3}}{4}$ & $-\frac{\sqrt{3}}{4}$\tabularnewline
	\hline 
	$\langle\Xi_{c}^{\prime+}|(V-A)_{\mu}|\Omega_{bc}^{0}\rangle$ & $-\frac{3}{4}$ & $\frac{1}{4}$\tabularnewline
	\hline 
	$\langle\Omega_{cc}^{+}|(V-A)_{\mu}|\Omega_{bc}^{0}\rangle$ & $\frac{3\sqrt{2}}{4}$ & $\frac{\sqrt{2}}{4}$\tabularnewline
	\hline 
	$\langle\Xi_{c}^{+}|(V-A)_{\mu}|\Omega_{bc}^{\prime0}\rangle$ & $-\frac{1}{4}$ & $\frac{3}{4}$\tabularnewline
	\hline 
	$\langle\Xi_{c}^{\prime+}|(V-A)_{\mu}|\Omega_{bc}^{\prime0}\rangle$ & $-\frac{\sqrt{3}}{4}$ & $-\frac{\sqrt{3}}{4}$\tabularnewline
	\hline 
	$\langle\Omega_{cc}^{+}|(V-A)_{\mu}|\Omega_{bc}^{\prime0}\rangle$ & $\frac{\sqrt{6}}{4}$ & $-\frac{\sqrt{6}}{4}$\tabularnewline
	\hline \hline
\end{tabular}
\end{table}
	
It should be noted that, in the above calculation, what we have obtained 
is simply the   transition matrix elements $\langle q_{1}[Q_{2}q]_{S}|(V-A)_{\mu}|Q_{1}[Q_{2}q]_{S}\rangle$
or $\langle q_{1}[Q_{2}q]_{A}|(V-A)_{\mu}|Q_{1}[Q_{2}q]_{A}\rangle$
with $S$ and $A$ denoting  a   scalar or axial vector diquark spectator, respectively. The hadronic transition matrix elements are actually linear combinations of the ones with the scalar or axial vector diquarks,
\begin{equation}
\langle B^{\prime}|(V-A)_{\mu}|B\rangle=c_{S}\langle q_{1}[Q_{2}q]_{S}|(V-A)_{\mu}|Q_{1}[Q_{2}q]_{S}\rangle+c_{A}\langle q_{1}[Q_{2}q]_{A}|(V-A)_{\mu}|Q_{1}[Q_{2}q]_{A}\rangle,\label{eq:csca}
\end{equation}
where the coefficients $c_{S,A}$ are determined by the wave functions of the initial and final states.

For the doubly charmed baryons, the wave functions are 
\begin{align}
\mathcal{B}_{cc}=\frac{1}{\sqrt{2}}\left[\left(-\frac{\sqrt{3}}{2}c^{1}(c^{2}q)_{S}+\frac{1}{2}c^{1}(c^{2}q)_{A}\right)+(c^{1}\leftrightarrow c^{2})\right],
\end{align}
with $q=u$, $d$ or $s$ for $\Xi_{cc}^{++}$, $\Xi_{cc}^{+}$ or $\Omega_{cc}^{+}$, respectively. The superscripts describe the symmetry between the two charm quarks. 
The ones of the doubly bottom baryons are replaced by $c\to b$. For the bottom-charm baryons, there are two sets of states, with $bc$ as a scalar or an axial vector diquarks. 
The wave functions of bottom-charm baryons with axial vector $bc$ diquark are
\begin{align}
\mathcal{B}_{bc} & =  -\frac{\sqrt{3}}{2}b(cq)_{S}+\frac{1}{2}b(cq)_{A}  =  -\frac{\sqrt{3}}{2}c(bq)_{S}+\frac{1}{2}c(bq)_{A},
\end{align}
while those with a scalar $bc$ diquark are given as
\begin{align}
\mathcal{B}_{bc}^{\prime} & =  -\frac{1}{2}b(cq)_{S}-\frac{\sqrt{3}}{2}b(cq)_{A}  = \frac{1}{2}c(bq)_{S}+\frac{\sqrt{3}}{2}c(bq)_{A} ,
\end{align}
with  $q=u$, $d$ or $s$ for $\Xi_{bc}^{(\prime)+}$, $\Xi_{bc}^{(\prime)0}$ or $\Omega_{bc}^{(\prime)0}$, respectively.

The wave functions of the anti-triplet singly charmed baryons are
\begin{align}
\Lambda_{c}^{+} & = -\frac{1}{2}d(cu)_{S}+\frac{\sqrt{3}}{2}d(cu)_{A}  = \frac{1}{2}u(cd)_{S}-\frac{\sqrt{3}}{2}u(cd)_{A},\nonumber \\
\Xi_{c}^{+} & =  -\frac{1}{2}s(cu)_{S}+\frac{\sqrt{3}}{2}s(cu)_{A}   = \frac{1}{2}u(cs)_{S}-\frac{\sqrt{3}}{2}u(cs)_{A} ,\\
\Xi_{c}^{0} & =  -\frac{1}{2}s(cd)_{S}+\frac{\sqrt{3}}{2}s(cd)_{A}  =  \frac{1}{2}d(cs)_{S}-\frac{\sqrt{3}}{2}d(cs)_{A}.\nonumber 
\end{align}
For the sextet of singly charmed baryons, the wave functions are
\begin{align}
\Sigma_{c}^{++} & =\frac{1}{\sqrt{2}}\left[\frac{\sqrt{3}}{2}u^{1}(cu^{2})_{S}+\frac{1}{2}u^{1}(cu^{2})_{A}+(u^{1}\leftrightarrow u^{2})\right],\nonumber \\
\Sigma_{c}^{+} & =  \frac{\sqrt{3}}{2}d(cu)_{S}+\frac{1}{2}d(cu)_{A}  =   \frac{\sqrt{3}}{2}u(cd)_{S}+\frac{1}{2}u(cd)_{A}, \\
\Sigma_{c}^{0} & =\frac{1}{\sqrt{2}}\left[\frac{\sqrt{3}}{2}d^{1}(cd^{2})_{S}+\frac{1}{2}d^{1}(cd^{2})_{A}+(d^{1}\leftrightarrow d^{2})\right],\nonumber \\
\Xi_{c}^{\prime+} & = \frac{\sqrt{3}}{2}s(cu)_{S}+\frac{1}{2}s(cu)_{A}  = \frac{\sqrt{3}}{2}u(cs)_{S}+\frac{1}{2}u(cs)_{A},\nonumber \\
\Xi_{c}^{\prime0} & = \frac{\sqrt{3}}{2}s(cd)_{S}+\frac{1}{2}s(cd)_{A}   = \frac{\sqrt{3}}{2}d(cs)_{S}+\frac{1}{2}d(cs)_{A},\nonumber \\
\Omega_{c}^{0} & =\frac{1}{\sqrt{2}}\left[\frac{\sqrt{3}}{2}s^{2}(cs^{1})_{S}+\frac{1}{2}s^{2}(cs^{1})_{A}+(s^{1}\leftrightarrow s^{2})\right].\nonumber
\end{align}
The ones of the singly bottom baryons are similar with the replacement of $c$ by $b$.  

With the wave functions given above, the mixing coefficients for the transition matrix elements in (\ref{eq:csca}) are given  in Tables \ref{Tab:mixing_cc}, \ref{Tab:mixing_bb}, \ref{Tab:mixing_bc_c} and \ref{Tab:mixing_bc_b}.

\subsection{Numerical results for transition form factors}
\label{subsec:ff}
	
\begin{table}[!]
\caption{The shape parameters $\beta$'s in the Gaussian-type wave functions Eq.~(\ref{Gaussian_type}) and the final baryon masses (in units of GeV).}
\label{Tab:Beta}
\begin{tabular}{c|c|c|c|c|c|c|c|c|c}
\hline\hline
~~~$\beta_{u[cq]}$ & $\beta_{d[cq]}$ & $\beta_{s[cq]}$ & $\beta_{c[cq]}$ & $\beta_{b[cq]}$& $\beta_{u[bq]}$ & $\beta_{d[bq]}$ & $\beta_{s[bq]}$ & $\beta_{c[bq]}$ & $\beta_{b[bq]}$~~~\tabularnewline
\hline
$0.470$ & $0.470$ & $0.535$ & $0.753$ & $0.886$ &$0.562$ & $0.562$ & $0.623$ & $0.886$ & $1.472$\tabularnewline
	\hline \end{tabular}\\\begin{tabular}{c|c|c|c|c|c|c|c|ccc} \hline
	~$m_{\Lambda_{c}^{+}}$~  & ~$m_{\Sigma_{c}^{++}}$~  & ~$m_{\Sigma_{c}^{+}}$~  & ~$m_{\Sigma_{c}^{0}}$~  & ~$m_{\Xi_{c}^{+}}$~  & ~$m_{\Xi_{c}^{\prime+}}$~  & ~$m_{\Xi_{c}^{0}}$~  & ~$m_{\Xi_{c}^{\prime0}}$~  & ~$m_{\Omega_{c}^{0}}$~ \tabularnewline
	\hline 
	$2.286$  & $2.454$  & $2.453$  & $2.454$  & $2.468$  & $2.576$  & $2.471$  & $2.578$  & $2.695$ \tabularnewline
	\hline 
	$m_{\Lambda_{b}^{0}}$  & $m_{\Sigma_{b}^{+}}$  & $m_{\Sigma_{b}^{0}}$  & $m_{\Sigma_{b}^{-}}$  & $m_{\Xi_{b}^{0}}$ & $m_{\Xi_{b}^{\prime0}}$ & $m_{\Xi_{b}^{-}}$  & $m_{\Xi_{b}^{\prime-}}$  & $m_{\Omega_{b}^{-}}$\tabularnewline
	\hline 
	$5.620$  & $5.811$  & $5.814$  & $5.816$  & $5.793$ & $5.935$ & $5.795$  & $5.935$  & $6.046$\tabularnewline
	\hline \hline
\end{tabular}
\end{table}

\begin{table}[!]
	\caption{Transition form factors of doubly charmed baryon decays with scalar ($0^{+}$) diquarks. The formula of (\ref{eq:main_fit_formula}) is adopted.}
	\label{Tab:ff_cc_scalar}
\begin{tabular}{c|c|c|c|c|c|c|c}
	\hline\hline
	$F$ &$F(0)$ &$m_{\rm{fit}}$ &$\delta$ &$F$ &$F(0)$ &$m_{\rm{fit}}$ &$\delta$\
	
	\\ \hline
	$f_{1}^{\Xi_{cc}^{++}\to\Lambda_{c}^{+}}$ &$0.653$ &$1.72$ &$0.27$ &$f_{2}^{
		
		\Xi_{cc}^{++}\to\Lambda_{c}^{+}}$ &$-0.738$ &$1.56$ &$0.32$ \\$g_{1}^{
		
		\Xi_{cc}^{++}\to\Lambda_{c}^{+}}$ &$0.533$ &$2.03$ &$0.38$ &$g_{2}^{
		
		\Xi_{cc}^{++}\to\Lambda_{c}^{+}}$ &$
	-0.053$ &$1.12$ &$1.10$ \\ \hline
	$f_{1}^{\Xi_{cc}^{++}\to\Sigma_{c}^{+}}$ &$0.653$ &$1.72$ &$0.27$ &$f_{2}^{
		
		\Xi_{cc}^{++}\to\Sigma_{c}^{+}}$ &$-0.738$ &$1.56$ &$0.32$ \\$g_{1}^{
		
		\Xi_{cc}^{++}\to\Sigma_{c}^{+}}$ &$0.533$ &$2.03$ &$0.38$ &$g_{2}^{
		
		\Xi_{cc}^{++}\to\Sigma_{c}^{+}}$ &$
	-0.053$ &$1.12$ &$1.10$ \\ \hline
	$f_{1}^{\Xi_{cc}^{++}\to\Xi_{c}^{+}}$ &$0.754$ &$1.84$ &$0.25$ &$f_{2}^{
		
		\Xi_{cc}^{++}\to\Xi_{c}^{+}}$ &$-0.782$ &$1.67$ &$0.30$ \\$g_{1}^{
		
		\Xi_{cc}^{++}\to\Xi_{c}^{+}}$ &$0.620$ &$2.16$ &$0.35$ &$g_{2}^{
		
		\Xi_{cc}^{++}\to\Xi_{c}^{+}}$ &$
	-0.080$ &$1.29$ &$0.52$ \\ \hline
	$f_{1}^{\Xi_{cc}^{++}\to\Xi_{c}^{\prime+}}$ &$0.754$ &$1.84$ &$0.25$ &$f_{2}^{
		
		\Xi_{cc}^{++}\to\Xi_{c}^{\prime+}}$ &$-0.782$ &$1.67$ &$0.30$ \\$g_{1}^{
		
		\Xi_{cc}^{++}\to\Xi_{c}^{\prime+}}$ &$0.620$ &$2.16$ &$0.35$ &$g_{2}^{
		
		\Xi_{cc}^{++}\to\Xi_{c}^{\prime+}}$ &$
	-0.080$ &$1.29$ &$0.52$ \\ \hline
	$f_{1}^{\Xi_{cc}^{+}\to\Sigma_{c}^{0}}$ &$0.653$ &$1.72$ &$0.27$ &$f_{2}^{
		
		\Xi_{cc}^{+}\to\Sigma_{c}^{0}}$ &$-0.738$ &$1.56$ &$0.32$ \\$g_{1}^{
		
		\Xi_{cc}^{+}\to\Sigma_{c}^{0}}$ &$0.533$ &$2.03$ &$0.38$ &$g_{2}^{
		
		\Xi_{cc}^{+}\to\Sigma_{c}^{0}}$ &$
	-0.053$ &$1.12$ &$1.10$ \\ \hline
	$f_{1}^{\Xi_{cc}^{+}\to\Xi_{c}^{0}}$ &$0.754$ &$1.84$ &$0.25$ &$f_{2}^{
		
		\Xi_{cc}^{+}\to\Xi_{c}^{0}}$ &$-0.782$ &$1.67$ &$0.30$ \\$g_{1}^{
		
		\Xi_{cc}^{+}\to\Xi_{c}^{0}}$ &$0.620$ &$2.16$ &$0.35$ &$g_{2}^{
		
		\Xi_{cc}^{+}\to\Xi_{c}^{0}}$ &$
	-0.080$ &$1.29$ &$0.52$ \\ \hline
	$f_{1}^{\Xi_{cc}^{+}\to\Xi_{c}^{\prime0}}$ &$0.754$ &$1.84$ &$0.25$ &$f_{2}^{
		
		\Xi_{cc}^{+}\to\Xi_{c}^{\prime0}}$ &$-0.782$ &$1.67$ &$0.30$ \\$g_{1}^{
		
		\Xi_{cc}^{+}\to\Xi_{c}^{\prime0}}$ &$0.620$ &$2.16$ &$0.35$ &$g_{2}^{
		
		\Xi_{cc}^{+}\to\Xi_{c}^{\prime0}}$ &$
	-0.080$ &$1.29$ &$0.52$ \\ \hline
	$f_{1}^{\Omega_{cc}^{+}\to\Xi_{c}^{0}}$ &$0.646$ &$1.68$ &$0.28$ &$f_{2}^{
		
		\Omega_{cc}^{+}\to\Xi_{c}^{0}}$ &$-0.770$ &$1.54$ &$0.33$ \\$g_{1}^{
		
		\Omega_{cc}^{+}\to\Xi_{c}^{0}}$ &$0.528$ &$1.99$ &$0.40$ &$g_{2}^{
		
		\Omega_{cc}^{+}\to\Xi_{c}^{0}}$ &$
	-0.060$ &$1.12$ &$1.02$ \\ \hline
	$f_{1}^{\Omega_{cc}^{+}\to\Xi_{c}^{\prime0}}$ &$0.646$ &$1.68$ &$0.28$ &
	
	$f_{2}^{\Omega_{cc}^{+}\to\Xi_{c}^{\prime0}}$ &$-0.770$ &$1.54$ &$0.33$ \\
	
	$g_{1}^{\Omega_{cc}^{+}\to\Xi_{c}^{\prime0}}$ &$0.528$ &$1.99$ &$0.40$ &
	
	$g_{2}^{\Omega_{cc}^{+}\to\Xi_{c}^{\prime0}}$ &$
	-0.060$ &$1.12$ &$1.02$ \\ \hline
	$f_{1}^{\Omega_{cc}^{+}\to\Omega_{c}^{0}}$ &$0.748$ &$1.80$ &$0.27$ &$f_{2}^{
		
		\Omega_{cc}^{+}\to\Omega_{c}^{0}}$ &$-0.819$ &$1.64$ &$0.32$ \\$g_{1}^{
		
		\Omega_{cc}^{+}\to\Omega_{c}^{0}}$ &$0.615$ &$2.11$ &$0.36$ &$g_{2}^{
		
		\Omega_{cc}^{+}\to\Omega_{c}^{0}}$ &$
	-0.088$ &$1.28$ &$0.52$ \\ \hline
	\hline
\end{tabular}
\end{table}
\begin{table}[!]
	\caption{Transition form factors of doubly charmed baryon decays with the axial vector $(1^{+})$ diquarks. The formula of (\ref{eq:auxiliary_fit_formula}) is adopted for the ones with asterisk, and that of (\ref{eq:main_fit_formula}) for all the others.}
	\label{Tab:ff_cc_axial}
\begin{tabular}{c|c|c|c|c|c|c|c}
	\hline\hline
	$F$ &$F(0)$ &$m_{\rm{fit}}$ &$\delta$ &$F$ &$F(0)$ &$m_{\rm{fit}}$ &$\delta$\
	
	\\ \hline
	$f_{1}^{\Xi_{cc}^{++}\to\Lambda_{c}^{+}}$ &$0.637$ &$1.49$ &$0.37$ &$f_{2}^{
		
		\Xi_{cc}^{++}\to\Lambda_{c}^{+}}$ &$0.725$ &$1.53$ &$0.32$ \\$g_{1}^{
		
		\Xi_{cc}^{++}\to\Lambda_{c}^{+}}$ &$-0.167$ &$1.99$ &$0.23$ &$g_{2}^{
		
		\Xi_{cc}^{++}\to\Lambda_{c}^{+}}$ &$
	-0.028^{*}$ &$2.03^{*}$ &$2.62^{*}$ \\ \hline
	$f_{1}^{\Xi_{cc}^{++}\to\Sigma_{c}^{+}}$ &$0.637$ &$1.49$ &$0.37$ &$f_{2}^{
		
		\Xi_{cc}^{++}\to\Sigma_{c}^{+}}$ &$0.725$ &$1.53$ &$0.32$ \\$g_{1}^{
		
		\Xi_{cc}^{++}\to\Sigma_{c}^{+}}$ &$-0.167$ &$1.99$ &$0.23$ &$g_{2}^{
		
		\Xi_{cc}^{++}\to\Sigma_{c}^{+}}$ &$
	-0.028^{*}$ &$2.03^{*}$ &$2.62^{*}$ \\ \hline
	$f_{1}^{\Xi_{cc}^{++}\to\Xi_{c}^{+}}$ &$0.739$ &$1.58$ &$0.36$ &$f_{2}^{
		
		\Xi_{cc}^{++}\to\Xi_{c}^{+}}$ &$0.801$ &$1.62$ &$0.31$ \\$g_{1}^{
		
		\Xi_{cc}^{++}\to\Xi_{c}^{+}}$ &$-0.198$ &$2.10$ &$0.21$ &$g_{2}^{
		
		\Xi_{cc}^{++}\to\Xi_{c}^{+}}$ &$
	-0.018^{*}$ &$1.62^{*}$ &$1.37^{*}$ \\ \hline
	$f_{1}^{\Xi_{cc}^{++}\to\Xi_{c}^{\prime+}}$ &$0.739$ &$1.58$ &$0.36$ &$f_{2}^{
		
		\Xi_{cc}^{++}\to\Xi_{c}^{\prime+}}$ &$0.801$ &$1.62$ &$0.31$ \\$g_{1}^{
		
		\Xi_{cc}^{++}\to\Xi_{c}^{\prime+}}$ &$-0.198$ &$2.10$ &$0.21$ &$g_{2}^{
		
		\Xi_{cc}^{++}\to\Xi_{c}^{\prime+}}$ &$
	-0.018^{*}$ &$1.62^{*}$ &$1.37^{*}$ \\ \hline
	$f_{1}^{\Xi_{cc}^{+}\to\Sigma_{c}^{0}}$ &$0.637$ &$1.49$ &$0.37$ &$f_{2}^{
		
		\Xi_{cc}^{+}\to\Sigma_{c}^{0}}$ &$0.725$ &$1.53$ &$0.32$ \\$g_{1}^{
		
		\Xi_{cc}^{+}\to\Sigma_{c}^{0}}$ &$-0.167$ &$1.99$ &$0.23$ &$g_{2}^{
		
		\Xi_{cc}^{+}\to\Sigma_{c}^{0}}$ &$
	-0.028^{*}$ &$2.03^{*}$ &$2.62^{*}$ \\ \hline
	$f_{1}^{\Xi_{cc}^{+}\to\Xi_{c}^{0}}$ &$0.739$ &$1.58$ &$0.36$ &$f_{2}^{
		
		\Xi_{cc}^{+}\to\Xi_{c}^{0}}$ &$0.801$ &$1.62$ &$0.31$ \\$g_{1}^{
		
		\Xi_{cc}^{+}\to\Xi_{c}^{0}}$ &$-0.198$ &$2.10$ &$0.21$ &$g_{2}^{
		
		\Xi_{cc}^{+}\to\Xi_{c}^{0}}$ &$
	-0.018^{*}$ &$1.62^{*}$ &$1.37^{*}$ \\ \hline
	$f_{1}^{\Xi_{cc}^{+}\to\Xi_{c}^{\prime0}}$ &$0.739$ &$1.58$ &$0.36$ &$f_{2}^{
		
		\Xi_{cc}^{+}\to\Xi_{c}^{\prime0}}$ &$0.801$ &$1.62$ &$0.31$ \\$g_{1}^{
		
		\Xi_{cc}^{+}\to\Xi_{c}^{\prime0}}$ &$-0.198$ &$2.10$ &$0.21$ &$g_{2}^{
		
		\Xi_{cc}^{+}\to\Xi_{c}^{\prime0}}$ &$
	-0.018^{*}$ &$1.62^{*}$ &$1.37^{*}$ \\ \hline
	$f_{1}^{\Omega_{cc}^{+}\to\Xi_{c}^{0}}$ &$0.632$ &$1.47$ &$0.38$ &$f_{2}^{
		
		\Omega_{cc}^{+}\to\Xi_{c}^{0}}$ &$0.734$ &$1.52$ &$0.33$ \\$g_{1}^{
		
		\Omega_{cc}^{+}\to\Xi_{c}^{0}}$ &$-0.165$ &$1.97$ &$0.27$ &$g_{2}^{
		
		\Omega_{cc}^{+}\to\Xi_{c}^{0}}$ &$
	-0.031^{*}$ &$2.32^{*}$ &$3.92^{*}$ \\ \hline
	$f_{1}^{\Omega_{cc}^{+}\to\Xi_{c}^{\prime0}}$ &$0.632$ &$1.47$ &$0.38$ &
	
	$f_{2}^{\Omega_{cc}^{+}\to\Xi_{c}^{\prime0}}$ &$0.734$ &$1.52$ &$0.33$ \\
	
	$g_{1}^{\Omega_{cc}^{+}\to\Xi_{c}^{\prime0}}$ &$-0.165$ &$1.97$ &$0.27$ &
	
	$g_{2}^{\Omega_{cc}^{+}\to\Xi_{c}^{\prime0}}$ &$
	-0.031^{*}$ &$2.32^{*}$ &$3.92^{*}$ \\ \hline
	$f_{1}^{\Omega_{cc}^{+}\to\Omega_{c}^{0}}$ &$0.735$ &$1.57$ &$0.37$ &$f_{2}^{
		
		\Omega_{cc}^{+}\to\Omega_{c}^{0}}$ &$0.812$ &$1.61$ &$0.32$ \\$g_{1}^{
		
		\Omega_{cc}^{+}\to\Omega_{c}^{0}}$ &$-0.196$ &$2.08$ &$0.24$ &$g_{2}^{
		
		\Omega_{cc}^{+}\to\Omega_{c}^{0}}$ &$
	-0.021^{*}$ &$1.79^{*}$ &$1.77^{*}$ \\ \hline
	\hline
\end{tabular} \end{table}
\begin{table}[!]
	\caption{Same as Table~\ref{Tab:ff_cc_scalar} but for the doubly bottom baryon decays with scalar ($0^{+}$) diquarks.}
	\label{Tab:ff_bb_scalar}
\begin{tabular}{c|c|c|c|c|c|c|c}
	\hline\hline
	$F$ &$F(0)$ &$m_{\rm{fit}}$ &$\delta$ &$F$ &$F(0)$ &$m_{\rm{fit}}$ &$\delta$\
	
	\\ \hline
	$f_{1}^{\Xi_{bb}^{0}\to\Sigma_{b}^{+}}$ &$0.084$ &$3.11$ &$0.80$ &$f_{2}^{
		
		\Xi_{bb}^{0}\to\Sigma_{b}^{+}}$ &$-0.106$ &$3.03$ &$0.88$ \\$g_{1}^{
		
		\Xi_{bb}^{0}\to\Sigma_{b}^{+}}$ &$0.078$ &$3.24$ &$0.80$ &$g_{2}^{
		
		\Xi_{bb}^{0}\to\Sigma_{b}^{+}}$ &$
	0.007$ &$5.65$ &$4.89$ \\ \hline
	$f_{1}^{\Xi_{bb}^{0}\to\Xi_{bc}^{+}}$ &$0.419$ &$3.76$ &$0.56$ &$f_{2}^{
		
		\Xi_{bb}^{0}\to\Xi_{bc}^{+}}$ &$-0.395$ &$3.61$ &$0.60$ \\$g_{1}^{
		
		\Xi_{bb}^{0}\to\Xi_{bc}^{+}}$ &$0.392$ &$3.91$ &$0.57$ &$g_{2}^{
		
		\Xi_{bb}^{0}\to\Xi_{bc}^{+}}$ &$
	0.009^{*}$ &$12.20^{*}$ &$41.70^{*}$ \\ \hline
	$f_{1}^{\Xi_{bb}^{0}\to\Xi_{bc}^{\prime+}}$ &$0.419$ &$3.76$ &$0.56$ &$f_{2}^{
		
		\Xi_{bb}^{0}\to\Xi_{bc}^{\prime+}}$ &$-0.395$ &$3.61$ &$0.60$ \\$g_{1}^{
		
		\Xi_{bb}^{0}\to\Xi_{bc}^{\prime+}}$ &$0.392$ &$3.91$ &$0.57$ &$g_{2}^{
		
		\Xi_{bb}^{0}\to\Xi_{bc}^{\prime+}}$ &$
	0.009^{*}$ &$12.20^{*}$ &$41.70^{*}$ \\ \hline
	$f_{1}^{\Xi_{bb}^{-}\to\Lambda_{b}^{0}}$ &$0.084$ &$3.11$ &$0.80$ &$f_{2}^{
		
		\Xi_{bb}^{-}\to\Lambda_{b}^{0}}$ &$-0.106$ &$3.03$ &$0.88$ \\$g_{1}^{
		
		\Xi_{bb}^{-}\to\Lambda_{b}^{0}}$ &$0.078$ &$3.24$ &$0.80$ &$g_{2}^{
		
		\Xi_{bb}^{-}\to\Lambda_{b}^{0}}$ &$
	0.007$ &$5.65$ &$4.89$ \\ \hline
	$f_{1}^{\Xi_{bb}^{-}\to\Sigma_{b}^{0}}$ &$0.084$ &$3.11$ &$0.80$ &$f_{2}^{
		
		\Xi_{bb}^{-}\to\Sigma_{b}^{0}}$ &$-0.106$ &$3.03$ &$0.88$ \\$g_{1}^{
		
		\Xi_{bb}^{-}\to\Sigma_{b}^{0}}$ &$0.078$ &$3.24$ &$0.80$ &$g_{2}^{
		
		\Xi_{bb}^{-}\to\Sigma_{b}^{0}}$ &$
	0.007$ &$5.65$ &$4.89$ \\ \hline
	$f_{1}^{\Xi_{bb}^{-}\to\Xi_{bc}^{0}}$ &$0.419$ &$3.76$ &$0.56$ &$f_{2}^{
		
		\Xi_{bb}^{-}\to\Xi_{bc}^{0}}$ &$-0.395$ &$3.61$ &$0.60$ \\$g_{1}^{
		
		\Xi_{bb}^{-}\to\Xi_{bc}^{0}}$ &$0.392$ &$3.91$ &$0.57$ &$g_{2}^{
		
		\Xi_{bb}^{-}\to\Xi_{bc}^{0}}$ &$
	0.009^{*}$ &$12.20^{*}$ &$41.70^{*}$ \\ \hline
	$f_{1}^{\Xi_{bb}^{-}\to\Xi_{bc}^{\prime0}}$ &$0.419$ &$3.76$ &$0.56$ &$f_{2}^{
		
		\Xi_{bb}^{-}\to\Xi_{bc}^{\prime0}}$ &$-0.395$ &$3.61$ &$0.60$ \\$g_{1}^{
		
		\Xi_{bb}^{-}\to\Xi_{bc}^{\prime0}}$ &$0.392$ &$3.91$ &$0.57$ &$g_{2}^{
		
		\Xi_{bb}^{-}\to\Xi_{bc}^{\prime0}}$ &$
	0.009^{*}$ &$12.20^{*}$ &$41.70^{*}$ \\ \hline
	$f_{1}^{\Omega_{bb}^{-}\to\Xi_{b}^{0}}$ &$0.082$ &$3.09$ &$0.82$ &$f_{2}^{
		
		\Omega_{bb}^{-}\to\Xi_{b}^{0}}$ &$-0.105$ &$3.02$ &$0.90$ \\$g_{1}^{
		
		\Omega_{bb}^{-}\to\Xi_{b}^{0}}$ &$0.076$ &$3.22$ &$0.81$ &$g_{2}^{
		
		\Omega_{bb}^{-}\to\Xi_{b}^{0}}$ &$
	0.007$ &$5.85$ &$5.79$ \\ \hline
	$f_{1}^{\Omega_{bb}^{-}\to\Xi_{b}^{\prime0}}$ &$0.082$ &$3.09$ &$0.82$ &
	
	$f_{2}^{\Omega_{bb}^{-}\to\Xi_{b}^{\prime0}}$ &$-0.105$ &$3.02$ &$0.90$ \\
	
	$g_{1}^{\Omega_{bb}^{-}\to\Xi_{b}^{\prime0}}$ &$0.076$ &$3.22$ &$0.81$ &
	
	$g_{2}^{\Omega_{bb}^{-}\to\Xi_{b}^{\prime0}}$ &$
	0.007$ &$5.85$ &$5.79$ \\ \hline
	$f_{1}^{\Omega_{bb}^{-}\to\Omega_{bc}^{0}}$ &$0.414$ &$3.73$ &$0.57$ &$f_{2}^{
		
		\Omega_{bb}^{-}\to\Omega_{bc}^{0}}$ &$-0.399$ &$3.59$ &$0.61$ \\$g_{1}^{
		
		\Omega_{bb}^{-}\to\Omega_{bc}^{0}}$ &$0.387$ &$3.88$ &$0.57$ &$g_{2}^{
		
		\Omega_{bb}^{-}\to\Omega_{bc}^{0}}$ &$
	0.007^{*}$ &$5.31^{*}$ &$2.84^{*}$ \\ \hline
	$f_{1}^{\Omega_{bb}^{-}\to\Omega_{bc}^{\prime0}}$ &$0.414$ &$3.73$ &$0.57$ &
	
	$f_{2}^{\Omega_{bb}^{-}\to\Omega_{bc}^{\prime0}}$ &$-0.399$ &$3.59$ &$0.61$ 
	
	\\$g_{1}^{\Omega_{bb}^{-}\to\Omega_{bc}^{\prime0}}$ &$0.387$ &$3.88$ &$0.57
	
	$ &$g_{2}^{\Omega_{bb}^{-}\to\Omega_{bc}^{\prime0}}$ &$
	0.007^{*}$ &$5.31^{*}$ &$2.84^{*}$ \\ \hline
	\hline
\end{tabular} \end{table}
\begin{table}
	\caption{Same as Table~\ref{Tab:ff_cc_axial} but for the doubly bottom baryons  with axial vector $(1^{+})$ diquarks.}
	\label{Tab:ff_bb_axial}
\begin{tabular}{c|c|c|c|c|c|c|c}
	\hline\hline
	$F$ &$F(0)$ &$m_{\rm{fit}}$ &$\delta$ &$F$ &$F(0)$ &$m_{\rm{fit}}$ &$\delta$\
	
	\\ \hline
	$f_{1}^{\Xi_{bb}^{0}\to\Sigma_{b}^{+}}$ &$0.083$ &$2.99$ &$0.97$ &$f_{2}^{
		
		\Xi_{bb}^{0}\to\Sigma_{b}^{+}}$ &$0.105$ &$3.03$ &$0.88$ \\$g_{1}^{
		
		\Xi_{bb}^{0}\to\Sigma_{b}^{+}}$ &$-0.019$ &$3.38$ &$0.75$ &$g_{2}^{
		
		\Xi_{bb}^{0}\to\Sigma_{b}^{+}}$ &$
	-0.026$ &$3.27$ &$0.86$ \\ \hline
	$f_{1}^{\Xi_{bb}^{0}\to\Xi_{bc}^{+}}$ &$0.414$ &$3.52$ &$0.64$ &$f_{2}^{
		
		\Xi_{bb}^{0}\to\Xi_{bc}^{+}}$ &$0.448$ &$3.59$ &$0.60$ \\$g_{1}^{
		
		\Xi_{bb}^{0}\to\Xi_{bc}^{+}}$ &$-0.116$ &$4.05$ &$0.55$ &$g_{2}^{
		
		\Xi_{bb}^{0}\to\Xi_{bc}^{+}}$ &$
	-0.063$ &$3.90$ &$0.60$ \\ \hline
	$f_{1}^{\Xi_{bb}^{0}\to\Xi_{bc}^{\prime+}}$ &$0.414$ &$3.52$ &$0.64$ &$f_{2}^{
		
		\Xi_{bb}^{0}\to\Xi_{bc}^{\prime+}}$ &$0.448$ &$3.59$ &$0.60$ \\$g_{1}^{
		
		\Xi_{bb}^{0}\to\Xi_{bc}^{\prime+}}$ &$-0.116$ &$4.05$ &$0.55$ &$g_{2}^{
		
		\Xi_{bb}^{0}\to\Xi_{bc}^{\prime+}}$ &$
	-0.063$ &$3.90$ &$0.60$ \\ \hline
	$f_{1}^{\Xi_{bb}^{-}\to\Lambda_{b}^{0}}$ &$0.083$ &$2.99$ &$0.97$ &$f_{2}^{
		
		\Xi_{bb}^{-}\to\Lambda_{b}^{0}}$ &$0.105$ &$3.03$ &$0.88$ \\$g_{1}^{
		
		\Xi_{bb}^{-}\to\Lambda_{b}^{0}}$ &$-0.019$ &$3.38$ &$0.75$ &$g_{2}^{
		
		\Xi_{bb}^{-}\to\Lambda_{b}^{0}}$ &$
	-0.026$ &$3.27$ &$0.86$ \\ \hline
	$f_{1}^{\Xi_{bb}^{-}\to\Sigma_{b}^{0}}$ &$0.083$ &$2.99$ &$0.97$ &$f_{2}^{
		
		\Xi_{bb}^{-}\to\Sigma_{b}^{0}}$ &$0.105$ &$3.03$ &$0.88$ \\$g_{1}^{
		
		\Xi_{bb}^{-}\to\Sigma_{b}^{0}}$ &$-0.019$ &$3.38$ &$0.75$ &$g_{2}^{
		
		\Xi_{bb}^{-}\to\Sigma_{b}^{0}}$ &$
	-0.026$ &$3.27$ &$0.86$ \\ \hline
	$f_{1}^{\Xi_{bb}^{-}\to\Xi_{bc}^{0}}$ &$0.414$ &$3.52$ &$0.64$ &$f_{2}^{
		
		\Xi_{bb}^{-}\to\Xi_{bc}^{0}}$ &$0.448$ &$3.59$ &$0.60$ \\$g_{1}^{
		
		\Xi_{bb}^{-}\to\Xi_{bc}^{0}}$ &$-0.116$ &$4.05$ &$0.55$ &$g_{2}^{
		
		\Xi_{bb}^{-}\to\Xi_{bc}^{0}}$ &$
	-0.063$ &$3.90$ &$0.60$ \\ \hline
	$f_{1}^{\Xi_{bb}^{-}\to\Xi_{bc}^{\prime0}}$ &$0.414$ &$3.52$ &$0.64$ &$f_{2}^{
		
		\Xi_{bb}^{-}\to\Xi_{bc}^{\prime0}}$ &$0.448$ &$3.59$ &$0.60$ \\$g_{1}^{
		
		\Xi_{bb}^{-}\to\Xi_{bc}^{\prime0}}$ &$-0.116$ &$4.05$ &$0.55$ &$g_{2}^{
		
		\Xi_{bb}^{-}\to\Xi_{bc}^{\prime0}}$ &$
	-0.063$ &$3.90$ &$0.60$ \\ \hline
	$f_{1}^{\Omega_{bb}^{-}\to\Xi_{b}^{0}}$ &$0.080$ &$2.98$ &$0.99$ &$f_{2}^{
		
		\Omega_{bb}^{-}\to\Xi_{b}^{0}}$ &$0.103$ &$3.02$ &$0.90$ \\$g_{1}^{
		
		\Omega_{bb}^{-}\to\Xi_{b}^{0}}$ &$-0.018$ &$3.36$ &$0.76$ &$g_{2}^{
		
		\Omega_{bb}^{-}\to\Xi_{b}^{0}}$ &$
	-0.026$ &$3.25$ &$0.88$ \\ \hline
	$f_{1}^{\Omega_{bb}^{-}\to\Xi_{b}^{\prime0}}$ &$0.080$ &$2.98$ &$0.99$ &
	
	$f_{2}^{\Omega_{bb}^{-}\to\Xi_{b}^{\prime0}}$ &$0.103$ &$3.02$ &$0.90$ \\
	
	$g_{1}^{\Omega_{bb}^{-}\to\Xi_{b}^{\prime0}}$ &$-0.018$ &$3.36$ &$0.76$ &
	
	$g_{2}^{\Omega_{bb}^{-}\to\Xi_{b}^{\prime0}}$ &$
	-0.026$ &$3.25$ &$0.88$ \\ \hline
	$f_{1}^{\Omega_{bb}^{-}\to\Omega_{bc}^{0}}$ &$0.410$ &$3.50$ &$0.65$ &$f_{2}^{
		
		\Omega_{bb}^{-}\to\Omega_{bc}^{0}}$ &$0.446$ &$3.57$ &$0.61$ \\$g_{1}^{
		
		\Omega_{bb}^{-}\to\Omega_{bc}^{0}}$ &$-0.115$ &$4.02$ &$0.55$ &$g_{2}^{
		
		\Omega_{bb}^{-}\to\Omega_{bc}^{0}}$ &$
	-0.063$ &$3.88$ &$0.61$ \\ \hline
	$f_{1}^{\Omega_{bb}^{-}\to\Omega_{bc}^{\prime0}}$ &$0.410$ &$3.50$ &$0.65$ &
	
	$f_{2}^{\Omega_{bb}^{-}\to\Omega_{bc}^{\prime0}}$ &$0.446$ &$3.57$ &$0.61$ 
	
	\\$g_{1}^{\Omega_{bb}^{-}\to\Omega_{bc}^{\prime0}}$ &$-0.115$ &$4.02$ &$0.55
	
	$ &$g_{2}^{\Omega_{bb}^{-}\to\Omega_{bc}^{\prime0}}$ &$
	-0.063$ &$3.88$ &$0.61$ \\ \hline
	\hline
\end{tabular} \end{table}
\begin{table}
	\caption{Same as Table~\ref{Tab:ff_cc_scalar} but for the charm decays of bottom-charm baryons  with scalar $(0^{+})$ diquarks.}
	\label{Tab:ff_bc_scalar}
\begin{tabular}{c|c|c|c|c|c|c|c}
	\hline\hline
	$F$ &$F(0)$ &$m_{\rm{fit}}$ &$\delta$ &$F$ &$F(0)$ &$m_{\rm{fit}}$ &$\delta$\
	
	\\ \hline
	$f_{1}^{\Xi_{bc}^{+}\to\Lambda_{b}^{0}}$ &$0.639$ &$1.52$ &$0.41$ &$f_{2}^{
		
		\Xi_{bc}^{+}\to\Lambda_{b}^{0}}$ &$-1.715$ &$1.47$ &$0.43$ \\$g_{1}^{
		
		\Xi_{bc}^{+}\to\Lambda_{b}^{0}}$ &$0.499$ &$1.84$ &$0.56$ &$g_{2}^{
		
		\Xi_{bc}^{+}\to\Lambda_{b}^{0}}$ &$
	-0.233$ &$1.12$ &$0.71$ \\ \hline
	$f_{1}^{\Xi_{bc}^{+}\to\Sigma_{b}^{0}}$ &$0.639$ &$1.52$ &$0.41$ &$f_{2}^{
		
		\Xi_{bc}^{+}\to\Sigma_{b}^{0}}$ &$-1.715$ &$1.47$ &$0.43$ \\$g_{1}^{
		
		\Xi_{bc}^{+}\to\Sigma_{b}^{0}}$ &$0.499$ &$1.84$ &$0.56$ &$g_{2}^{
		
		\Xi_{bc}^{+}\to\Sigma_{b}^{0}}$ &$
	-0.233$ &$1.12$ &$0.71$ \\ \hline
	$f_{1}^{\Xi_{bc}^{+}\to\Xi_{b}^{0}}$ &$0.725$ &$1.60$ &$0.40$ &$f_{2}^{
		
		\Xi_{bc}^{+}\to\Xi_{b}^{0}}$ &$-1.809$ &$1.54$ &$0.42$ \\$g_{1}^{
		
		\Xi_{bc}^{+}\to\Xi_{b}^{0}}$ &$0.571$ &$1.92$ &$0.52$ &$g_{2}^{
		
		\Xi_{bc}^{+}\to\Xi_{b}^{0}}$ &$
	-0.270$ &$1.20$ &$0.57$ \\ \hline
	$f_{1}^{\Xi_{bc}^{+}\to\Xi_{b}^{\prime0}}$ &$0.725$ &$1.60$ &$0.40$ &$f_{2}^{
		
		\Xi_{bc}^{+}\to\Xi_{b}^{\prime0}}$ &$-1.809$ &$1.54$ &$0.42$ \\$g_{1}^{
		
		\Xi_{bc}^{+}\to\Xi_{b}^{\prime0}}$ &$0.571$ &$1.92$ &$0.52$ &$g_{2}^{
		
		\Xi_{bc}^{+}\to\Xi_{b}^{\prime0}}$ &$
	-0.270$ &$1.20$ &$0.57$ \\ \hline
	$f_{1}^{\Xi_{bc}^{0}\to\Sigma_{b}^{-}}$ &$0.639$ &$1.52$ &$0.41$ &$f_{2}^{
		
		\Xi_{bc}^{0}\to\Sigma_{b}^{-}}$ &$-1.715$ &$1.47$ &$0.43$ \\$g_{1}^{
		
		\Xi_{bc}^{0}\to\Sigma_{b}^{-}}$ &$0.499$ &$1.84$ &$0.56$ &$g_{2}^{
		
		\Xi_{bc}^{0}\to\Sigma_{b}^{-}}$ &$
	-0.233$ &$1.12$ &$0.71$ \\ \hline
	$f_{1}^{\Xi_{bc}^{0}\to\Xi_{b}^{-}}$ &$0.725$ &$1.60$ &$0.40$ &$f_{2}^{
		
		\Xi_{bc}^{0}\to\Xi_{b}^{-}}$ &$-1.809$ &$1.54$ &$0.42$ \\$g_{1}^{
		
		\Xi_{bc}^{0}\to\Xi_{b}^{-}}$ &$0.571$ &$1.92$ &$0.52$ &$g_{2}^{
		
		\Xi_{bc}^{0}\to\Xi_{b}^{-}}$ &$
	-0.270$ &$1.20$ &$0.57$ \\ \hline
	$f_{1}^{\Xi_{bc}^{0}\to\Xi_{b}^{\prime-}}$ &$0.725$ &$1.60$ &$0.40$ &$f_{2}^{
		
		\Xi_{bc}^{0}\to\Xi_{b}^{\prime-}}$ &$-1.809$ &$1.54$ &$0.42$ \\$g_{1}^{
		
		\Xi_{bc}^{0}\to\Xi_{b}^{\prime-}}$ &$0.571$ &$1.92$ &$0.52$ &$g_{2}^{
		
		\Xi_{bc}^{0}\to\Xi_{b}^{\prime-}}$ &$
	-0.270$ &$1.20$ &$0.57$ \\ \hline
	$f_{1}^{\Omega_{bc}^{0}\to\Xi_{b}^{-}}$ &$0.638$ &$1.51$ &$0.42$ &$f_{2}^{
		
		\Omega_{bc}^{0}\to\Xi_{b}^{-}}$ &$-1.732$ &$1.46$ &$0.44$ \\$g_{1}^{
		
		\Omega_{bc}^{0}\to\Xi_{b}^{-}}$ &$0.498$ &$1.83$ &$0.56$ &$g_{2}^{
		
		\Omega_{bc}^{0}\to\Xi_{b}^{-}}$ &$
	-0.238$ &$1.12$ &$0.71$ \\ \hline
	$f_{1}^{\Omega_{bc}^{0}\to\Xi_{b}^{\prime-}}$ &$0.638$ &$1.51$ &$0.42$ &
	
	$f_{2}^{\Omega_{bc}^{0}\to\Xi_{b}^{\prime-}}$ &$-1.732$ &$1.46$ &$0.44$ \\
	
	$g_{1}^{\Omega_{bc}^{0}\to\Xi_{b}^{\prime-}}$ &$0.498$ &$1.83$ &$0.56$ &
	
	$g_{2}^{\Omega_{bc}^{0}\to\Xi_{b}^{\prime-}}$ &$
	-0.238$ &$1.12$ &$0.71$ \\ \hline
	$f_{1}^{\Omega_{bc}^{0}\to\Omega_{b}^{-}}$ &$0.723$ &$1.60$ &$0.40$ &$f_{2}^{
		
		\Omega_{bc}^{0}\to\Omega_{b}^{-}}$ &$-1.828$ &$1.54$ &$0.42$ \\$g_{1}^{
		
		\Omega_{bc}^{0}\to\Omega_{b}^{-}}$ &$0.570$ &$1.91$ &$0.52$ &$g_{2}^{
		
		\Omega_{bc}^{0}\to\Omega_{b}^{-}}$ &$
	-0.275$ &$1.20$ &$0.57$ \\ \hline
	\hline
\end{tabular}
\end{table}
\begin{table}
	\caption{Same as Table~\ref{Tab:ff_cc_axial} but for the charm decays of bottom-charm baryons with axial vector $(1^{+})$ diquarks.}
	\label{Tab:ff_bc_axial}
\begin{tabular}{c|c|c|c|c|c|c|c}
	\hline\hline
	$F$ &$F(0)$ &$m_{\rm{fit}}$ &$\delta$ &$F$ &$F(0)$ &$m_{\rm{fit}}$ &$\delta$\
	
	\\ \hline
	$f_{1}^{\Xi_{bc}^{+}\to\Lambda_{b}^{0}}$ &$0.637$ &$1.44$ &$0.45$ &$f_{2}^{
		
		\Xi_{bc}^{+}\to\Lambda_{b}^{0}}$ &$1.027$ &$1.47$ &$0.43$ \\$g_{1}^{
		
		\Xi_{bc}^{+}\to\Lambda_{b}^{0}}$ &$-0.160$ &$1.89$ &$0.54$ &$g_{2}^{
		
		\Xi_{bc}^{+}\to\Lambda_{b}^{0}}$ &$
	0.006^{*}$ &$0.28^{*}$ &$0.08^{*}$ \\ \hline
	$f_{1}^{\Xi_{bc}^{+}\to\Sigma_{b}^{0}}$ &$0.637$ &$1.44$ &$0.45$ &$f_{2}^{
		
		\Xi_{bc}^{+}\to\Sigma_{b}^{0}}$ &$1.027$ &$1.47$ &$0.43$ \\$g_{1}^{
		
		\Xi_{bc}^{+}\to\Sigma_{b}^{0}}$ &$-0.160$ &$1.89$ &$0.54$ &$g_{2}^{
		
		\Xi_{bc}^{+}\to\Sigma_{b}^{0}}$ &$
	0.006^{*}$ &$0.28^{*}$ &$0.08^{*}$ \\ \hline
	$f_{1}^{\Xi_{bc}^{+}\to\Xi_{b}^{0}}$ &$0.723$ &$1.52$ &$0.44$ &$f_{2}^{
		
		\Xi_{bc}^{+}\to\Xi_{b}^{0}}$ &$1.111$ &$1.55$ &$0.42$ \\$g_{1}^{
		
		\Xi_{bc}^{+}\to\Xi_{b}^{0}}$ &$-0.185$ &$1.96$ &$0.50$ &$g_{2}^{
		
		\Xi_{bc}^{+}\to\Xi_{b}^{0}}$ &$
	0.019$ &$0.21$ &$-0.06$ \\ \hline
	$f_{1}^{\Xi_{bc}^{+}\to\Xi_{b}^{\prime0}}$ &$0.723$ &$1.52$ &$0.44$ &$f_{2}^{
		
		\Xi_{bc}^{+}\to\Xi_{b}^{\prime0}}$ &$1.111$ &$1.55$ &$0.42$ \\$g_{1}^{
		
		\Xi_{bc}^{+}\to\Xi_{b}^{\prime0}}$ &$-0.185$ &$1.96$ &$0.50$ &$g_{2}^{
		
		\Xi_{bc}^{+}\to\Xi_{b}^{\prime0}}$ &$
	0.019$ &$0.21$ &$-0.06$ \\ \hline
	$f_{1}^{\Xi_{bc}^{0}\to\Sigma_{b}^{-}}$ &$0.637$ &$1.44$ &$0.45$ &$f_{2}^{
		
		\Xi_{bc}^{0}\to\Sigma_{b}^{-}}$ &$1.027$ &$1.47$ &$0.43$ \\$g_{1}^{
		
		\Xi_{bc}^{0}\to\Sigma_{b}^{-}}$ &$-0.160$ &$1.89$ &$0.54$ &$g_{2}^{
		
		\Xi_{bc}^{0}\to\Sigma_{b}^{-}}$ &$
	0.006^{*}$ &$0.28^{*}$ &$0.08^{*}$ \\ \hline
	$f_{1}^{\Xi_{bc}^{0}\to\Xi_{b}^{-}}$ &$0.723$ &$1.52$ &$0.44$ &$f_{2}^{
		
		\Xi_{bc}^{0}\to\Xi_{b}^{-}}$ &$1.111$ &$1.55$ &$0.42$ \\$g_{1}^{
		
		\Xi_{bc}^{0}\to\Xi_{b}^{-}}$ &$-0.185$ &$1.96$ &$0.50$ &$g_{2}^{
		
		\Xi_{bc}^{0}\to\Xi_{b}^{-}}$ &$
	0.019$ &$0.21$ &$-0.06$ \\ \hline
	$f_{1}^{\Xi_{bc}^{0}\to\Xi_{b}^{\prime-}}$ &$0.723$ &$1.52$ &$0.44$ &$f_{2}^{
		
		\Xi_{bc}^{0}\to\Xi_{b}^{\prime-}}$ &$1.111$ &$1.55$ &$0.42$ \\$g_{1}^{
		
		\Xi_{bc}^{0}\to\Xi_{b}^{\prime-}}$ &$-0.185$ &$1.96$ &$0.50$ &$g_{2}^{
		
		\Xi_{bc}^{0}\to\Xi_{b}^{\prime-}}$ &$
	0.019$ &$0.21$ &$-0.06$ \\ \hline
	$f_{1}^{\Omega_{bc}^{0}\to\Xi_{b}^{-}}$ &$0.636$ &$1.44$ &$0.46$ &$f_{2}^{
		
		\Omega_{bc}^{0}\to\Xi_{b}^{-}}$ &$1.028$ &$1.47$ &$0.43$ \\$g_{1}^{
		
		\Omega_{bc}^{0}\to\Xi_{b}^{-}}$ &$-0.160$ &$1.88$ &$0.54$ &$g_{2}^{
		
		\Omega_{bc}^{0}\to\Xi_{b}^{-}}$ &$
	0.008$ &$0.31$ &$-0.13$ \\ \hline
	$f_{1}^{\Omega_{bc}^{0}\to\Xi_{b}^{\prime-}}$ &$0.636$ &$1.44$ &$0.46$ &
	
	$f_{2}^{\Omega_{bc}^{0}\to\Xi_{b}^{\prime-}}$ &$1.028$ &$1.47$ &$0.43$ \\
	
	$g_{1}^{\Omega_{bc}^{0}\to\Xi_{b}^{\prime-}}$ &$-0.160$ &$1.88$ &$0.54$ &
	
	$g_{2}^{\Omega_{bc}^{0}\to\Xi_{b}^{\prime-}}$ &$
	0.008$ &$0.31$ &$-0.13$ \\ \hline
	$f_{1}^{\Omega_{bc}^{0}\to\Omega_{b}^{-}}$ &$0.721$ &$1.51$ &$0.44$ &$f_{2}^{
		
		\Omega_{bc}^{0}\to\Omega_{b}^{-}}$ &$1.112$ &$1.54$ &$0.42$ \\$g_{1}^{
		
		\Omega_{bc}^{0}\to\Omega_{b}^{-}}$ &$-0.185$ &$1.95$ &$0.50$ &$g_{2}^{
		
		\Omega_{bc}^{0}\to\Omega_{b}^{-}}$ &$
	0.021$ &$0.32$ &$-0.23$ \\ \hline
	\hline
\end{tabular}
\end{table}
\begin{table}
	\caption{Same as Table~\ref{Tab:ff_cc_scalar} but for the $b$ decays of bottom-charm baryons  with scalar $(0^{+})$ diquarks.}
	\label{Tab:ff_bc_scalar_2}
\begin{tabular}{c|c|c|c|c|c|c|c}
	\hline\hline
	$F$ &$F(0)$ &$m_{\rm{fit}}$ &$\delta$ &$F$ &$F(0)$ &$m_{\rm{fit}}$ &$\delta$\
	
	\\ \hline
	$f_{1}^{\Xi_{bc}^{+}\to\Sigma_{c}^{++}}$ &$0.136$ &$3.48$ &$0.58$ &$f_{2}^{
		
		\Xi_{bc}^{+}\to\Sigma_{c}^{++}}$ &$-0.081$ &$3.25$ &$0.64$ \\$g_{1}^{
		
		\Xi_{bc}^{+}\to\Sigma_{c}^{++}}$ &$0.130$ &$3.59$ &$0.59$ &$g_{2}^{
		
		\Xi_{bc}^{+}\to\Sigma_{c}^{++}}$ &$
	-0.009$ &$2.95$ &$0.98$ \\ \hline
	$f_{1}^{\Xi_{bc}^{+}\to\Xi_{cc}^{++}}$ &$0.550$ &$4.45$ &$0.43$ &$f_{2}^{
		
		\Xi_{bc}^{+}\to\Xi_{cc}^{++}}$ &$-0.230$ &$4.07$ &$0.47$ \\$g_{1}^{
		
		\Xi_{bc}^{+}\to\Xi_{cc}^{++}}$ &$0.530$ &$4.57$ &$0.44$ &$g_{2}^{
		
		\Xi_{bc}^{+}\to\Xi_{cc}^{++}}$ &$
	-0.043$ &$3.90$ &$0.48$ \\ \hline
	$f_{1}^{\Xi_{bc}^{0}\to\Lambda_{c}^{+}}$ &$0.136$ &$3.48$ &$0.58$ &$f_{2}^{
		
		\Xi_{bc}^{0}\to\Lambda_{c}^{+}}$ &$-0.081$ &$3.25$ &$0.64$ \\$g_{1}^{
		
		\Xi_{bc}^{0}\to\Lambda_{c}^{+}}$ &$0.130$ &$3.59$ &$0.59$ &$g_{2}^{
		
		\Xi_{bc}^{0}\to\Lambda_{c}^{+}}$ &$
	-0.009$ &$2.95$ &$0.98$ \\ \hline
	$f_{1}^{\Xi_{bc}^{0}\to\Sigma_{c}^{+}}$ &$0.136$ &$3.48$ &$0.58$ &$f_{2}^{
		
		\Xi_{bc}^{0}\to\Sigma_{c}^{+}}$ &$-0.081$ &$3.25$ &$0.64$ \\$g_{1}^{
		
		\Xi_{bc}^{0}\to\Sigma_{c}^{+}}$ &$0.130$ &$3.59$ &$0.59$ &$g_{2}^{
		
		\Xi_{bc}^{0}\to\Sigma_{c}^{+}}$ &$
	-0.009$ &$2.95$ &$0.98$ \\ \hline
	$f_{1}^{\Xi_{bc}^{0}\to\Xi_{cc}^{+}}$ &$0.550$ &$4.45$ &$0.43$ &$f_{2}^{
		
		\Xi_{bc}^{0}\to\Xi_{cc}^{+}}$ &$-0.230$ &$4.07$ &$0.47$ \\$g_{1}^{
		
		\Xi_{bc}^{0}\to\Xi_{cc}^{+}}$ &$0.530$ &$4.57$ &$0.44$ &$g_{2}^{
		
		\Xi_{bc}^{0}\to\Xi_{cc}^{+}}$ &$
	-0.043$ &$3.90$ &$0.48$ \\ \hline
	$f_{1}^{\Omega_{bc}^{0}\to\Xi_{c}^{+}}$ &$0.123$ &$3.39$ &$0.61$ &$f_{2}^{
		
		\Omega_{bc}^{0}\to\Xi_{c}^{+}}$ &$-0.077$ &$3.19$ &$0.69$ \\$g_{1}^{
		
		\Omega_{bc}^{0}\to\Xi_{c}^{+}}$ &$0.118$ &$3.49$ &$0.63$ &$g_{2}^{
		
		\Omega_{bc}^{0}\to\Xi_{c}^{+}}$ &$
	-0.009$ &$2.92$ &$1.06$ \\ \hline
	$f_{1}^{\Omega_{bc}^{0}\to\Xi_{c}^{\prime+}}$ &$0.123$ &$3.39$ &$0.61$ &
	
	$f_{2}^{\Omega_{bc}^{0}\to\Xi_{c}^{\prime+}}$ &$-0.077$ &$3.19$ &$0.69$ \\
	
	$g_{1}^{\Omega_{bc}^{0}\to\Xi_{c}^{\prime+}}$ &$0.118$ &$3.49$ &$0.63$ &
	
	$g_{2}^{\Omega_{bc}^{0}\to\Xi_{c}^{\prime+}}$ &$
	-0.009$ &$2.92$ &$1.06$ \\ \hline
	$f_{1}^{\Omega_{bc}^{0}\to\Omega_{cc}^{+}}$ &$0.531$ &$4.33$ &$0.45$ &$f_{2}^{
		
		\Omega_{bc}^{0}\to\Omega_{cc}^{+}}$ &$-0.231$ &$3.98$ &$0.49$ \\$g_{1}^{
		
		\Omega_{bc}^{0}\to\Omega_{cc}^{+}}$ &$0.511$ &$4.44$ &$0.46$ &$g_{2}^{
		
		\Omega_{bc}^{0}\to\Omega_{cc}^{+}}$ &$
	-0.045$ &$3.81$ &$0.51$ \\ \hline
	\hline
\end{tabular}
\end{table}
\begin{table}
	\caption{Same as Table~\ref{Tab:ff_cc_axial} but for the $b$ decays of bottom-charm baryons  with axial vector $(1^{+})$ diquarks.}
	\label{Tab:ff_bc_axial_2}
\begin{tabular}{c|c|c|c|c|c|c|c}
	\hline\hline
	$F$ &$F(0)$ &$m_{\rm{fit}}$ &$\delta$ &$F$ &$F(0)$ &$m_{\rm{fit}}$ &$\delta$\
	
	\\ \hline
	$f_{1}^{\Xi_{bc}^{+}\to\Sigma_{c}^{++}}$ &$0.125$ &$3.11$ &$0.79$ &$f_{2}^{
		
		\Xi_{bc}^{+}\to\Sigma_{c}^{++}}$ &$0.150$ &$3.20$ &$0.65$ \\$g_{1}^{
		
		\Xi_{bc}^{+}\to\Sigma_{c}^{++}}$ &$-0.022$ &$4.18$ &$0.58$ &$g_{2}^{
		
		\Xi_{bc}^{+}\to\Sigma_{c}^{++}}$ &$
	-0.039$ &$3.50$ &$0.66$ \\ \hline
	$f_{1}^{\Xi_{bc}^{+}\to\Xi_{cc}^{++}}$ &$0.527$ &$3.78$ &$0.55$ &$f_{2}^{
		
		\Xi_{bc}^{+}\to\Xi_{cc}^{++}}$ &$0.525$ &$3.91$ &$0.48$ \\$g_{1}^{
		
		\Xi_{bc}^{+}\to\Xi_{cc}^{++}}$ &$-0.146$ &$4.76$ &$0.38$ &$g_{2}^{
		
		\Xi_{bc}^{+}\to\Xi_{cc}^{++}}$ &$
	-0.060$ &$4.50$ &$0.51$ \\ \hline
	$f_{1}^{\Xi_{bc}^{0}\to\Lambda_{c}^{+}}$ &$0.125$ &$3.11$ &$0.79$ &$f_{2}^{
		
		\Xi_{bc}^{0}\to\Lambda_{c}^{+}}$ &$0.150$ &$3.20$ &$0.65$ \\$g_{1}^{
		
		\Xi_{bc}^{0}\to\Lambda_{c}^{+}}$ &$-0.022$ &$4.18$ &$0.58$ &$g_{2}^{
		
		\Xi_{bc}^{0}\to\Lambda_{c}^{+}}$ &$
	-0.039$ &$3.50$ &$0.66$ \\ \hline
	$f_{1}^{\Xi_{bc}^{0}\to\Sigma_{c}^{+}}$ &$0.125$ &$3.11$ &$0.79$ &$f_{2}^{
		
		\Xi_{bc}^{0}\to\Sigma_{c}^{+}}$ &$0.150$ &$3.20$ &$0.65$ \\$g_{1}^{
		
		\Xi_{bc}^{0}\to\Sigma_{c}^{+}}$ &$-0.022$ &$4.18$ &$0.58$ &$g_{2}^{
		
		\Xi_{bc}^{0}\to\Sigma_{c}^{+}}$ &$
	-0.039$ &$3.50$ &$0.66$ \\ \hline
	$f_{1}^{\Xi_{bc}^{0}\to\Xi_{cc}^{+}}$ &$0.527$ &$3.78$ &$0.55$ &$f_{2}^{
		
		\Xi_{bc}^{0}\to\Xi_{cc}^{+}}$ &$0.525$ &$3.91$ &$0.48$ \\$g_{1}^{
		
		\Xi_{bc}^{0}\to\Xi_{cc}^{+}}$ &$-0.146$ &$4.76$ &$0.38$ &$g_{2}^{
		
		\Xi_{bc}^{0}\to\Xi_{cc}^{+}}$ &$
	-0.060$ &$4.50$ &$0.51$ \\ \hline
	$f_{1}^{\Omega_{bc}^{0}\to\Xi_{c}^{+}}$ &$0.114$ &$3.07$ &$0.83$ &$f_{2}^{
		
		\Omega_{bc}^{0}\to\Xi_{c}^{+}}$ &$0.137$ &$3.14$ &$0.69$ \\$g_{1}^{
		
		\Omega_{bc}^{0}\to\Xi_{c}^{+}}$ &$-0.020$ &$4.07$ &$0.64$ &$g_{2}^{
		
		\Omega_{bc}^{0}\to\Xi_{c}^{+}}$ &$
	-0.036$ &$3.40$ &$0.69$ \\ \hline
	$f_{1}^{\Omega_{bc}^{0}\to\Xi_{c}^{\prime+}}$ &$0.114$ &$3.07$ &$0.83$ &
	
	$f_{2}^{\Omega_{bc}^{0}\to\Xi_{c}^{\prime+}}$ &$0.137$ &$3.14$ &$0.69$ \\
	
	$g_{1}^{\Omega_{bc}^{0}\to\Xi_{c}^{\prime+}}$ &$-0.020$ &$4.07$ &$0.64$ &
	
	$g_{2}^{\Omega_{bc}^{0}\to\Xi_{c}^{\prime+}}$ &$
	-0.036$ &$3.40$ &$0.69$ \\ \hline
	$f_{1}^{\Omega_{bc}^{0}\to\Omega_{cc}^{+}}$ &$0.511$ &$3.72$ &$0.57$ &$f_{2}^{
		
		\Omega_{bc}^{0}\to\Omega_{cc}^{+}}$ &$0.509$ &$3.84$ &$0.50$ \\$g_{1}^{
		
		\Omega_{bc}^{0}\to\Omega_{cc}^{+}}$ &$-0.141$ &$4.66$ &$0.42$ &$g_{2}^{
		
		\Omega_{bc}^{0}\to\Omega_{cc}^{+}}$ &$
	-0.062$ &$4.31$ &$0.51$ \\ \hline
	\hline
\end{tabular}
\end{table}

The masses of quarks (in units of GeV) are used as~\cite{Lu:2007sg,Wang:2007sxa,Wang:2008xt,Wang:2008ci,Wang:2009mi,Chen:2009qk,Li:2010bb,Verma:2011yw,Shi:2016gqt} 
\begin{eqnarray}
 m_u=m_d= 0.25, \;\; m_s=0.37, \;\; m_c=1.4, \;\; m_b=4.8. 
\end{eqnarray}
$m_{[ci]}$ and $m_{[bj]}$ are approximated respectively
by the sum of $m_{c}+m_{i}$ and $m_{b}+m_{j}$ with
$i,j=u,d,s$.

The $\beta$ parameters in the wave functions of the doubly and singly heavy flavor
baryons are approximately the same as those of the corresponding mesons, since the heavy-light diquark behaves a color anti-triplet just like an heavy anti-quark. Taking $\Xi_{cc}^{++}$ as an example, $\beta_{c[cu]}\approx\beta_{c\bar{c}}$ taken from the decay constants of $\eta_{c}$.
The $\beta$ parameters are then obtained by following  Eq.~(2.17) of \cite{Cheng:2003sm}, with the decay constants as \cite{Carrasco:2014poa,Becirevic:2013bsa,Wang:2010npa}
\begin{equation}
f_{D}=207.4{\rm MeV},\quad f_{D_{s}}=247.2{\rm GeV},\quad f_{\eta_{c}}=387{\rm MeV},\quad f_{\Upsilon}=715{\rm MeV}.
\end{equation}
The values of the $\beta$ parameters are then listed in Table~\ref{Tab:Beta}.
Other $\beta$'s are  taken directly from \cite{Shi:2016gqt}.	The masses of singly heavy flavor baryons are also collected in Table  \ref{Tab:Beta}~\cite{Olive:2016xmw,Thakkar:2016dna}.

With the above inputs, the form factors with the scalar or axial vector diquarks in Eqs.~(\ref{eq:ff_scalar}) and (\ref{eq:ff_axial}) can be obtained. To access the $q^2$ distribution of the form factors, we adopt the following parametrized form  
\begin{equation}
F(q^{2})=\frac{F(0)}{1-\frac{q^{2}}{m_{{\rm fit}}^{2}}+\delta\left(\frac{q^{2}}{m_{{\rm fit}}^{2}}\right)^{2}},\label{eq:main_fit_formula}
\end{equation}
where the $F(0)$ is the form factor at $q^2=0$. The $m_{\rm fit}$ and $\delta$ are two parameters to be fitted from numerical results. 
For the form factor $g_2$, the above formula may lead to an imaginary result for $m_{\rm fit}$ and in this case we adopt the modified form as: 
\begin{equation}
F(q^{2})=\frac{F(0)}{1+\frac{q^{2}}{m_{{\rm fit}}^{2}}+\delta\left(\frac{q^{2}}{m_{{\rm fit}}^{2}}\right)^{2}}.\label{eq:auxiliary_fit_formula}
\end{equation}
The results for  form factors with scalar diquark spectators  
are given in Tables \ref{Tab:ff_cc_scalar}, \ref{Tab:ff_bb_scalar}, \ref{Tab:ff_bc_scalar} and \ref{Tab:ff_bc_scalar_2}, while those with axial vector diquarks are shown in Tables \ref{Tab:ff_cc_axial}, \ref{Tab:ff_bb_axial}, \ref{Tab:ff_bc_axial} and \ref{Tab:ff_bc_axial_2}. With the results of these form factors, the physical hadronic transition matrix elements can be obtained through Eq.~(\ref{eq:csca}).

\section{Semi-leptonic decays  }
\label{sec:semileptonic}

\subsection{Semi-leptonic $B\to B'\ell\bar\nu$ decay widths}
\label{app:semi_results}

\begin{table}
	\caption{The $cc$ sector: decay widths, branching ratios and $\Gamma_{L}/\Gamma_{T}$'s for semi-leptonic decays, with lepton mass neglected.}
	\label{Tab:SemiLep_cc}
\begin{tabular}{l|c|c|c}
	\hline\hline
	channels &$\Gamma/\text{~GeV}$ &${\cal B}$ &$\Gamma_{L}/\Gamma_{T}$\\ \hline
	$\Xi_{cc}^{++}\to\Lambda_{c}^{+}l^{+}\nu_{l}$ &$1.05\times 10^{-14}$ &$
	
	4.81\times 10^{-3}$ &$8.52$\\
	$\Xi_{cc}^{++}\to\Sigma_{c}^{+}l^{+}\nu_{l}$ &$9.60\times 10^{-15}$ &$
	
	4.38\times 10^{-3}$ &$1.28$\\
	$\Xi_{cc}^{++}\to\Xi_{c}^{+}l^{+}\nu_{l}$ &$1.15\times 10^{-13}$ &$
	
	5.25\times 10^{-2}$ &$9.99$\\
	$\Xi_{cc}^{++}\to\Xi_{c}^{\prime+}l^{+}\nu_{l}$ &$1.28\times 10^{-13}$ &$
	
	5.84\times 10^{-2}$ &$1.42$\\
	$\Xi_{cc}^{+}\to\Sigma_{c}^{0}l^{+}\nu_{l}$ &$1.91\times 10^{-14}$ &$
	
	2.91\times 10^{-3}$ &$1.28$\\
	$\Xi_{cc}^{+}\to\Xi_{c}^{0}l^{+}\nu_{l}$ &$1.14\times 10^{-13}$ &$
	
	1.73\times 10^{-2}$ &$9.99$\\
	$\Xi_{cc}^{+}\to\Xi_{c}^{\prime0}l^{+}\nu_{l}$ &$1.27\times 10^{-13}$ &$
	
	1.93\times 10^{-2}$ &$1.42$\\
	$\Omega_{cc}^{+}\to\Xi_{c}^{0}l^{+}\nu_{l}$ &$8.06\times 10^{-15}$ &$
	
	3.31\times 10^{-3}$ &$8.84$\\
	$\Omega_{cc}^{+}\to\Xi_{c}^{\prime0}l^{+}\nu_{l}$ &$9.34\times 10^{-15}$ &$
	
	3.83\times 10^{-3}$ &$1.28$\\
	$\Omega_{cc}^{+}\to\Omega_{c}^{0}l^{+}\nu_{l}$ &$2.55\times 10^{-13}$ &$
	
	1.05\times 10^{-1}$ &$1.42$\\
	\hline\hline
\end{tabular}
\end{table}
\begin{table}
	\caption{The $bb$ sector: decay widths, branching ratios and $\Gamma_{L}/\Gamma_{T}$'s for semi-leptonic decays, with lepton mass neglected.}
	\label{Tab:SemiLep_bb}
\begin{tabular}{l|c|c|c}
	\hline\hline
	channels &$\Gamma/\text{~GeV}$ &${\cal B}$ &$\Gamma_{L}/\Gamma_{T}$\\ \hline
	$\Xi_{bb}^{0}\to\Sigma_{b}^{+}l^{-}\bar{\nu}_{l}$ &$6.67\times 10^{-17}$ &$
	
	3.75\times 10^{-5}$ &$1.32$\\
	$\Xi_{bb}^{0}\to\Xi_{bc}^{+}l^{-}\bar{\nu}_{l}$ &$3.30\times 10^{-14}$ &$
	
	1.86\times 10^{-2}$ &$2.32$\\
	$\Xi_{bb}^{0}\to\Xi_{bc}^{\prime+}l^{-}\bar{\nu}_{l}$ &$1.45\times 10^{-14}
	
	$ &$8.13\times 10^{-3}$ &$0.91$\\
	$\Xi_{bb}^{-}\to\Lambda_{b}^{0}l^{-}\bar{\nu}_{l}$ &$1.58\times 10^{-17}$ &$
	
	8.91\times 10^{-6}$ &$8.62$\\
	$\Xi_{bb}^{-}\to\Sigma_{b}^{0}l^{-}\bar{\nu}_{l}$ &$3.33\times 10^{-17}$ &$
	
	1.87\times 10^{-5}$ &$1.32$\\
	$\Xi_{bb}^{-}\to\Xi_{bc}^{0}l^{-}\bar{\nu}_{l}$ &$3.30\times 10^{-14}$ &$
	
	1.86\times 10^{-2}$ &$2.32$\\
	$\Xi_{bb}^{-}\to\Xi_{bc}^{\prime0}l^{-}\bar{\nu}_{l}$ &$1.45\times 10^{-14}
	
	$ &$8.13\times 10^{-3}$ &$0.91$\\
	$\Omega_{bb}^{-}\to\Xi_{b}^{0}l^{-}\bar{\nu}_{l}$ &$1.43\times 10^{-17}$ &$
	
	1.74\times 10^{-5}$ &$8.76$\\
	$\Omega_{bb}^{-}\to\Xi_{b}^{\prime0}l^{-}\bar{\nu}_{l}$ &$3.10\times 10^{-17}
	
	$ &$3.77\times 10^{-5}$ &$1.34$\\
	$\Omega_{bb}^{-}\to\Omega_{bc}^{0}l^{-}\bar{\nu}_{l}$ &$3.69\times 10^{-14}
	
	$ &$4.49\times 10^{-2}$ &$2.30$\\
	$\Omega_{bb}^{-}\to\Omega_{bc}^{\prime0}l^{-}\bar{\nu}_{l}$ &$
	
	1.62\times 10^{-14}$ &$1.98\times 10^{-2}$ &$0.91$\\
	\hline\hline
\end{tabular}
\end{table}
\begin{table}
	\caption{The $bc$ sector with the $c$ quark decay and an axial vector $bc$ diquark in the initial state: decay widths, branching ratios and $\Gamma_{L}/\Gamma_{T}$'s for semi-leptonic decays, with lepton mass neglected.}
	\label{Tab:SemiLep_bc_c}
\begin{tabular}{l|c|c|c}
	\hline\hline
	channels &$\Gamma/\text{~GeV}$ &${\cal B}$ &$\Gamma_{L}/\Gamma_{T}$\\ \hline
	$\Xi_{bc}^{+}\to\Lambda_{b}^{0}l^{+}\nu_{l}$ &$6.85\times 10^{-15}$ &$
	
	2.54\times 10^{-3}$ &$
	\text{ 10.3}$\\
	$\Xi_{bc}^{+}\to\Sigma_{b}^{0}l^{+}\nu_{l}$ &$4.63\times 10^{-15}$ &$
	
	1.72\times 10^{-3}$ &$
	\text{ 1.37}$\\
	$\Xi_{bc}^{+}\to\Xi_{b}^{0}l^{+}\nu_{l}$ &$7.13\times 10^{-14}$ &$
	
	2.64\times 10^{-2}$ &$
	\text{ 11.7}$\\
	$\Xi_{bc}^{+}\to\Xi_{b}^{\prime0}l^{+}\nu_{l}$ &$5.86\times 10^{-14}$ &$
	
	2.18\times 10^{-2}$ &$
	\text{ 1.49}$\\
	$\Xi_{bc}^{0}\to\Sigma_{b}^{-}l^{+}\nu_{l}$ &$9.18\times 10^{-15}$ &$
	
	1.30\times 10^{-3}$ &$
	\text{ 1.37}$\\
	$\Xi_{bc}^{0}\to\Xi_{b}^{-}l^{+}\nu_{l}$ &$7.06\times 10^{-14}$ &$
	
	9.98\times 10^{-3}$ &$
	\text{ 11.7}$\\
	$\Xi_{bc}^{0}\to\Xi_{b}^{\prime-}l^{+}\nu_{l}$ &$5.86\times 10^{-14}$ &$
	
	8.29\times 10^{-3}$ &$
	\text{ 1.49}$\\
	$\Omega_{bc}^{0}\to\Xi_{b}^{-}l^{+}\nu_{l}$ &$3.97\times 10^{-15}$ &$
	
	1.33\times 10^{-3}$ &$
	\text{ 11.0}$\\
	$\Omega_{bc}^{0}\to\Xi_{b}^{\prime-}l^{+}\nu_{l}$ &$3.32\times 10^{-15}$ &$
	
	1.11\times 10^{-3}$ &$
	\text{ 1.42}$\\
	$\Omega_{bc}^{0}\to\Omega_{b}^{-}l^{+}\nu_{l}$ &$8.66\times 10^{-14}$ &$
	
	2.90\times 10^{-2}$ &$
	\text{ 1.52}$\\
	\hline\hline
\end{tabular}
\end{table}
\begin{table}
	\caption{The $bc$ sector with the $c$ quark decay and a scalar $bc$ diquark in the initial state: decay widths, branching ratios and $\Gamma_{L}/\Gamma_{T}$'s for semi-leptonic decays, with lepton mass neglected. We have assumed $m_{B_{i}^{\prime}}=m_{B_{i}}$ and $\tau_{B_{i}^{\prime}}=\tau_{B_{i}}$, i.e. the only difference between $B_{i}^{\prime}\to B_{f}$ and $B_{i}\to B_{f}$ is the mixing coefficients.}
	\label{Tab:SemiLep_bc_c_primed}
\begin{tabular}{l|c|c|c}
	\hline\hline
	channels &$\Gamma/\text{~GeV}$ &${\cal B}$ &$\Gamma_{L}/\Gamma_{T}$\\ \hline
	$\Xi_{bc}^{\prime+}\to\Lambda_{b}^{0}l^{+}\nu_{l}$ &$5.36\times 10^{-15}$ &$
	
	1.99\times 10^{-3}$ &$
	\text{ 1.79}$\\
	$\Xi_{bc}^{\prime+}\to\Sigma_{b}^{0}l^{+}\nu_{l}$ &$2.78\times 10^{-15}$ &$
	
	1.03\times 10^{-3}$ &$
	\text{ 11.4}$\\
	$\Xi_{bc}^{\prime+}\to\Xi_{b}^{0}l^{+}\nu_{l}$ &$5.64\times 10^{-14}$ &$
	
	2.09\times 10^{-2}$ &$
	\text{ 2.03}$\\
	$\Xi_{bc}^{\prime+}\to\Xi_{b}^{\prime0}l^{+}\nu_{l}$ &$3.50\times 10^{-14}$ &$
	
	1.30\times 10^{-2}$ &$
	\text{ 12.0}$\\
	$\Xi_{bc}^{\prime0}\to\Sigma_{b}^{-}l^{+}\nu_{l}$ &$5.51\times 10^{-15}$ &$
	
	7.79\times 10^{-4}$ &$
	\text{ 11.4}$\\
	$\Xi_{bc}^{\prime0}\to\Xi_{b}^{-}l^{+}\nu_{l}$ &$5.59\times 10^{-14}$ &$
	
	7.91\times 10^{-3}$ &$
	\text{ 2.02}$\\
	$\Xi_{bc}^{\prime0}\to\Xi_{b}^{\prime-}l^{+}\nu_{l}$ &$3.50\times 10^{-14}$ &$
	
	4.95\times 10^{-3}$ &$
	\text{ 12.0}$\\
	$\Omega_{bc}^{\prime0}\to\Xi_{b}^{-}l^{+}\nu_{l}$ &$3.10\times 10^{-15}$ &$
	
	1.04\times 10^{-3}$ &$
	\text{ 1.92}$\\
	$\Omega_{bc}^{\prime0}\to\Xi_{b}^{\prime-}l^{+}\nu_{l}$ &$2.00\times 10^{-15}
	
	$ &$6.68\times 10^{-4}$ &$
	\text{ 11.5}$\\
	$\Omega_{bc}^{\prime0}\to\Omega_{b}^{-}l^{+}\nu_{l}$ &$5.18\times 10^{-14}$ &$
	
	1.73\times 10^{-2}$ &$
	\text{ 12.2}$\\
	\hline\hline
\end{tabular}
\end{table}
\begin{table}
	\caption{The $bc$ sector with the $b$ quark decay and an axial vector $bc$ diquark in the initial state: decay widths, branching ratios and $\Gamma_{L}/\Gamma_{T}$'s for semi-leptonic decays, with lepton mass neglected.}
	\label{Tab:SemiLep_bc_b}
\begin{tabular}{l|c|c|c}
	\hline\hline
	channels &$\Gamma/\text{~GeV}$ &${\cal B}$ &$\Gamma_{L}/\Gamma_{T}$\\ \hline
	$\Xi_{bc}^{+}\to\Sigma_{c}^{++}l^{-}\bar{\nu}_{l}$ &$9.48\times 10^{-17}$ &$
	
	3.52\times 10^{-5}$ &$
	\text{ 1.15}$\\
	$\Xi_{bc}^{+}\to\Xi_{cc}^{++}l^{-}\bar{\nu}_{l}$ &$4.50\times 10^{-14}$ &$
	
	1.67\times 10^{-2}$ &$
	\text{ 2.48}$\\
	$\Xi_{bc}^{0}\to\Lambda_{c}^{+}l^{-}\bar{\nu}_{l}$ &$1.84\times 10^{-17}$ &$
	
	2.60\times 10^{-6}$ &$
	\text{ 5.96}$\\
	$\Xi_{bc}^{0}\to\Sigma_{c}^{+}l^{-}\bar{\nu}_{l}$ &$4.74\times 10^{-17}$ &$
	
	6.71\times 10^{-6}$ &$
	\text{ 1.15}$\\
	$\Xi_{bc}^{0}\to\Xi_{cc}^{+}l^{-}\bar{\nu}_{l}$ &$4.50\times 10^{-14}$ &$
	
	6.36\times 10^{-3}$ &$
	\text{ 2.48}$\\
	$\Omega_{bc}^{0}\to\Xi_{c}^{+}l^{-}\bar{\nu}_{l}$ &$1.34\times 10^{-17}$ &$
	
	4.47\times 10^{-6}$ &$
	\text{ 6.34}$\\
	$\Omega_{bc}^{0}\to\Xi_{c}^{\prime+}l^{-}\bar{\nu}_{l}$ &$3.47\times 10^{-17}
	
	$ &$1.16\times 10^{-5}$ &$
	\text{ 1.19}$\\
	$\Omega_{bc}^{0}\to\Omega_{cc}^{+}l^{-}\bar{\nu}_{l}$ &$3.94\times 10^{-14}
	
	$ &$1.32\times 10^{-2}$ &$
	\text{ 2.49}$\\
	\hline\hline
\end{tabular} \end{table}
\begin{table}
	\caption{The $bc$ sector with the $b$ quark decay and a scalar $bc$ diquark in the initial state: decay widths, branching ratios and $\Gamma_{L}/\Gamma_{T}$'s for semi-leptonic decays, with lepton mass neglected. We have assumed $m_{B_{i}^{\prime}}=m_{B_{i}}$ and $\tau_{B_{i}^{\prime}}=\tau_{B_{i}}$, i.e. the only difference between $B_{i}^{\prime}\to B_{f}$ and $B_{i}\to B_{f}$ is the mixing coefficients.}
	\label{Tab:SemiLep_bc_b_primed}
\begin{tabular}{l|c|c|c}
	\hline\hline
	channels &$\Gamma/\text{~GeV}$ &${\cal B}$ &$\Gamma_{L}/\Gamma_{T}$\\ \hline
	$\Xi_{bc}^{\prime+}\to\Sigma_{c}^{++}l^{-}\bar{\nu}_{l}$ &$3.28\times 10^{-17}
	
	$ &$1.22\times 10^{-5}$ &$
	\text{ 5.87}$\\
	$\Xi_{bc}^{\prime+}\to\Xi_{cc}^{++}l^{-}\bar{\nu}_{l}$ &$1.91\times 10^{-14}
	
	$ &$7.09\times 10^{-3}$ &$
	\text{ 0.95}$\\
	$\Xi_{bc}^{\prime0}\to\Lambda_{c}^{+}l^{-}\bar{\nu}_{l}$ &$1.71\times 10^{-17}
	
	$ &$2.41\times 10^{-6}$ &$
	\text{ 0.79}$\\
	$\Xi_{bc}^{\prime0}\to\Sigma_{c}^{+}l^{-}\bar{\nu}_{l}$ &$1.64\times 10^{-17}
	
	$ &$2.32\times 10^{-6}$ &$
	\text{ 5.87}$\\
	$\Xi_{bc}^{\prime0}\to\Xi_{cc}^{+}l^{-}\bar{\nu}_{l}$ &$1.91\times 10^{-14}
	
	$ &$2.70\times 10^{-3}$ &$
	\text{ 0.95}$\\
	$\Omega_{bc}^{\prime0}\to\Xi_{c}^{+}l^{-}\bar{\nu}_{l}$ &$1.21\times 10^{-17}
	
	$ &$4.04\times 10^{-6}$ &$
	\text{ 0.84}$\\
	$\Omega_{bc}^{\prime0}\to\Xi_{c}^{\prime+}l^{-}\bar{\nu}_{l}$ &$
	
	1.24\times 10^{-17}$ &$4.15\times 10^{-6}$ &$
	\text{ 6.26}$\\
	$\Omega_{bc}^{\prime0}\to\Omega_{cc}^{+}l^{-}\bar{\nu}_{l}$ &$
	
	1.67\times 10^{-14}$ &$5.59\times 10^{-3}$ &$
	\text{ 0.95}$\\
	\hline\hline
\end{tabular} \end{table}

The effective electroweak Hamiltonian   
reads
\begin{eqnarray}
{\cal H}_{{\rm eff}}&=&\frac{G_{F}}{\sqrt{2}}\bigg(V_{cs}^{*}[\bar{s}\gamma_{\mu}(1-\gamma_{5})c][\bar{\nu}\gamma^{\mu}(1-\gamma_{5})l]+V_{cd}^{*}[\bar{d}\gamma_{\mu}(1-\gamma_{5})c][\bar{\nu}\gamma^{\mu}(1-\gamma_{5})l] \bigg)\nonumber\\
&& + \frac{G_{F}}{\sqrt{2}}\bigg(V_{cb}[\bar{c}\gamma_{\mu}(1-\gamma_{5})b][\bar{l}\gamma^{\mu}(1-\gamma_{5})\nu]+V_{ub}[\bar{u}\gamma_{\mu}(1-\gamma_{5})b][\bar{l}\gamma^{\mu}(1-\gamma_{5})\nu] \bigg),
\end{eqnarray}
where the $G_{F}$ and $V_{cs,cd,ub,cb}$ are Fermi constant and   Cabibbo-Kobayashi-Maskawa (CKM)
matrix element, respectively. Leptonic parts can be computed in  perturbation theory while hadronic contributions are paraemtrized in terms of form factors.

The $B\to B^{\prime}$ form factors   are parametrized in   Eq.~\eqref{eq:weakMatrix},  and the helicity amplitudes of the vector current are related
to these form factors through the following expressions:
\begin{align}
H_{\frac{1}{2},0}^{V} & =-i\frac{\sqrt{Q_{-}}}{\sqrt{q^{2}}}\left((M+M^{\prime})f_{1}-\frac{q^{2}}{M}f_{2}\right),\nonumber\\
H_{\frac{1}{2},1}^{V} & =i\sqrt{2Q_{-}}\left(-f_{1}+\frac{M+M^{\prime}}{M}f_{2}\right),\nonumber\\
H_{\frac{1}{2},0}^{A} & =-i\frac{\sqrt{Q_{+}}}{\sqrt{q^{2}}}\left((M-M^{\prime})g_{1}+\frac{q^{2}}{M}g_{2}\right),\nonumber\\
H_{\frac{1}{2},1}^{A} & =i\sqrt{2Q_{+}}\left(-g_{1}-\frac{M-M^{\prime}}{M}g_{2}\right),
\end{align}
where $Q_{\pm}=2(P\cdot P^{\prime}\pm MM^{\prime})=2MM^{\prime}(\omega\pm1)$. The parameter   $\omega\equiv\frac{P\cdot P^{\prime}}{MM^{\prime}}=\frac{M^{2}+M^{\prime2}-q^{2}}{2MM^{\prime}}$
ranges from 1 to $\omega_{\rm{max}}=\frac{1}{2}(\frac{M}{M^{\prime}}+\frac{M^{\prime}}{M})$.  The $M$ and $M^{\prime}$  is the  mass  for the initial and final baryon. The negative  helicity  amplitudes   are derived  as 
\begin{equation}
H_{-\lambda^{\prime},-\lambda_{W}}^{V}=H_{\lambda^{\prime},\lambda_{W}}^{V}\quad\text{and}\quad H_{-\lambda^{\prime},-\lambda_{W}}^{A}=-H_{\lambda^{\prime},\lambda_{W}}^{A}.
\end{equation}
The helicity
amplitudes for the left-handed current  are obtained as
\begin{equation}
H_{\lambda^{\prime},\lambda_{W}}=H_{\lambda^{\prime},\lambda_{W}}^{V}-H_{\lambda^{\prime},\lambda_{W}}^{A}.
\end{equation}

The differential decay width for the $B\to B^{\prime}l\bar\nu$ decay is written as 
\begin{equation}
\frac{d\Gamma}{d\omega}=\frac{d\Gamma_{L}}{d\omega}+\frac{d\Gamma_{T}}{d\omega},\label{eq:decay-width}
\end{equation}
with the longitudinal and transverse polarizations: 
\begin{align}
\frac{d\Gamma_{L}}{d\omega} & =\frac{G_{F}^{2}|V_{CKM}|^{2}}{(2\pi)^{3}}\frac{q^{2}pM^{\prime}}{12M}(|H_{\frac{1}{2},0}|^{2}+|H_{-\frac{1}{2},0}|^{2}),\label{eq:longi}\\
\frac{d\Gamma_{T}}{d\omega} & =\frac{G_{F}^{2}|V_{CKM}|^{2}}{(2\pi)^{3}}\frac{q^{2}pM^{\prime}}{12M}(|H_{\frac{1}{2},1}|^{2}+|H_{-\frac{1}{2},-1}|^{2}),\label{eq:trans}
\end{align}
where $p=M^{\prime}\sqrt{\omega^{2}-1}$ is the  three-momentum of $B^{\prime}$
in the  $B$ rest frame. Integrating over the parameter $\omega$, we obtain 
the total decay width
\begin{equation}
\Gamma=\int_{1}^{\omega_{{\rm max}}}d\omega\frac{d\Gamma}{d\omega}. \label{eq:inte_decay_width}
\end{equation}
One can also study the ratio of the longitudinal to transverse decay rates $\Gamma_{L}/\Gamma_{T}$.

It is also  possible to express the differential decay width  as \begin{equation}
\frac{d\Gamma}{dq^{2}}=\frac{d\Gamma_{L}}{dq^{2}}+\frac{d\Gamma_{T}}{dq^{2}},
\end{equation}
where the $q^2$ is the lepton pair invariant mass. The polarized decay widths are given as 
\begin{align}
\frac{d\Gamma_{L}}{dq^{2}} & =\frac{G_{F}^{2}|V_{CKM}|^{2}}{(2\pi)^{3}}\frac{q^{2}p}{24M^{2}}(|H_{\frac{1}{2},0}|^{2}+|H_{-\frac{1}{2},0}|^{2}),\label{eq:longi-1}\\
\frac{d\Gamma_{T}}{dq^{2}} & =\frac{G_{F}^{2}|V_{CKM}|^{2}}{(2\pi)^{3}}\frac{q^{2}p}{24M^{2}}(|H_{\frac{1}{2},1}|^{2}+|H_{-\frac{1}{2},-1}|^{2}),\label{eq:trans-1}
\end{align}
where $p=\sqrt{Q_{+}Q_{-}}/2M$ and  $Q_{\pm}=(M\pm M^{\prime})^{2}-q^{2}$.
The total width is derived as: 
\begin{equation}
\Gamma=\int_{0}^{(M-M^{\prime})^{2}}dq^{2}\frac{d\Gamma}{dq^{2}}.
\end{equation}

\subsection{Numerical results }

For the numerical calculation, we will use the results for   Fermi constant and CKM matrix elements from particle data group~\cite{Olive:2016xmw}:
\begin{align}
& G_{F}=1.166\times10^{-5}{\rm GeV}^{-2},\nonumber \\
& |V_{ud}|=0.974,\quad|V_{us}|=0.225,\quad|V_{ub}|=0.00357,\nonumber \\
& |V_{cd}|=0.225,\quad|V_{cs}|=0.974,\quad|V_{cb}|=0.0411.\label{eq:GFCKM}
\end{align}

The integrated partial decay widths and the relevant  branching ratios and $\Gamma_{L}/\Gamma_{T}$'s are given in Tables \ref{Tab:SemiLep_cc}, \ref{Tab:SemiLep_bb}, \ref{Tab:SemiLep_bc_c}, \ref{Tab:SemiLep_bc_c_primed}, \ref{Tab:SemiLep_bc_b} and \ref{Tab:SemiLep_bc_b_primed}, respectively.

For a comparison, we   quote the experimental data on a few  semi-leptonic $D$ and $B$ decays into different final state in the following~\cite{Olive:2016xmw}: 
\begin{eqnarray}
 {\cal B}(D^+\to \bar K^0 e^+\nu_e) &=& (8.82\pm 0.13)\%, \nonumber\\
 {\cal B}(D^+\to \pi^0 e^+\nu_e) &=& (4.05\pm0.18)\times 10^{-3}, \nonumber\\
 {\cal B}(D^0\to K^- e^+\nu_e) &=& (3.530\pm 0.028)\%, \nonumber\\
 {\cal B}(D^0\to \pi^- e^+\nu_e) &=& (2.91\pm0.04)\times 10^{-3}, \nonumber\\
 {\cal B}(\Lambda_{c}^{+}\to \Lambda e^+\nu_e) &=& (3.6\pm 0.4)\%, \nonumber\\
 {\cal B}(  B^0\to D^- \ell^+\nu_\ell )&=& (2.19\pm0.12)\%, \nonumber\\
 {\cal B}(  B^0\to \pi^- \ell^+\nu_\ell )&=& (1.45\pm0.05)\times 10^{-4}. 
 \end{eqnarray}

A few remarks are given in order.
\begin{itemize}

\item When presenting the numerical result for branching fractions, we have used the lifetimes as given in Table~\ref{Tab:para_doubly_heavy}, but as we have pointed out that there exist  large uncertainties in the  lifetimes. So we have also presented the results for decay widths.

\item 
The $\Xi_{cc}^{++}\to \Xi_{c}^+l^+\nu_l$ and  $\Xi_{cc}^{++}\to \Xi_{c}^{\prime+}l^+\nu_l$  are induced by the $c\to s$ transition. Their  branching fractions,  at a few percent level, are comparable to those of $D\to Ke^+\nu_e$.

\item The branching ratio for $\Xi_{cc}^{++}\to \Lambda_{c}^+l^+\nu_l$ is suppressed due to the CKM matrix element $V_{cd}$,  which is also comparable to those of the  $D\to \pi e^+\nu_e$ mode.  

\item The branching ratios for  $b\to c$ transitions are typically at the order $10^{-2}$ to $10^{-3}$, while the $b\to u$ transition is highly suppressed due to the smallness of $|V_{ub}|$.

\item 
In the calculation carried above, we have neglected the form factors $f_{3}(q^{2})$ and $g_{3}(q^{2})$.
For semi-leptonic decays, the contribution from $f_{3}$
or $g_{3}$ is proportional to $m_{l}$, thus it is safe to drop them
if $l=e,\mu$. As for $l=\tau$,    the transverse decay width $d\Gamma_{T}/dq^{2}$
in Eq.~(\ref{eq:trans}) remains unchanged while the longitudinal decay width $d\Gamma_{L}/dq^{2}$
in Eq.~(\ref{eq:longi}) should be re-calculated: 
\begin{equation}
\frac{d\Gamma_{L}}{dq^{2}}=\frac{G_{F}^{2}|V_{cb}|^{2}p\ q^{2}\ (1-\hat{m}_{l}^{2})^{2}}{384\pi^{3}M^{2}}\left((2+\hat{m}_{l}^{2})(|H_{-\frac{1}{2},0}|^{2}+|H_{\frac{1}{2},0}|^{2})+3\hat{m}_{l}^{2}(|H_{-\frac{1}{2},t}|^{2}+|H_{\frac{1}{2},t}|^{2})\right),
\end{equation}
where $\hat{m}_{l}\equiv {m_{l}}/{\sqrt{q^{2}}}$ and $H_{\pm\frac{1}{2},t}$ are given by
\begin{align}
H_{\frac{1}{2},t}^{V} & =-i\frac{\sqrt{Q_{+}}}{\sqrt{q^{2}}}\left((M-M^{\prime})f_{1}+\frac{q^{2}}{M}f_{3}\right)=H_{-\frac{1}{2},t}^{V},\nonumber \\
H_{\frac{1}{2},t}^{A} & =-i\frac{\sqrt{Q_{-}}}{\sqrt{q^{2}}}\left((M+M^{\prime})g_{1}-\frac{q^{2}}{M}g_{3}\right)=-H_{-\frac{1}{2},t}^{A}.
\end{align}

\end{itemize} 

\subsection{SU(3) analysis}

Recently, an analysis of weak decays of doubly-heavy baryons based on flavor symmetry is available in Ref.~\cite{Wang:2017azm}. In the SU(3) symmetry limit,  there exist the a number of relations
among these semileptonic decay widths, which we are going to examine in the following. 

\begin{itemize}
	
\item $cc$ sector
\begin{align*}
\Gamma(\Xi_{cc}^{++}\to\Lambda_{c}^{+}l^{+}\nu) & =\Gamma(\Omega_{cc}^{+}\to\Xi_{c}^{0}l^{+}\nu),\\
\Gamma(\Xi_{cc}^{++}\to\Xi_{c}^{+}l^{+}\nu) & =\Gamma(\Xi_{cc}^{+}\to\Xi_{c}^{0}l^{+}\nu),\\
\Gamma(\Xi_{cc}^{++}\to\Sigma_{c}^{+}l^{+}\nu) & =\frac{1}{2}\Gamma(\Xi_{cc}^{+}\to\Sigma_{c}^{0}l^{+}\nu)=\Gamma(\Omega_{cc}^{+}\to\Xi_{c}^{\prime0}l^{+}\nu),\\
\Gamma(\Xi_{cc}^{++}\to\Xi_{c}^{\prime+}l^{+}\nu) & =\Gamma(\Xi_{cc}^{+}\to\Xi_{c}^{\prime0}l^{+}\nu)=\frac{1}{2}\Gamma(\Omega_{cc}^{+}\to\Omega_{c}^{0}l^{+}\nu),\\
\Gamma(\Xi_{cc}^{+}\to\Sigma_{c}^{0}l^{+}\nu) & =2\Gamma(\Omega_{cc}^{+}\to\Xi_{c}^{\prime0}l^{+}\nu),
\end{align*}

\item $bb$ sector
\begin{align*}
\Gamma(\Xi_{bb}^{0}\to\Xi_{bc}^{+}l^{-}\bar{\nu}) & =\Gamma(\Xi_{bb}^{-}\to\Xi_{bc}^{0}l^{-}\bar{\nu})=\Gamma(\Omega_{bb}^{-}\to\Omega_{bc}^{0}l^{-}\bar{\nu}),\\
\Gamma(\Xi_{bb}^{-}\to\Lambda_{b}^{0}l^{-}\bar{\nu}) & =\Gamma(\Omega_{bb}^{-}\to\Xi_{b}^{0}l^{-}\bar{\nu}),\\
\Gamma(\Xi_{bb}^{0}\to\Sigma_{b}^{+}l^{-}\bar{\nu}) & =2\Gamma(\Xi_{bb}^{-}\to\Sigma_{b}^{0}l^{-}\bar{\nu})=2\Gamma(\Omega_{bb}^{-}\to\Xi_{b}^{\prime0}l^{-}\bar{\nu}),
\end{align*}

\item $bc$ sector with the $c$ quark decay
\begin{align*}
\Gamma(\Xi_{bc}^{+}\to\Lambda_{b}^{0}l^{+}\nu) & =\Gamma(\Omega_{bc}^{0}\to\Xi_{b}^{-}l^{+}\nu),\\
\Gamma(\Xi_{bc}^{+}\to\Xi_{b}^{0}l^{+}\nu) & =\Gamma(\Xi_{bc}^{0}\to\Xi_{b}^{-}l^{+}\nu),\\
\Gamma(\Xi_{bc}^{+}\to\Sigma_{b}^{0}l^{+}\nu) & =\frac{1}{2}\Gamma(\Xi_{bc}^{0}\to\Sigma_{b}^{-}l^{+}\nu)=\Gamma(\Omega_{bc}^{0}\to\Xi_{b}^{\prime-}l^{+}\nu),\\
\Gamma(\Xi_{bc}^{+}\to\Xi_{b}^{\prime0}l^{+}\nu) & =\Gamma(\Xi_{bc}^{0}\to\Xi_{b}^{\prime-}l^{+}\nu)=\frac{1}{2}\Gamma(\Omega_{bc}^{0}\to\Omega_{b}^{-}l^{+}\nu),
\end{align*}

\item $bc$ sector with the $b$ quark decay
\begin{align*}
\Gamma(\Xi_{bc}^{+}\to\Xi_{cc}^{++}l^{-}\bar{\nu}) & =\Gamma(\Xi_{bc}^{0}\to\Xi_{cc}^{+}l^{-}\bar{\nu})=\Gamma(\Omega_{bc}^{0}\to\Omega_{cc}^{+}l^{-}\bar{\nu}),\\
\Gamma(\Xi_{bc}^{+}\to\Sigma_{c}^{++}l^{-}\bar{\nu}) & =2\Gamma(\Xi_{bc}^{0}\to\Sigma_{c}^{+}l^{-}\bar{\nu})=2\Gamma(\Omega_{bc}^{0}\to\Xi_{c}^{\prime+}l^{-}\bar{\nu}).
\end{align*}

\end{itemize}

Comparing the above equations predicted by SU(3) symmetry with the corresponding results in this work, we have the following remarks.

\begin{itemize}
	\item SU(3) symmetry is respected very well in most cases, except for the following ones
	\begin{align}
	\Gamma(\Xi_{cc}^{++}\to\Lambda_{c}^{+}l^{+}\nu) &= \Gamma(\Omega_{cc}^{+}\to\Xi_{c}^{0}l^{+}\nu),\nonumber\\ \Gamma(\Xi_{bc}^{+}\to\Lambda_{b}^{0}l^{+}\nu)  &= \Gamma(\Omega_{bc}^{0}\to\Xi_{b}^{-}l^{+}\nu),\nonumber\\ \Gamma(\Xi_{bc}^{+}\to\Sigma_{b}^{0}l^{+}\nu) &= \Gamma(\Omega_{bc}^{0}\to\Xi_{b}^{\prime-}l^{+}\nu),\nonumber\\ \Gamma(\Xi_{bc}^{+}\to\Xi_{b}^{\prime0}l^{+}\nu) &= \frac{1}{2}\Gamma(\Omega_{bc}^{0}\to\Omega_{b}^{-}l^{+}\nu),\nonumber\\ \Gamma(\Xi_{bc}^{0}\to\Sigma_{c}^{+}l^{-}\bar{\nu}) &= \Gamma(\Omega_{bc}^{0}\to\Xi_{c}^{\prime+}l^{-}\bar{\nu}).\label{eq:su3_breaking}
\end{align}
These five relations  are broken considerably: larger  than 20\% but still less than 50\% using the definition of $({\rm Max}[\Gamma_{{\rm LHS}},\Gamma_{{\rm RHS}}]-{\rm Min}[\Gamma_{{\rm LHS}},\Gamma_{{\rm RHS}}])/{\rm Max}[\Gamma_{{\rm LHS}},\Gamma_{{\rm RHS}}]$.  

\item Since the mass difference between the $u$ and $d$ quark has been neglected in this work, the isospin symmetry is well respected. But since the strange quark is much heavier,  the SU(3) relations for the channels involving   $u,d$ quark and $s$ quark can be sizably broken. All relations given in  Eq.~(\ref{eq:su3_breaking})  are of this type. 

\item The first   4 relations  in Eq.~(\ref{eq:su3_breaking}) involve the $c$ quark decay but  the last one involves the $b$ quark decay.  It indicates that the $c$ quark decay modes tend to break SU(3) symmetry easily. This can be understood since  the  phase space of the $c$ quark decay is smaller, and thus the decay amplitude  is more  sensitive to the mass   of  the initial and final baryons.

\end{itemize}

\subsection{The loosely bounded diquark  approximation and shape parameter uncertainty  }

In the above calculations, we have made the assumption that the  two spectators, a heavy  and a light quark, are treated as a system. Then   baryons are composed of this {\it loosely bounded} system and the quark involved in the weak transition. We approximated  this  system as a heavy quark, and obtained   the $\beta$ parameter in the light-front wave function by comparing the   doubly heavy baryons with  heavy quarkonia, and singly heavy baryons with heavy-light mesons.

From a phenomenological viewpoint,  namely since the size of the loosely bounded system  is mainly determined by the size of the light quark, it might also be applicable to  derive the shape parameter $\beta$ under the  approximation that the heavy-light spectator is treated a light system.  Using these new parameters $\beta$,  we have found that the form factors for the doubly charmed baryons are not significantly  changed since the charm quark mass is not very large.  

\begin{figure}[!]
\begin{center}
\includegraphics[scale=0.6]{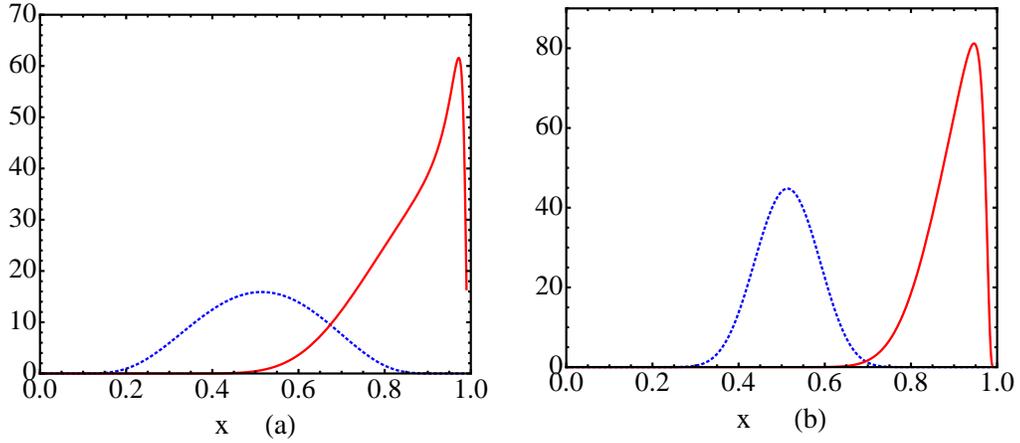}
\end{center}
\caption{Light-front wave functions for the $\Xi_{bbu}^0$ (dotted curves) and $\Sigma^+_{buu}$(solid curves). Here $x$ is the momentum fraction.  In the left panel, we have used $\beta_{\Xi^0_{bbu}}=1.472$ GeV and $\beta_{\Sigma^+_{buu}}=0.562$ GeV; while in the right panel, the parameters are used as: $\beta_{\Xi_{bbu}^0}=0.562$ GeV and $\beta_{\Sigma^+_{buu}}=0.318$ GeV.   The form factor can be viewed as the overlap of wave functions of the $\Xi_{bbu}^0$ and $\Sigma^+_{buu}$, and thus the left (right) panel leads to a larger(smaller) form factors. When showing these results, we have used   $k_T^2=0.1 {\rm GeV}^2$, and we have found similar results for  other transverse momentum $k_T$.   }\label{fig:LFWF}
\end{figure}

We show the form factors for the transition of $\Xi_{bbu}^0\to \Sigma_{buu}^+$ in Tab.~\ref{tab:form_factor_compare}, where for the left  columns, $\beta_{\Xi_{bbu}^0}=1.472$ GeV and $\beta_{\Sigma_{buu}^{+}}=0.562$ GeV
have been used; while the results in the right columns, $\beta_{\Xi_{bbu}^0}=0.562$ GeV and $\beta_{\Sigma^+(buu)}=0.318$ GeV have been adopted. From this table,   we can find that using the new   parameters $\beta$, the form factors are tiny.  This feature can be understood when we show the light-front wave functions in Fig.~\ref{fig:LFWF}. The light-front wave function for the $\Xi^0_{bbu}$ (dotted curves) and $\Sigma^+_{buu}$(solid curves) are shown in this figure. In the left panel, we have used $\beta_{\Xi^0_{bbu}}=1.472$ GeV and $\beta_{\Sigma^+_{buu}}=0.562$ GeV; while in the right panel, the parameters are used as: $\beta_{\Xi^0_{bbu}}=0.562$ GeV and $\beta_{\Sigma^+_{buu}}=0.318$ GeV.   Transition  form factors  can be viewed as the overlap of wave functions of the $\Xi^0_{bbu}$ and $\Sigma^+_{buu}$, and apparently  thus the left (right) panel leads to a larger(smaller) form factors. When showing these results, we have used   $k_T^2=0.1 {\rm GeV}^2$, and we have found similar results for  other transverse momentum $k_T$.

\begin{table}
\caption{Form factors $f_{1,2}$
and $g_{1,2}$ for the $\Xi_{bb}^{0}\to\Sigma_{b}^{+}$ process.  For the left  columns, $\beta_{\Xi_{bb}^0}=1.472$ GeV and $\beta_{\Sigma_{b}^{+}}=0.562$ GeV
have been used; while the results in the right columns, $\beta_{\Xi_{bb}^0}=0.562$ GeV and $\beta_{\Sigma_{b}^{+}}=0.318$ GeV
have been adopted. The fit formulae of (\ref{eq:main_fit_formula})
and (\ref{eq:auxiliary_fit_formula}) are adopted  to access the $q^2$ distribution. }\label{tab:form_factor_compare}
\begin{tabular}{cccc||cccc}
\multicolumn{4}{c}{$\beta_{\Xi_{bb}^{0}}=1.472$,  $\beta_{\Sigma_{b}^{+}}=0.562$}   & \multicolumn{4}{c}{$\beta_{\Xi_{bb}^{0}}=0.562$,  $\beta_{\Sigma_{b}^{+}}=0.318$}\\
\hline 
$F$ & $F(0)$ & $m_{{\rm fit}}$ & $\delta$   & $F$ & $F(0)$ & $m_{{\rm fit}}$ & $\delta$\\
\hline 
$f_{1}^{\Xi_{bb}^{0}\to\Sigma_{b}^{+}}$ & $0.084$ & $3.11$ & $0.80$   & $f_{1}^{\Xi_{bb}^{0}\to\Sigma_{b}^{+}}$ & $0.00012$ & $0.429$ & $0.048$\\
$g_{1}^{\Xi_{bb}^{0}\to\Sigma_{b}^{+}}$ & $0.078$ & $3.24$ & $0.80$   & $g_{1}^{\Xi_{bb}^{0}\to\Sigma_{b}^{+}}$ & $0.00012$ & $0.450$ & $0.053$\\
$f_{2}^{\Xi_{bb}^{0}\to\Sigma_{b}^{+}}$ & $-0.106$ & $3.03$ & $0.88$   & $f_{2}^{\Xi_{bb}^{0}\to\Sigma_{b}^{+}}$ & $-0.00017$ & $0.404$ & $0.041$\\
$g_{2}^{\Xi_{bb}^{0}\to\Sigma_{b}^{+}}$ & $0.007$ & $5.65$ & $4.89$   & $g_{2}^{\Xi_{bb}^{0}\to\Sigma_{b}^{+}}$ & $-0.00004$ & $0.340$ & $0.028$\\
\hline 
\end{tabular}
\end{table}

For the $B\to \pi$ transition, the form factors at $q^2=0$ are suppressed by ${\Lambda_{\rm QCD}}/m_b$ in the heavy quark limit, and for example we   have~\cite{Ali:2007ff}:  
\begin{eqnarray}
 F_1^{B\to \pi}(q^2=0)= (0.23\pm0.05). 
\end{eqnarray}
Though a   QCD analysis of $\Xi_{bbu}^0\to \Sigma^+_{buu}$ transition is not available, we may expect similar power suppressions for these form factors, and the form factors might be of the order 0.1.  Thus  under the approximation of the two spectators as a loosely connected system,  it may be reasonable to use the   parameters $\beta_{\Xi^0_{bbu}}=1.472$ GeV and $\beta_{\Sigma^+_{buu}}=0.562$ GeV as an effective parameter.

\section{Non-leptonic decays} 
\label{sec:nonleptonic}

\begin{table}
	\caption{Partial decay widths (in units of GeV) and branching ratios for non-leptonic decays of the doubly charmed baryons.}
	\label{Tab:NonLep_cc}
\begin{tabular}{l|c|c|l|c|c}
	\hline \hline
	channels  & $\Gamma/\text{~GeV}$  & ${\cal B}$  & channels  & $\Gamma/\text{~GeV}$\
	& ${\cal B}$ \tabularnewline
	\hline 
	$\Xi_{cc}^{++}\to\Lambda_{c}^{+}\pi^{+}$  & $8.87\times10^{-15}$  & $4.05\times10^{-3}$ & $\Xi_{cc}^{++}\to\Lambda_{c}^{+}\rho^{+}$  & $2.32\times10^{-14}$  & $1.06\times10^{-2}$ \tabularnewline
	$\Xi_{cc}^{++}\to\Lambda_{c}^{+}a_{1}^{+}$  & $1.02\times10^{-14}$  & $4.66\times10^{-3}$ & $\Xi_{cc}^{++}\to\Lambda_{c}^{+}K^{+}$  & $7.79\times10^{-16}$  & $3.55\times10^{-4}$ \tabularnewline
	$\Xi_{cc}^{++}\to\Lambda_{c}^{+}K^{*+}$  & $1.09\times10^{-15}$  & $4.98\times10^{-4}$ &  &  & \tabularnewline
	\hline 
	$\Xi_{cc}^{++}\to\Sigma_{c}^{+}\pi^{+}$  & $5.75\times10^{-15}$  & $2.62\times10^{-3}$ & $\Xi_{cc}^{++}\to\Sigma_{c}^{+}\rho^{+}$  & $2.47\times10^{-14}$  & $1.13\times10^{-2}$ \tabularnewline
	$\Xi_{cc}^{++}\to\Sigma_{c}^{+}K^{*+}$  & $1.28\times10^{-15}$  & $5.83\times10^{-4}$ & $\Xi_{cc}^{++}\to\Sigma_{c}^{+}K^{+}$  & $4.22\times10^{-16}$  & $1.92\times10^{-4}$ \tabularnewline
	\hline 
	$\Xi_{cc}^{++}\to\Xi_{c}^{+}\pi^{+}$  & $1.57\times10^{-13}$  & $7.14\times10^{-2}$ & $\Xi_{cc}^{++}\to\Xi_{c}^{+}\rho^{+}$  & $3.03\times10^{-13}$  & $1.38\times10^{-1}$ \tabularnewline
	$\Xi_{cc}^{++}\to\Xi_{c}^{+}K^{*+}$  & $1.19\times10^{-14}$  & $5.44\times10^{-3}$ & $\Xi_{cc}^{++}\to\Xi_{c}^{+}K^{+}$  & $1.31\times10^{-14}$  & $5.97\times10^{-3}$ \tabularnewline
	\hline 
	$\Xi_{cc}^{++}\to\Xi_{c}^{\prime+}\pi^{+}$  & $1.10\times10^{-13}$  & $5.00\times10^{-2}$ & $\Xi_{cc}^{++}\to\Xi_{c}^{\prime+}\rho^{+}$  & $4.12\times10^{-13}$  & $1.88\times10^{-1}$ \tabularnewline
	$\Xi_{cc}^{++}\to\Xi_{c}^{\prime+}K^{*+}$  & $1.87\times10^{-14}$  & $8.54\times10^{-3}$ & $\Xi_{cc}^{++}\to\Xi_{c}^{\prime+}K^{+}$  & $7.48\times10^{-15}$  & $3.41\times10^{-3}$ \tabularnewline
	\hline 
	$\Xi_{cc}^{+}\to\Sigma_{c}^{0}\pi^{+}$  & $1.15\times10^{-14}$  & $1.74\times10^{-3}$ & $\Xi_{cc}^{+}\to\Sigma_{c}^{0}\rho^{+}$  & $4.93\times10^{-14}$  & $7.49\times10^{-3}$ \tabularnewline
	$\Xi_{cc}^{+}\to\Sigma_{c}^{0}K^{*+}$  & $2.55\times10^{-15}$  & $3.88\times10^{-4}$ & $\Xi_{cc}^{+}\to\Sigma_{c}^{0}K^{+}$  & $8.41\times10^{-16}$  & $1.28\times10^{-4}$ \tabularnewline
	\hline 
	$\Xi_{cc}^{+}\to\Xi_{c}^{0}\pi^{+}$  & $1.56\times10^{-13}$  & $2.36\times10^{-2}$ & $\Xi_{cc}^{+}\to\Xi_{c}^{0}\rho^{+}$  & $2.99\times10^{-13}$  & $4.55\times10^{-2}$ \tabularnewline
	$\Xi_{cc}^{+}\to\Xi_{c}^{0}K^{*+}$  & $1.18\times10^{-14}$  & $1.79\times10^{-3}$ & $\Xi_{cc}^{+}\to\Xi_{c}^{0}K^{+}$  & $1.30\times10^{-14}$  & $1.98\times10^{-3}$ \tabularnewline
	\hline 
	$\Xi_{cc}^{+}\to\Xi_{c}^{\prime0}\pi^{+}$  & $1.09\times10^{-13}$  & $1.66\times10^{-2}$ & $\Xi_{cc}^{+}\to\Xi_{c}^{\prime0}\rho^{+}$  & $4.10\times10^{-13}$  & $6.23\times10^{-2}$ \tabularnewline
	$\Xi_{cc}^{+}\to\Xi_{c}^{\prime0}K^{*+}$  & $1.86\times10^{-14}$  & $2.82\times10^{-3}$ & $\Xi_{cc}^{+}\to\Xi_{c}^{\prime0}K^{+}$  & $7.44\times10^{-15}$  & $1.13\times10^{-3}$ \tabularnewline
	\hline 
	$\Omega_{cc}^{+}\to\Xi_{c}^{0}\pi^{+}$  & $7.86\times10^{-15}$  & $3.22\times10^{-3}$ & $\Omega_{cc}^{+}\to\Xi_{c}^{0}\rho^{+}$  & $1.93\times10^{-14}$  & $7.93\times10^{-3}$ \tabularnewline
	$\Omega_{cc}^{+}\to\Xi_{c}^{0}a_{1}^{+}$  & $3.77\times10^{-15}$  & $1.55\times10^{-3}$ & $\Omega_{cc}^{+}\to\Xi_{c}^{0}K^{+}$  & $6.88\times10^{-16}$  & $2.82\times10^{-4}$ \tabularnewline
	$\Omega_{cc}^{+}\to\Xi_{c}^{0}K^{*+}$  & $8.76\times10^{-16}$  & $3.59\times10^{-4}$ &  &  & \tabularnewline
	\hline 
	$\Omega_{cc}^{+}\to\Xi_{c}^{\prime0}\pi^{+}$  & $5.63\times10^{-15}$  & $2.31\times10^{-3}$ & $\Omega_{cc}^{+}\to\Xi_{c}^{\prime0}\rho^{+}$  & $2.44\times10^{-14}$  & $1.00\times10^{-2}$ \tabularnewline
	$\Omega_{cc}^{+}\to\Xi_{c}^{\prime0}K^{*+}$  & $1.26\times10^{-15}$  & $5.18\times10^{-4}$ & $\Omega_{cc}^{+}\to\Xi_{c}^{\prime0}K^{+}$  & $4.14\times10^{-16}$  & $1.70\times10^{-4}$ \tabularnewline
	\hline 
	$\Omega_{cc}^{+}\to\Omega_{c}^{0}\pi^{+}$  & $2.18\times10^{-13}$  & $8.95\times10^{-2}$ & $\Omega_{cc}^{+}\to\Omega_{c}^{0}\rho^{+}$  & $8.27\times10^{-13}$  & $3.39\times10^{-1}$ \tabularnewline
	$\Omega_{cc}^{+}\to\Omega_{c}^{0}K^{*+}$  & $3.76\times10^{-14}$  & $1.54\times10^{-2}$ & $\Omega_{cc}^{+}\to\Omega_{c}^{0}K^{+}$  & $1.49\times10^{-14}$  & $6.13\times10^{-3}$ \tabularnewline
	\hline \hline
\end{tabular}
\end{table}

\begin{table}
	\caption{Partial decay widths and branching ratios for non-leptonic decays of $\Xi_{bb}^{0}$ and $\Xi_{bb}^{-}$.}
	\label{Tab:NonLep_bb_part1}
\begin{tabular}{l|c|c|l|c|c}
	\hline\hline
	channels &$\Gamma/\text{~GeV}$ &${\cal B}$ &channels &$\Gamma/\text{~GeV}$\
	
	&${\cal B}$ \\\hline
	$\Xi_{bb}^{0}\to\Sigma_{b}^{+}\pi^{-}$ &$2.35\times 10^{-18}$ &$
	
	1.32\times 10^{-6}$&$\Xi_{bb}^{0}\to\Sigma_{b}^{+}\rho^{-}$ &$
	
	7.66\times 10^{-18}$ &$4.31\times 10^{-6}$ \\
	$\Xi_{bb}^{0}\to\Sigma_{b}^{+}a_{1}^{-}$ &$1.18\times 10^{-17}$ &$
	
	6.66\times 10^{-6}$&$\Xi_{bb}^{0}\to\Sigma_{b}^{+}K^{-}$ &$
	
	1.94\times 10^{-19}$ &$1.09\times 10^{-7}$ \\
	$\Xi_{bb}^{0}\to\Sigma_{b}^{+}K^{*-}$ &$4.08\times 10^{-19}$ &$
	
	2.29\times 10^{-7}$&$\Xi_{bb}^{0}\to\Sigma_{b}^{+}D^{-}$ &$
	
	4.38\times 10^{-19}$ &$2.46\times 10^{-7}$ \\
	$\Xi_{bb}^{0}\to\Sigma_{b}^{+}D^{*-}$ &$8.80\times 10^{-19}$ &$
	
	4.95\times 10^{-7}$&$\Xi_{bb}^{0}\to\Sigma_{b}^{+}D_{s}^{-}$ &$
	
	1.18\times 10^{-17}$ &$6.66\times 10^{-6}$ \\
	$\Xi_{bb}^{0}\to\Sigma_{b}^{+}D_{s}^{*-}$ &$2.19\times 10^{-17}$ &$
	
	1.23\times 10^{-5}$& & & \\
	\hline
	$\Xi_{bb}^{0}\to\Xi_{bc}^{+}\pi^{-}$ &$3.24\times 10^{-15}$ &$
	
	1.82\times 10^{-3}$&$\Xi_{bb}^{0}\to\Xi_{bc}^{+}\rho^{-}$ &$
	
	9.36\times 10^{-15}$ &$5.26\times 10^{-3}$ \\
	$\Xi_{bb}^{0}\to\Xi_{bc}^{+}a_{1}^{-}$ &$1.22\times 10^{-14}$ &$
	
	6.87\times 10^{-3}$&$\Xi_{bb}^{0}\to\Xi_{bc}^{+}K^{-}$ &$2.64\times 10^{-16}
	
	$ &$1.48\times 10^{-4}$ \\
	$\Xi_{bb}^{0}\to\Xi_{bc}^{+}K^{*-}$ &$4.80\times 10^{-16}$ &$
	
	2.70\times 10^{-4}$&$\Xi_{bb}^{0}\to\Xi_{bc}^{+}D^{-}$ &$5.09\times 10^{-16}
	
	$ &$2.86\times 10^{-4}$ \\
	$\Xi_{bb}^{0}\to\Xi_{bc}^{+}D^{*-}$ &$6.32\times 10^{-16}$ &$
	
	3.55\times 10^{-4}$&$\Xi_{bb}^{0}\to\Xi_{bc}^{+}D_{s}^{-}$ &$
	
	1.36\times 10^{-14}$ &$7.67\times 10^{-3}$ \\
	$\Xi_{bb}^{0}\to\Xi_{bc}^{+}D_{s}^{*-}$ &$1.50\times 10^{-14}$ &$
	
	8.46\times 10^{-3}$& & & \\
	\hline
	$\Xi_{bb}^{0}\to\Xi_{bc}^{\prime+}\pi^{-}$ &$6.39\times 10^{-16}$ &$
	
	3.59\times 10^{-4}$&$\Xi_{bb}^{0}\to\Xi_{bc}^{\prime+}\rho^{-}$ &$
	
	2.13\times 10^{-15}$ &$1.20\times 10^{-3}$ \\
	$\Xi_{bb}^{0}\to\Xi_{bc}^{\prime+}a_{1}^{-}$ &$3.33\times 10^{-15}$ &$
	
	1.87\times 10^{-3}$&$\Xi_{bb}^{0}\to\Xi_{bc}^{\prime+}K^{-}$ &$
	
	5.11\times 10^{-17}$ &$2.87\times 10^{-5}$ \\
	$\Xi_{bb}^{0}\to\Xi_{bc}^{\prime+}K^{*-}$ &$1.14\times 10^{-16}$ &$
	
	6.41\times 10^{-5}$&$\Xi_{bb}^{0}\to\Xi_{bc}^{\prime+}D^{-}$ &$
	
	7.10\times 10^{-17}$ &$3.99\times 10^{-5}$ \\
	$\Xi_{bb}^{0}\to\Xi_{bc}^{\prime+}D^{*-}$ &$2.49\times 10^{-16}$ &$
	
	1.40\times 10^{-4}$&$\Xi_{bb}^{0}\to\Xi_{bc}^{\prime+}D_{s}^{-}$ &$
	
	1.81\times 10^{-15}$ &$1.02\times 10^{-3}$ \\
	$\Xi_{bb}^{0}\to\Xi_{bc}^{\prime+}D_{s}^{*-}$ &$6.24\times 10^{-15}$ &$
	
	3.51\times 10^{-3}$& & & \\
	\hline
	$\Xi_{bb}^{-}\to\Lambda_{b}^{0}\pi^{-}$ &$1.31\times 10^{-18}$ &$
	
	7.34\times 10^{-7}$&$\Xi_{bb}^{-}\to\Lambda_{b}^{0}\rho^{-}$ &$
	
	3.91\times 10^{-18}$ &$2.20\times 10^{-6}$ \\
	$\Xi_{bb}^{-}\to\Lambda_{b}^{0}a_{1}^{-}$ &$5.34\times 10^{-18}$ &$
	
	3.00\times 10^{-6}$&$\Xi_{bb}^{-}\to\Lambda_{b}^{0}K^{-}$ &$
	
	1.09\times 10^{-19}$ &$6.14\times 10^{-8}$ \\
	$\Xi_{bb}^{-}\to\Lambda_{b}^{0}K^{*-}$ &$2.02\times 10^{-19}$ &$
	
	1.14\times 10^{-7}$&$\Xi_{bb}^{-}\to\Lambda_{b}^{0}D^{-}$ &$
	
	2.73\times 10^{-19}$ &$1.53\times 10^{-7}$ \\
	$\Xi_{bb}^{-}\to\Lambda_{b}^{0}D^{*-}$ &$2.84\times 10^{-19}$ &$
	
	1.60\times 10^{-7}$&$\Xi_{bb}^{-}\to\Lambda_{b}^{0}D_{s}^{-}$ &$
	
	7.39\times 10^{-18}$ &$4.16\times 10^{-6}$ \\
	$\Xi_{bb}^{-}\to\Lambda_{b}^{0}D_{s}^{*-}$ &$6.70\times 10^{-18}$ &$
	
	3.77\times 10^{-6}$& & & \\
	\hline
	$\Xi_{bb}^{-}\to\Sigma_{b}^{0}\pi^{-}$ &$1.17\times 10^{-18}$ &$
	
	6.59\times 10^{-7}$&$\Xi_{bb}^{-}\to\Sigma_{b}^{0}\rho^{-}$ &$
	
	3.82\times 10^{-18}$ &$2.15\times 10^{-6}$ \\
	$\Xi_{bb}^{-}\to\Sigma_{b}^{0}a_{1}^{-}$ &$5.91\times 10^{-18}$ &$
	
	3.32\times 10^{-6}$&$\Xi_{bb}^{-}\to\Sigma_{b}^{0}K^{-}$ &$
	
	9.68\times 10^{-20}$ &$5.44\times 10^{-8}$ \\
	$\Xi_{bb}^{-}\to\Sigma_{b}^{0}K^{*-}$ &$2.04\times 10^{-19}$ &$
	
	1.15\times 10^{-7}$&$\Xi_{bb}^{-}\to\Sigma_{b}^{0}D^{-}$ &$
	
	2.18\times 10^{-19}$ &$1.23\times 10^{-7}$ \\
	$\Xi_{bb}^{-}\to\Sigma_{b}^{0}D^{*-}$ &$4.39\times 10^{-19}$ &$
	
	2.47\times 10^{-7}$&$\Xi_{bb}^{-}\to\Sigma_{b}^{0}D_{s}^{-}$ &$
	
	5.91\times 10^{-18}$ &$3.32\times 10^{-6}$ \\
	$\Xi_{bb}^{-}\to\Sigma_{b}^{0}D_{s}^{*-}$ &$1.10\times 10^{-17}$ &$
	
	6.16\times 10^{-6}$& & & \\
	\hline
	$\Xi_{bb}^{-}\to\Xi_{bc}^{0}\pi^{-}$ &$3.24\times 10^{-15}$ &$
	
	1.82\times 10^{-3}$&$\Xi_{bb}^{-}\to\Xi_{bc}^{0}\rho^{-}$ &$
	
	9.36\times 10^{-15}$ &$5.26\times 10^{-3}$ \\
	$\Xi_{bb}^{-}\to\Xi_{bc}^{0}a_{1}^{-}$ &$1.22\times 10^{-14}$ &$
	
	6.87\times 10^{-3}$&$\Xi_{bb}^{-}\to\Xi_{bc}^{0}K^{-}$ &$2.64\times 10^{-16}
	
	$ &$1.48\times 10^{-4}$ \\
	$\Xi_{bb}^{-}\to\Xi_{bc}^{0}K^{*-}$ &$4.80\times 10^{-16}$ &$
	
	2.70\times 10^{-4}$&$\Xi_{bb}^{-}\to\Xi_{bc}^{0}D^{-}$ &$5.09\times 10^{-16}
	
	$ &$2.86\times 10^{-4}$ \\
	$\Xi_{bb}^{-}\to\Xi_{bc}^{0}D^{*-}$ &$6.32\times 10^{-16}$ &$
	
	3.55\times 10^{-4}$&$\Xi_{bb}^{-}\to\Xi_{bc}^{0}D_{s}^{-}$ &$
	
	1.36\times 10^{-14}$ &$7.67\times 10^{-3}$ \\
	$\Xi_{bb}^{-}\to\Xi_{bc}^{0}D_{s}^{*-}$ &$1.50\times 10^{-14}$ &$
	
	8.46\times 10^{-3}$& & & \\
	\hline
	$\Xi_{bb}^{-}\to\Xi_{bc}^{\prime0}\pi^{-}$ &$6.39\times 10^{-16}$ &$
	
	3.59\times 10^{-4}$&$\Xi_{bb}^{-}\to\Xi_{bc}^{\prime0}\rho^{-}$ &$
	
	2.13\times 10^{-15}$ &$1.20\times 10^{-3}$ \\
	$\Xi_{bb}^{-}\to\Xi_{bc}^{\prime0}a_{1}^{-}$ &$3.33\times 10^{-15}$ &$
	
	1.87\times 10^{-3}$&$\Xi_{bb}^{-}\to\Xi_{bc}^{\prime0}K^{-}$ &$
	
	5.11\times 10^{-17}$ &$2.87\times 10^{-5}$ \\
	$\Xi_{bb}^{-}\to\Xi_{bc}^{\prime0}K^{*-}$ &$1.14\times 10^{-16}$ &$
	
	6.41\times 10^{-5}$&$\Xi_{bb}^{-}\to\Xi_{bc}^{\prime0}D^{-}$ &$
	
	7.10\times 10^{-17}$ &$3.99\times 10^{-5}$ \\
	$\Xi_{bb}^{-}\to\Xi_{bc}^{\prime0}D^{*-}$ &$2.49\times 10^{-16}$ &$
	
	1.40\times 10^{-4}$&$\Xi_{bb}^{-}\to\Xi_{bc}^{\prime0}D_{s}^{-}$ &$
	
	1.81\times 10^{-15}$ &$1.02\times 10^{-3}$ \\
	$\Xi_{bb}^{-}\to\Xi_{bc}^{\prime0}D_{s}^{*-}$ &$6.24\times 10^{-15}$ &$
	
	3.51\times 10^{-3}$& & & \\
	\hline
 \hline
\end{tabular}
\end{table}
\begin{table}
	\caption{Partial decay widths and branching ratios for non-leptonic decays of $\Omega_{bb}^{-}$.}
	\label{Tab:NonLep_bb_part2}
\begin{tabular}{l|c|c|l|c|c}
	\hline \hline
	channels  & $\Gamma/\text{~GeV}$  & ${\cal B}$  & channels  & $\Gamma/\text{~GeV}$  & ${\cal B}$ \tabularnewline
	\hline 	$\Omega_{bb}^{-}\to\Xi_{b}^{0}\pi^{-}$ &$1.22\times 10^{-18}$ &$
	
	1.49\times 10^{-6}$&$\Omega_{bb}^{-}\to\Xi_{b}^{0}\rho^{-}$ &$
	
	3.66\times 10^{-18}$ &$4.46\times 10^{-6}$ \\
	$\Omega_{bb}^{-}\to\Xi_{b}^{0}a_{1}^{-}$ &$5.00\times 10^{-18}$ &$
	
	6.08\times 10^{-6}$&$\Omega_{bb}^{-}\to\Xi_{b}^{0}K^{-}$ &$
	
	1.02\times 10^{-19}$ &$1.24\times 10^{-7}$ \\
	$\Omega_{bb}^{-}\to\Xi_{b}^{0}K^{*-}$ &$1.90\times 10^{-19}$ &$
	
	2.31\times 10^{-7}$&$\Omega_{bb}^{-}\to\Xi_{b}^{0}D^{-}$ &$
	
	2.54\times 10^{-19}$ &$3.09\times 10^{-7}$ \\
	$\Omega_{bb}^{-}\to\Xi_{b}^{0}D^{*-}$ &$2.63\times 10^{-19}$ &$
	
	3.20\times 10^{-7}$&$\Omega_{bb}^{-}\to\Xi_{b}^{0}D_{s}^{-}$ &$
	
	6.87\times 10^{-18}$ &$8.36\times 10^{-6}$ \\
	$\Omega_{bb}^{-}\to\Xi_{b}^{0}D_{s}^{*-}$ &$6.18\times 10^{-18}$ &$
	
	7.51\times 10^{-6}$& & & \\
	\hline
	$\Omega_{bb}^{-}\to\Xi_{b}^{\prime0}\pi^{-}$ &$1.13\times 10^{-18}$ &$
	
	1.37\times 10^{-6}$&$\Omega_{bb}^{-}\to\Xi_{b}^{\prime0}\rho^{-}$ &$
	
	3.68\times 10^{-18}$ &$4.47\times 10^{-6}$ \\
	$\Omega_{bb}^{-}\to\Xi_{b}^{\prime0}a_{1}^{-}$ &$5.69\times 10^{-18}$ &$
	
	6.92\times 10^{-6}$&$\Omega_{bb}^{-}\to\Xi_{b}^{\prime0}K^{-}$ &$
	
	9.31\times 10^{-20}$ &$1.13\times 10^{-7}$ \\
	$\Omega_{bb}^{-}\to\Xi_{b}^{\prime0}K^{*-}$ &$1.96\times 10^{-19}$ &$
	
	2.38\times 10^{-7}$&$\Omega_{bb}^{-}\to\Xi_{b}^{\prime0}D^{-}$ &$
	
	2.10\times 10^{-19}$ &$2.56\times 10^{-7}$ \\
	$\Omega_{bb}^{-}\to\Xi_{b}^{\prime0}D^{*-}$ &$4.21\times 10^{-19}$ &$
	
	5.12\times 10^{-7}$&$\Omega_{bb}^{-}\to\Xi_{b}^{\prime0}D_{s}^{-}$ &$
	
	5.68\times 10^{-18}$ &$6.91\times 10^{-6}$ \\
	$\Omega_{bb}^{-}\to\Xi_{b}^{\prime0}D_{s}^{*-}$ &$1.05\times 10^{-17}$ &$
	
	1.27\times 10^{-5}$& & & \\
	\hline
	$\Omega_{bb}^{-}\to\Omega_{bc}^{0}\pi^{-}$ &$3.37\times 10^{-15}$ &$
	
	4.10\times 10^{-3}$&$\Omega_{bb}^{-}\to\Omega_{bc}^{0}\rho^{-}$ &$
	
	9.77\times 10^{-15}$ &$1.19\times 10^{-2}$ \\
	$\Omega_{bb}^{-}\to\Omega_{bc}^{0}a_{1}^{-}$ &$1.28\times 10^{-14}$ &$
	
	1.56\times 10^{-2}$&$\Omega_{bb}^{-}\to\Omega_{bc}^{0}K^{-}$ &$
	
	2.75\times 10^{-16}$ &$3.34\times 10^{-4}$ \\
	$\Omega_{bb}^{-}\to\Omega_{bc}^{0}K^{*-}$ &$5.02\times 10^{-16}$ &$
	
	6.10\times 10^{-4}$&$\Omega_{bb}^{-}\to\Omega_{bc}^{0}D^{-}$ &$
	
	5.41\times 10^{-16}$ &$6.58\times 10^{-4}$ \\
	$\Omega_{bb}^{-}\to\Omega_{bc}^{0}D^{*-}$ &$6.74\times 10^{-16}$ &$
	
	8.19\times 10^{-4}$&$\Omega_{bb}^{-}\to\Omega_{bc}^{0}D_{s}^{-}$ &$
	
	1.45\times 10^{-14}$ &$1.77\times 10^{-2}$ \\
	$\Omega_{bb}^{-}\to\Omega_{bc}^{0}D_{s}^{*-}$ &$1.61\times 10^{-14}$ &$
	
	1.96\times 10^{-2}$& & & \\
	\hline
	$\Omega_{bb}^{-}\to\Omega_{bc}^{\prime0}\pi^{-}$ &$6.66\times 10^{-16}$ &$
	
	8.10\times 10^{-4}$&$\Omega_{bb}^{-}\to\Omega_{bc}^{\prime0}\rho^{-}$ &$
	
	2.22\times 10^{-15}$ &$2.70\times 10^{-3}$ \\
	$\Omega_{bb}^{-}\to\Omega_{bc}^{\prime0}a_{1}^{-}$ &$3.46\times 10^{-15}$ &$
	
	4.20\times 10^{-3}$&$\Omega_{bb}^{-}\to\Omega_{bc}^{\prime0}K^{-}$ &$
	
	5.33\times 10^{-17}$ &$6.49\times 10^{-5}$ \\
	$\Omega_{bb}^{-}\to\Omega_{bc}^{\prime0}K^{*-}$ &$1.18\times 10^{-16}$ &$
	
	1.44\times 10^{-4}$&$\Omega_{bb}^{-}\to\Omega_{bc}^{\prime0}D^{-}$ &$
	
	7.70\times 10^{-17}$ &$9.36\times 10^{-5}$ \\
	$\Omega_{bb}^{-}\to\Omega_{bc}^{\prime0}D^{*-}$ &$2.60\times 10^{-16}$ &$
	
	3.16\times 10^{-4}$&$\Omega_{bb}^{-}\to\Omega_{bc}^{\prime0}D_{s}^{-}$ &$
	
	1.98\times 10^{-15}$ &$2.41\times 10^{-3}$ \\
	$\Omega_{bb}^{-}\to\Omega_{bc}^{\prime0}D_{s}^{*-}$ &$6.53\times 10^{-15}$ &$
	
	7.95\times 10^{-3}$& & & \\
	\hline
	\hline
\end{tabular}
\end{table}

In the following, we study two-body non-leptonic decays of doubly heavy baryons, $B\to B^{\prime}M$ with $M$ as a pseudoscalar ($P$), vector ($V$) or axial vector ($A$) meson. As to a first systematic analysis, we only consider the tree current-current operators in the effective Hamiltonian, taking $c$ quark decays as an example,
\begin{equation}
{\cal H}_{W}=\frac{G_{F}}{\sqrt{2}}V_{uq_{1}}V_{cq_{2}}^{*}(C_{1}O_{1}+C_{2}O_{2}),
\end{equation}
where $O_{1}=(\bar{q}_{2}c)_{V-A}(\bar u{q}_{1})_{V-A}$,
$O_{2}=(\bar u c)_{V-A}(\bar{q}_{2}q_{1})_{V-A}$, $C_{i}(\mu)$
denote the corresponding Wilson coefficients, $q_{1,2}=d\ {\rm or}\ s$. It is similar for  the $b\to c/u$ decays.
In general, the transition amplitude of $B\to B^{\prime}M$ can be
written as
\begin{align}\label{eq:amp}
{\cal M}(B\to B^{\prime}P) & =i\bar{u}_{B^{\prime}}(A+B\gamma_{5})u_{B},\nonumber \\
{\cal M}(B\to B^{\prime}V(A)) & =\epsilon^{*\mu}\bar{u}_{B^{\prime}}\left(A_{1}\gamma_{\mu}\gamma_{5}+A_{2}\frac{P_{\mu}^{\prime}}{M}\gamma_{5}+B_{1}\gamma_{\mu}+B_{2}\frac{P_{\mu}^{\prime}}{M}\right)u_{B},
\end{align}
where $\epsilon^{\mu}$ is the polarization vector of the final vector
or axial vector mesons. In the heavy hadron decays, it has manifested that the factorization hypothesis works well in the heavy quark limit~\cite{Beneke:1999br,Bauer:2000ew,Bauer:2000yr,Beneke:2003pa,Keum:2000ph,Keum:2000wi,Lu:2000em,Lu:2000hj,Kurimoto:2001zj}.
The above decay amplitudes in the factorization approach are expressed as
\begin{align}
A & =-\lambda f_{P}(M-M^{\prime})f_{1}(m^{2}),\nonumber \\
B & =-\lambda f_{P}(M+M^{\prime})g_{1}(m^{2}),\nonumber \\
A_{1} & =-\lambda f_{V}m\left[g_{1}(m^{2})+g_{2}(m^{2})\frac{M-M^{\prime}}{M}\right],\nonumber \\
A_{2} & =-2\lambda f_{V}mg_{2}(m^{2}),\nonumber \\
B_{1} & =\lambda f_{V}m\left[f_{1}(m^{2})-f_{2}(m^{2})\frac{M+M^{\prime}}{M}\right],\nonumber \\
B_{2} & =2\lambda f_{V}mf_{2}(m^{2}),\label{eq:AB}
\end{align}
where $\lambda=\frac{G_{F}}{\sqrt{2}}V_{CKM}V_{q_{1}q_{2}}^{*}a_{1}$
with $a_{1}=C_{1}(\mu_{c})+C_{2}(\mu_{c})/3=1.07$ \cite{Li:2012cfa}, $M(M')$ is the mass of the initial (final) baryon and $m$ is the mass of the emitted meson. For the decay modes with emitted an axial vector meson, $A_{1,2}$ and $B_{1,2}$ in Eq.~\eqref{eq:AB} are modified with the replacement of $f_{V}$ by $-f_{A}$. $f_{P,V,A}$ are the decay constants of pseudoscalar, vector and axial vector mesons, respectively, defined as
\begin{align}
\langle P(P)|A_{\mu}|0\rangle & =-if_{P}P_{\mu},\nonumber \\ 
\langle V(P,\epsilon)|V_{\mu}|0\rangle & =f_{V}M_{V}\epsilon_{\mu}^{*}, \\
\langle A(P,\epsilon)|A_{\mu}|0\rangle & =f_{A}M_{A}\epsilon_{\mu}^{*}.\nonumber
\end{align}

The decay width for the $B\to B^{\prime}P$ is given as 
\begin{equation}
\Gamma=\frac{p}{8\pi}\left(\frac{(M+M^{\prime})^{2}-m^{2}}{M^{2}}|A|^{2}+\frac{(M-M^{\prime})^{2}-m^{2}}{M^{2}}|B|^{2}\right),
\end{equation}
where $p$ is the magnitude of the three-momentum of the final-state particles in the rest frame of initial state.
For $B\to B^{\prime}V(A)$ decay, the decay width is
\begin{equation}
\Gamma=\frac{p(E^{\prime}+M^{\prime})}{4\pi M}\left(2(|S|^{2}+|P_{2}|^{2})+\frac{E^{2}}{m^{2}}(|S+D|^{2}+|P_{1}|^{2})\right),
\end{equation}
where $E(E')$ is the energy of final-state meson (baryon), and
\begin{align*}
S & =-A_{1},\\
P_{1} & =-\frac{p}{E}\left(\frac{M+M^{\prime}}{E^{\prime}+M^{\prime}}B_{1}+B_{2}\right),\\
P_{2} & =\frac{p}{E^{\prime}+M^{\prime}}B_{1},\\
D & =-\frac{p^{2}}{E(E^{\prime}+M^{\prime})}(A_{1}-A_{2}).
\end{align*}
The values of the CKM matrix elements and the masses of the relevant mesons and baryons are taken from \cite{Olive:2016xmw}.
The decay constants are \cite{Cheng:2003sm,Shi:2016gqt,Carrasco:2014poa}
\begin{align}
f_{\pi} & =130.4{\rm MeV},\quad f_{\rho}=216{\rm MeV},\quad f_{a_{1}}=238{\rm MeV},\quad f_{K}=160{\rm MeV},\quad f_{K^{*}}=210{\rm MeV},\nonumber \\
f_{D} & =207.4{\rm MeV},\quad f_{D^{*}}=220{\rm MeV},\quad f_{D_{s}}=247.2{\rm MeV},\quad f_{D_{s}^{*}}=247.2{\rm MeV}.
\end{align}

\begin{table}
	\caption{Partial decay widths and branching ratios for non-leptonic charm decays of the bottom-charm baryons with axial vector $bc$ diquark.}
	\label{Tab:NonLep_bc_c}
\begin{tabular}{l|c|c|l|c|c}
	\hline \hline
	channels  & $\Gamma/\text{~GeV}$  & ${\cal B}$  & channels  & $\Gamma/\text{~GeV}$ & ${\cal B}$ \tabularnewline
	\hline 
	$\Xi_{bc}^{+}\to\Lambda_{b}^{0}\pi^{+}$  & $5.74\times10^{-15}$  & $2.13\times10^{-3}$ & $\Xi_{bc}^{+}\to\Lambda_{b}^{0}\rho^{+}$  & $1.55\times10^{-14}$  & $5.77\times10^{-3}$ \tabularnewline
	$\Xi_{bc}^{+}\to\Lambda_{b}^{0}a_{1}^{+}$  & $5.85\times10^{-15}$  & $2.17\times10^{-3}$ & $\Xi_{bc}^{+}\to\Lambda_{b}^{0}K^{+}$  & $5.21\times10^{-16}$  & $1.93\times10^{-4}$ \tabularnewline
	$\Xi_{bc}^{+}\to\Lambda_{b}^{0}K^{*+}$  & $7.32\times10^{-16}$  & $2.71\times10^{-4}$ &  &  & \tabularnewline
	\hline 
	$\Xi_{bc}^{+}\to\Sigma_{b}^{0}\pi^{+}$  & $3.08\times10^{-15}$  & $1.14\times10^{-3}$ & $\Xi_{bc}^{+}\to\Sigma_{b}^{0}\rho^{+}$  & $1.30\times10^{-14}$  & $4.81\times10^{-3}$ \tabularnewline
	$\Xi_{bc}^{+}\to\Sigma_{b}^{0}K^{*+}$  & $6.50\times10^{-16}$  & $2.41\times10^{-4}$ & $\Xi_{bc}^{+}\to\Sigma_{b}^{0}K^{+}$  & $2.32\times10^{-16}$  & $8.62\times10^{-5}$ \tabularnewline
	\hline 
	$\Xi_{bc}^{+}\to\Xi_{b}^{0}\pi^{+}$  & $9.42\times10^{-14}$  & $3.49\times10^{-2}$ & $\Xi_{bc}^{+}\to\Xi_{b}^{0}\rho^{+}$  & $1.91\times10^{-13}$  & $7.09\times10^{-2}$ \tabularnewline
	$\Xi_{bc}^{+}\to\Xi_{b}^{0}K^{*+}$  & $7.55\times10^{-15}$  & $2.80\times10^{-3}$ & $\Xi_{bc}^{+}\to\Xi_{b}^{0}K^{+}$  & $8.16\times10^{-15}$  & $3.03\times10^{-3}$ \tabularnewline
	\hline 
	$\Xi_{bc}^{+}\to\Xi_{b}^{\prime0}\pi^{+}$  & $5.47\times10^{-14}$  & $2.03\times10^{-2}$ & $\Xi_{bc}^{+}\to\Xi_{b}^{\prime0}\rho^{+}$  & $2.01\times10^{-13}$  & $7.44\times10^{-2}$ \tabularnewline
	$\Xi_{bc}^{+}\to\Xi_{b}^{\prime0}K^{*+}$  & $8.53\times10^{-15}$  & $3.16\times10^{-3}$ & $\Xi_{bc}^{+}\to\Xi_{b}^{\prime0}K^{+}$  & $3.82\times10^{-15}$  & $1.42\times10^{-3}$ \tabularnewline
	\hline 
	$\Xi_{bc}^{0}\to\Sigma_{b}^{-}\pi^{+}$  & $6.13\times10^{-15}$  & $8.66\times10^{-4}$ & $\Xi_{bc}^{0}\to\Sigma_{b}^{-}\rho^{+}$  & $2.58\times10^{-14}$  & $3.64\times10^{-3}$ \tabularnewline
	$\Xi_{bc}^{0}\to\Sigma_{b}^{-}K^{*+}$  & $1.29\times10^{-15}$  & $1.82\times10^{-4}$ & $\Xi_{bc}^{0}\to\Sigma_{b}^{-}K^{+}$  & $4.62\times10^{-16}$  & $6.53\times10^{-5}$ \tabularnewline
	\hline 
	$\Xi_{bc}^{0}\to\Xi_{b}^{-}\pi^{+}$  & $9.38\times10^{-14}$  & $1.33\times10^{-2}$ & $\Xi_{bc}^{0}\to\Xi_{b}^{-}\rho^{+}$  & $1.90\times10^{-13}$  & $2.68\times10^{-2}$ \tabularnewline
	$\Xi_{bc}^{0}\to\Xi_{b}^{-}K^{*+}$  & $7.47\times10^{-15}$  & $1.06\times10^{-3}$ & $\Xi_{bc}^{0}\to\Xi_{b}^{-}K^{+}$  & $8.12\times10^{-15}$  & $1.15\times10^{-3}$ \tabularnewline
	\hline 
	$\Xi_{bc}^{0}\to\Xi_{b}^{\prime-}\pi^{+}$  & $5.47\times10^{-14}$  & $7.73\times10^{-3}$ & $\Xi_{bc}^{0}\to\Xi_{b}^{\prime-}\rho^{+}$  & $2.01\times10^{-13}$  & $2.83\times10^{-2}$ \tabularnewline
	$\Xi_{bc}^{0}\to\Xi_{b}^{\prime-}K^{*+}$  & $8.53\times10^{-15}$  & $1.21\times10^{-3}$ & $\Xi_{bc}^{0}\to\Xi_{b}^{\prime-}K^{+}$  & $3.82\times10^{-15}$  & $5.40\times10^{-4}$ \tabularnewline
	\hline 
	$\Omega_{bc}^{0}\to\Xi_{b}^{-}\pi^{+}$  & $4.42\times10^{-15}$  & $1.48\times10^{-3}$ & $\Omega_{bc}^{0}\to\Xi_{b}^{-}\rho^{+}$  & $1.03\times10^{-14}$  & $3.46\times10^{-3}$ \tabularnewline
	$\Omega_{bc}^{0}\to\Xi_{b}^{-}K^{*+}$  & $4.42\times10^{-16}$  & $1.48\times10^{-4}$ & $\Omega_{bc}^{0}\to\Xi_{b}^{-}K^{+}$  & $3.95\times10^{-16}$  & $1.32\times10^{-4}$ \tabularnewline
	\hline 
	$\Omega_{bc}^{0}\to\Xi_{b}^{\prime-}\pi^{+}$  & $2.60\times10^{-15}$  & $8.69\times10^{-4}$ & $\Omega_{bc}^{0}\to\Xi_{b}^{\prime-}\rho^{+}$  & $1.05\times10^{-14}$  & $3.50\times10^{-3}$ \tabularnewline
	$\Omega_{bc}^{0}\to\Xi_{b}^{\prime-}K^{*+}$  & $4.92\times10^{-16}$  & $1.65\times10^{-4}$ & $\Omega_{bc}^{0}\to\Xi_{b}^{\prime-}K^{+}$  & $1.90\times10^{-16}$  & $6.37\times10^{-5}$ \tabularnewline
	\hline 
	$\Omega_{bc}^{0}\to\Omega_{b}^{-}\pi^{+}$  & $9.29\times10^{-14}$  & $3.11\times10^{-2}$ & $\Omega_{bc}^{0}\to\Omega_{b}^{-}\rho^{+}$  & $3.17\times10^{-13}$  & $1.06\times10^{-1}$ \tabularnewline
	$\Omega_{bc}^{0}\to\Omega_{b}^{-}K^{*+}$  & $1.11\times10^{-14}$  & $3.71\times10^{-3}$ & $\Omega_{bc}^{0}\to\Omega_{b}^{-}K^{+}$  & $6.26\times10^{-15}$  & $2.09\times10^{-3}$ \tabularnewline
	\hline \hline
\end{tabular}
\end{table}
\begin{table}
	\caption{Partial decay widths and branching ratios for non-leptonic charm decays of the bottom-charm baryons with scalar $bc$ diquark.}
	\label{Tab:NonLep_bc_c_primed}
\begin{tabular}{l|c|c|l|c|c}
	\hline \hline
	channels  & $\Gamma/\text{~GeV}$  & ${\cal B}$  & channels  & $\Gamma/\text{~GeV}$ & ${\cal B}$ \tabularnewline
	\hline 
	$\Xi_{bc}^{\prime+}\to\Lambda_{b}^{0}\pi^{+}$  & $2.81\times10^{-15}$  & $1.04\times10^{-3}$ & $\Xi_{bc}^{\prime+}\to\Lambda_{b}^{0}\rho^{+}$  & $1.06\times10^{-14}$  & $3.95\times10^{-3}$ \tabularnewline
	$\Xi_{bc}^{\prime+}\to\Lambda_{b}^{0}a_{1}^{+}$  & $8.83\times10^{-15}$  & $3.27\times10^{-3}$ & $\Xi_{bc}^{\prime+}\to\Lambda_{b}^{0}K^{+}$  & $2.43\times10^{-16}$  & $9.02\times10^{-5}$ \tabularnewline
	$\Xi_{bc}^{\prime+}\to\Lambda_{b}^{0}K^{*+}$  & $5.59\times10^{-16}$  & $2.07\times10^{-4}$ &  &  & \tabularnewline
	\hline 
	$\Xi_{bc}^{\prime+}\to\Sigma_{b}^{0}\pi^{+}$  & $3.72\times10^{-15}$  & $1.38\times10^{-3}$ & $\Xi_{bc}^{\prime+}\to\Sigma_{b}^{0}\rho^{+}$  & $7.62\times10^{-15}$  & $2.83\times10^{-3}$ \tabularnewline
	$\Xi_{bc}^{\prime+}\to\Sigma_{b}^{0}K^{*+}$  & $2.96\times10^{-16}$  & $1.10\times10^{-4}$ & $\Xi_{bc}^{\prime+}\to\Sigma_{b}^{0}K^{+}$  & $3.27\times10^{-16}$  & $1.21\times10^{-4}$ \tabularnewline
	\hline 
	$\Xi_{bc}^{\prime+}\to\Xi_{b}^{0}\pi^{+}$  & $4.66\times10^{-14}$  & $1.73\times10^{-2}$ & $\Xi_{bc}^{\prime+}\to\Xi_{b}^{0}\rho^{+}$  & $1.51\times10^{-13}$  & $5.61\times10^{-2}$ \tabularnewline
	$\Xi_{bc}^{\prime+}\to\Xi_{b}^{0}K^{*+}$  & $7.22\times10^{-15}$  & $2.68\times10^{-3}$ & $\Xi_{bc}^{\prime+}\to\Xi_{b}^{0}K^{+}$  & $3.80\times10^{-15}$  & $1.41\times10^{-3}$ \tabularnewline
	\hline 
	$\Xi_{bc}^{\prime+}\to\Xi_{b}^{\prime0}\pi^{+}$  & $6.54\times10^{-14}$  & $2.43\times10^{-2}$ & $\Xi_{bc}^{\prime+}\to\Xi_{b}^{\prime0}\rho^{+}$  & $9.26\times10^{-14}$  & $3.44\times10^{-2}$ \tabularnewline
	$\Xi_{bc}^{\prime+}\to\Xi_{b}^{\prime0}K^{*+}$  & $2.57\times10^{-15}$  & $9.52\times10^{-4}$ & $\Xi_{bc}^{\prime+}\to\Xi_{b}^{\prime0}K^{+}$  & $5.47\times10^{-15}$  & $2.03\times10^{-3}$ \tabularnewline
	\hline 
	$\Xi_{bc}^{\prime0}\to\Sigma_{b}^{-}\pi^{+}$  & $7.41\times10^{-15}$  & $1.05\times10^{-3}$ & $\Xi_{bc}^{\prime0}\to\Sigma_{b}^{-}\rho^{+}$  & $1.51\times10^{-14}$  & $2.14\times10^{-3}$ \tabularnewline
	$\Xi_{bc}^{\prime0}\to\Sigma_{b}^{-}K^{*+}$  & $5.85\times10^{-16}$  & $8.27\times10^{-5}$ & $\Xi_{bc}^{\prime0}\to\Sigma_{b}^{-}K^{+}$  & $6.51\times10^{-16}$  & $9.21\times10^{-5}$ \tabularnewline
	\hline 
	$\Xi_{bc}^{\prime0}\to\Xi_{b}^{-}\pi^{+}$  & $4.64\times10^{-14}$  & $6.56\times10^{-3}$ & $\Xi_{bc}^{\prime0}\to\Xi_{b}^{-}\rho^{+}$  & $1.50\times10^{-13}$  & $2.12\times10^{-2}$ \tabularnewline
	$\Xi_{bc}^{\prime0}\to\Xi_{b}^{-}K^{*+}$  & $7.16\times10^{-15}$  & $1.01\times10^{-3}$ & $\Xi_{bc}^{\prime0}\to\Xi_{b}^{-}K^{+}$  & $3.78\times10^{-15}$  & $5.35\times10^{-4}$ \tabularnewline
	\hline 
	$\Xi_{bc}^{\prime0}\to\Xi_{b}^{\prime-}\pi^{+}$  & $6.54\times10^{-14}$  & $9.25\times10^{-3}$ & $\Xi_{bc}^{\prime0}\to\Xi_{b}^{\prime-}\rho^{+}$  & $9.26\times10^{-14}$  & $1.31\times10^{-2}$ \tabularnewline
	$\Xi_{bc}^{\prime0}\to\Xi_{b}^{\prime-}K^{*+}$  & $2.57\times10^{-15}$  & $3.63\times10^{-4}$ & $\Xi_{bc}^{\prime0}\to\Xi_{b}^{\prime-}K^{+}$  & $5.47\times10^{-15}$  & $7.73\times10^{-4}$ \tabularnewline
	\hline 
	$\Omega_{bc}^{\prime0}\to\Xi_{b}^{-}\pi^{+}$  & $2.17\times10^{-15}$  & $7.24\times10^{-4}$ & $\Omega_{bc}^{\prime0}\to\Xi_{b}^{-}\rho^{+}$  & $7.65\times10^{-15}$  & $2.56\times10^{-3}$ \tabularnewline
	$\Omega_{bc}^{\prime0}\to\Xi_{b}^{-}K^{*+}$  & $3.81\times10^{-16}$  & $1.27\times10^{-4}$ & $\Omega_{bc}^{\prime0}\to\Xi_{b}^{-}K^{+}$  & $1.83\times10^{-16}$  & $6.12\times10^{-5}$ \tabularnewline
	\hline 
	$\Omega_{bc}^{\prime0}\to\Xi_{b}^{\prime-}\pi^{+}$  & $3.15\times10^{-15}$  & $1.05\times10^{-3}$ & $\Omega_{bc}^{\prime0}\to\Xi_{b}^{\prime-}\rho^{+}$  & $5.54\times10^{-15}$  & $1.85\times10^{-3}$ \tabularnewline
	$\Omega_{bc}^{\prime0}\to\Xi_{b}^{\prime-}K^{*+}$  & $1.88\times10^{-16}$  & $6.28\times10^{-5}$ & $\Omega_{bc}^{\prime0}\to\Xi_{b}^{\prime-}K^{+}$  & $2.73\times10^{-16}$  & $9.12\times10^{-5}$ \tabularnewline
	\hline 
	$\Omega_{bc}^{\prime0}\to\Omega_{b}^{-}\pi^{+}$  & $1.11\times10^{-13}$  & $3.72\times10^{-2}$ & $\Omega_{bc}^{\prime0}\to\Omega_{b}^{-}\rho^{+}$  & $1.25\times10^{-13}$  & $4.19\times10^{-2}$ \tabularnewline
	$\Omega_{bc}^{\prime0}\to\Omega_{b}^{-}K^{*+}$  & $2.41\times10^{-15}$  & $8.05\times10^{-4}$ & $\Omega_{bc}^{\prime0}\to\Omega_{b}^{-}K^{+}$  & $9.13\times10^{-15}$  & $3.05\times10^{-3}$ \tabularnewline
	\hline \hline
\end{tabular}
\end{table}
\begin{table}
	\caption{Partial decay widths and branching ratios for non-leptonic decays: the $bc$ sector with the $b$ quark decay and an axial vector $bc$ diquark in the initial state.}
	\label{Tab:NonLep_bc_b}
\begin{tabular}{l|c|c|l|c|c}
	\hline\hline
	channels &$\Gamma/\text{~GeV}$ &${\cal B}$ &channels &$\Gamma/\text{~GeV}$\
	
	&${\cal B}$ \\\hline
	$\Xi_{bc}^{+}\to\Sigma_{c}^{++}\pi^{-}$ &$2.24\times 10^{-18}$ &$
	
	8.29\times 10^{-7}$&$\Xi_{bc}^{+}\to\Sigma_{c}^{++}\rho^{-}$ &$
	
	7.06\times 10^{-18}$ &$2.62\times 10^{-6}$ \\
	$\Xi_{bc}^{+}\to\Sigma_{c}^{++}a_{1}^{-}$ &$1.05\times 10^{-17}$ &$
	
	3.89\times 10^{-6}$&$\Xi_{bc}^{+}\to\Sigma_{c}^{++}K^{-}$ &$
	
	1.83\times 10^{-19}$ &$6.79\times 10^{-8}$ \\
	$\Xi_{bc}^{+}\to\Sigma_{c}^{++}K^{*-}$ &$3.72\times 10^{-19}$ &$
	
	1.38\times 10^{-7}$&$\Xi_{bc}^{+}\to\Sigma_{c}^{++}D^{-}$ &$
	
	3.92\times 10^{-19}$ &$1.45\times 10^{-7}$ \\
	$\Xi_{bc}^{+}\to\Sigma_{c}^{++}D^{*-}$ &$7.69\times 10^{-19}$ &$
	
	2.85\times 10^{-7}$&$\Xi_{bc}^{+}\to\Sigma_{c}^{++}D_{s}^{-}$ &$
	
	1.07\times 10^{-17}$ &$3.96\times 10^{-6}$ \\
	$\Xi_{bc}^{+}\to\Sigma_{c}^{++}D_{s}^{*-}$ &$1.95\times 10^{-17}$ &$
	
	7.21\times 10^{-6}$& & & \\
	\hline
	$\Xi_{bc}^{+}\to\Xi_{cc}^{++}\pi^{-}$ &$4.65\times 10^{-15}$ &$
	
	1.72\times 10^{-3}$&$\Xi_{bc}^{+}\to\Xi_{cc}^{++}\rho^{-}$ &$
	
	1.31\times 10^{-14}$ &$4.87\times 10^{-3}$ \\
	$\Xi_{bc}^{+}\to\Xi_{cc}^{++}a_{1}^{-}$ &$1.66\times 10^{-14}$ &$
	
	6.14\times 10^{-3}$&$\Xi_{bc}^{+}\to\Xi_{cc}^{++}K^{-}$ &$
	
	3.75\times 10^{-16}$ &$1.39\times 10^{-4}$ \\
	$\Xi_{bc}^{+}\to\Xi_{cc}^{++}K^{*-}$ &$6.68\times 10^{-16}$ &$
	
	2.48\times 10^{-4}$&$\Xi_{bc}^{+}\to\Xi_{cc}^{++}D^{-}$ &$
	
	6.54\times 10^{-16}$ &$2.43\times 10^{-4}$ \\
	$\Xi_{bc}^{+}\to\Xi_{cc}^{++}D^{*-}$ &$7.97\times 10^{-16}$ &$
	
	2.96\times 10^{-4}$&$\Xi_{bc}^{+}\to\Xi_{cc}^{++}D_{s}^{-}$ &$
	
	1.74\times 10^{-14}$ &$6.45\times 10^{-3}$ \\
	$\Xi_{bc}^{+}\to\Xi_{cc}^{++}D_{s}^{*-}$ &$1.89\times 10^{-14}$ &$
	
	6.99\times 10^{-3}$& & & \\
	\hline
	$\Xi_{bc}^{0}\to\Lambda_{c}^{+}\pi^{-}$ &$1.13\times 10^{-18}$ &$
	
	1.60\times 10^{-7}$&$\Xi_{bc}^{0}\to\Lambda_{c}^{+}\rho^{-}$ &$
	
	3.31\times 10^{-18}$ &$4.68\times 10^{-7}$ \\
	$\Xi_{bc}^{0}\to\Lambda_{c}^{+}a_{1}^{-}$ &$4.42\times 10^{-18}$ &$
	
	6.24\times 10^{-7}$&$\Xi_{bc}^{0}\to\Lambda_{c}^{+}K^{-}$ &$
	
	9.36\times 10^{-20}$ &$1.32\times 10^{-8}$ \\
	$\Xi_{bc}^{0}\to\Lambda_{c}^{+}K^{*-}$ &$1.70\times 10^{-19}$ &$
	
	2.41\times 10^{-8}$&$\Xi_{bc}^{0}\to\Lambda_{c}^{+}D^{-}$ &$
	
	2.27\times 10^{-19}$ &$3.21\times 10^{-8}$ \\
	$\Xi_{bc}^{0}\to\Lambda_{c}^{+}D^{*-}$ &$2.42\times 10^{-19}$ &$
	
	3.42\times 10^{-8}$&$\Xi_{bc}^{0}\to\Lambda_{c}^{+}D_{s}^{-}$ &$
	
	6.23\times 10^{-18}$ &$8.80\times 10^{-7}$ \\
	$\Xi_{bc}^{0}\to\Lambda_{c}^{+}D_{s}^{*-}$ &$5.82\times 10^{-18}$ &$
	
	8.22\times 10^{-7}$& & & \\
	\hline
	$\Xi_{bc}^{0}\to\Sigma_{c}^{+}\pi^{-}$ &$1.12\times 10^{-18}$ &$
	
	1.58\times 10^{-7}$&$\Xi_{bc}^{0}\to\Sigma_{c}^{+}\rho^{-}$ &$
	
	3.53\times 10^{-18}$ &$4.99\times 10^{-7}$ \\
	$\Xi_{bc}^{0}\to\Sigma_{c}^{+}a_{1}^{-}$ &$5.24\times 10^{-18}$ &$
	
	7.41\times 10^{-7}$&$\Xi_{bc}^{0}\to\Sigma_{c}^{+}K^{-}$ &$
	
	9.16\times 10^{-20}$ &$1.29\times 10^{-8}$ \\
	$\Xi_{bc}^{0}\to\Sigma_{c}^{+}K^{*-}$ &$1.86\times 10^{-19}$ &$
	
	2.63\times 10^{-8}$&$\Xi_{bc}^{0}\to\Sigma_{c}^{+}D^{-}$ &$
	
	1.96\times 10^{-19}$ &$2.77\times 10^{-8}$ \\
	$\Xi_{bc}^{0}\to\Sigma_{c}^{+}D^{*-}$ &$3.85\times 10^{-19}$ &$
	
	5.44\times 10^{-8}$&$\Xi_{bc}^{0}\to\Sigma_{c}^{+}D_{s}^{-}$ &$
	
	5.34\times 10^{-18}$ &$7.55\times 10^{-7}$ \\
	$\Xi_{bc}^{0}\to\Sigma_{c}^{+}D_{s}^{*-}$ &$9.73\times 10^{-18}$ &$
	
	1.38\times 10^{-6}$& & & \\
	\hline
	$\Xi_{bc}^{0}\to\Xi_{cc}^{+}\pi^{-}$ &$4.65\times 10^{-15}$ &$
	
	6.57\times 10^{-4}$&$\Xi_{bc}^{0}\to\Xi_{cc}^{+}\rho^{-}$ &$
	
	1.31\times 10^{-14}$ &$1.86\times 10^{-3}$ \\
	$\Xi_{bc}^{0}\to\Xi_{cc}^{+}a_{1}^{-}$ &$1.66\times 10^{-14}$ &$
	
	2.34\times 10^{-3}$&$\Xi_{bc}^{0}\to\Xi_{cc}^{+}K^{-}$ &$3.75\times 10^{-16}
	
	$ &$5.30\times 10^{-5}$ \\
	$\Xi_{bc}^{0}\to\Xi_{cc}^{+}K^{*-}$ &$6.68\times 10^{-16}$ &$
	
	9.45\times 10^{-5}$&$\Xi_{bc}^{0}\to\Xi_{cc}^{+}D^{-}$ &$6.54\times 10^{-16}
	
	$ &$9.24\times 10^{-5}$ \\
	$\Xi_{bc}^{0}\to\Xi_{cc}^{+}D^{*-}$ &$7.97\times 10^{-16}$ &$
	
	1.13\times 10^{-4}$&$\Xi_{bc}^{0}\to\Xi_{cc}^{+}D_{s}^{-}$ &$
	
	1.74\times 10^{-14}$ &$2.46\times 10^{-3}$ \\
	$\Xi_{bc}^{0}\to\Xi_{cc}^{+}D_{s}^{*-}$ &$1.89\times 10^{-14}$ &$
	
	2.67\times 10^{-3}$& & & \\
	\hline
	$\Omega_{bc}^{0}\to\Xi_{c}^{+}\pi^{-}$ &$9.10\times 10^{-19}$ &$
	
	3.04\times 10^{-7}$&$\Omega_{bc}^{0}\to\Xi_{c}^{+}\rho^{-}$ &$
	
	2.67\times 10^{-18}$ &$8.93\times 10^{-7}$ \\
	$\Omega_{bc}^{0}\to\Xi_{c}^{+}a_{1}^{-}$ &$3.57\times 10^{-18}$ &$
	
	1.19\times 10^{-6}$&$\Omega_{bc}^{0}\to\Xi_{c}^{+}K^{-}$ &$
	
	7.55\times 10^{-20}$ &$2.52\times 10^{-8}$ \\
	$\Omega_{bc}^{0}\to\Xi_{c}^{+}K^{*-}$ &$1.38\times 10^{-19}$ &$
	
	4.61\times 10^{-8}$&$\Omega_{bc}^{0}\to\Xi_{c}^{+}D^{-}$ &$
	
	1.84\times 10^{-19}$ &$6.15\times 10^{-8}$ \\
	$\Omega_{bc}^{0}\to\Xi_{c}^{+}D^{*-}$ &$1.93\times 10^{-19}$ &$
	
	6.47\times 10^{-8}$&$\Omega_{bc}^{0}\to\Xi_{c}^{+}D_{s}^{-}$ &$
	
	5.03\times 10^{-18}$ &$1.68\times 10^{-6}$ \\
	$\Omega_{bc}^{0}\to\Xi_{c}^{+}D_{s}^{*-}$ &$4.63\times 10^{-18}$ &$
	
	1.55\times 10^{-6}$& & & \\
	\hline
	$\Omega_{bc}^{0}\to\Xi_{c}^{\prime+}\pi^{-}$ &$9.08\times 10^{-19}$ &$
	
	3.04\times 10^{-7}$&$\Omega_{bc}^{0}\to\Xi_{c}^{\prime+}\rho^{-}$ &$
	
	2.88\times 10^{-18}$ &$9.64\times 10^{-7}$ \\
	$\Omega_{bc}^{0}\to\Xi_{c}^{\prime+}a_{1}^{-}$ &$4.31\times 10^{-18}$ &$
	
	1.44\times 10^{-6}$&$\Omega_{bc}^{0}\to\Xi_{c}^{\prime+}K^{-}$ &$
	
	7.44\times 10^{-20}$ &$2.49\times 10^{-8}$ \\
	$\Omega_{bc}^{0}\to\Xi_{c}^{\prime+}K^{*-}$ &$1.52\times 10^{-19}$ &$
	
	5.09\times 10^{-8}$&$\Omega_{bc}^{0}\to\Xi_{c}^{\prime+}D^{-}$ &$
	
	1.62\times 10^{-19}$ &$5.40\times 10^{-8}$ \\
	$\Omega_{bc}^{0}\to\Xi_{c}^{\prime+}D^{*-}$ &$3.19\times 10^{-19}$ &$
	
	1.07\times 10^{-7}$&$\Omega_{bc}^{0}\to\Xi_{c}^{\prime+}D_{s}^{-}$ &$
	
	4.41\times 10^{-18}$ &$1.47\times 10^{-6}$ \\
	$\Omega_{bc}^{0}\to\Xi_{c}^{\prime+}D_{s}^{*-}$ &$8.07\times 10^{-18}$ &$
	
	2.70\times 10^{-6}$& & & \\
	\hline
	$\Omega_{bc}^{0}\to\Omega_{cc}^{+}\pi^{-}$ &$4.20\times 10^{-15}$ &$
	
	1.40\times 10^{-3}$&$\Omega_{bc}^{0}\to\Omega_{cc}^{+}\rho^{-}$ &$
	
	1.19\times 10^{-14}$ &$3.98\times 10^{-3}$ \\
	$\Omega_{bc}^{0}\to\Omega_{cc}^{+}a_{1}^{-}$ &$1.50\times 10^{-14}$ &$
	
	5.03\times 10^{-3}$&$\Omega_{bc}^{0}\to\Omega_{cc}^{+}K^{-}$ &$
	
	3.39\times 10^{-16}$ &$1.13\times 10^{-4}$ \\
	$\Omega_{bc}^{0}\to\Omega_{cc}^{+}K^{*-}$ &$6.06\times 10^{-16}$ &$
	
	2.02\times 10^{-4}$&$\Omega_{bc}^{0}\to\Omega_{cc}^{+}D^{-}$ &$
	
	5.95\times 10^{-16}$ &$1.99\times 10^{-4}$ \\
	$\Omega_{bc}^{0}\to\Omega_{cc}^{+}D^{*-}$ &$7.24\times 10^{-16}$ &$
	
	2.42\times 10^{-4}$&$\Omega_{bc}^{0}\to\Omega_{cc}^{+}D_{s}^{-}$ &$
	
	1.58\times 10^{-14}$ &$5.29\times 10^{-3}$ \\
	$\Omega_{bc}^{0}\to\Omega_{cc}^{+}D_{s}^{*-}$ &$1.71\times 10^{-14}$ &$
	
	5.72\times 10^{-3}$& & & \\
	\hline
	\hline
\end{tabular}
\end{table}
\begin{table}
	\caption{Partial decay widths and branching ratios for non-leptonic decays: the $bc$ sector with the $b$ quark decay and a scalar $bc$ diquark in the initial state. We have assumed $m_{B_{i}^{\prime}}=m_{B_{i}}$ and $\tau_{B_{i}^{\prime}}=\tau_{B_{i}}$, i.e. the only difference between $B_{i}^{\prime}\to B_{f}$ and $B_{i}\to B_{f}$ is the mixing coefficients.}
	\label{Tab:NonLep_bc_b_primed}
\begin{tabular}{l|c|c|l|c|c}
	\hline\hline
	channels &$\Gamma/\text{~GeV}$ &${\cal B}$ &channels &$\Gamma/\text{~GeV}$\
	
	&${\cal B}$ \\\hline
	$\Xi_{bc}^{\prime+}\to\Sigma_{c}^{++}\pi^{-}$ &$2.14\times 10^{-18}$ &$
	
	7.92\times 10^{-7}$&$\Xi_{bc}^{\prime+}\to\Sigma_{c}^{++}\rho^{-}$ &$
	
	6.24\times 10^{-18}$ &$2.31\times 10^{-6}$ \\
	$\Xi_{bc}^{\prime+}\to\Sigma_{c}^{++}a_{1}^{-}$ &$8.29\times 10^{-18}$ &$
	
	3.08\times 10^{-6}$&$\Xi_{bc}^{\prime+}\to\Sigma_{c}^{++}K^{-}$ &$
	
	1.77\times 10^{-19}$ &$6.56\times 10^{-8}$ \\
	$\Xi_{bc}^{\prime+}\to\Sigma_{c}^{++}K^{*-}$ &$3.21\times 10^{-19}$ &$
	
	1.19\times 10^{-7}$&$\Xi_{bc}^{\prime+}\to\Sigma_{c}^{++}D^{-}$ &$
	
	4.26\times 10^{-19}$ &$1.58\times 10^{-7}$ \\
	$\Xi_{bc}^{\prime+}\to\Sigma_{c}^{++}D^{*-}$ &$4.48\times 10^{-19}$ &$
	
	1.66\times 10^{-7}$&$\Xi_{bc}^{\prime+}\to\Sigma_{c}^{++}D_{s}^{-}$ &$
	
	1.17\times 10^{-17}$ &$4.33\times 10^{-6}$ \\
	$\Xi_{bc}^{\prime+}\to\Sigma_{c}^{++}D_{s}^{*-}$ &$1.07\times 10^{-17}$ &$
	
	3.98\times 10^{-6}$& & & \\
	\hline
	$\Xi_{bc}^{\prime+}\to\Xi_{cc}^{++}\pi^{-}$ &$9.34\times 10^{-16}$ &$
	
	3.47\times 10^{-4}$&$\Xi_{bc}^{\prime+}\to\Xi_{cc}^{++}\rho^{-}$ &$
	
	3.00\times 10^{-15}$ &$1.11\times 10^{-3}$ \\
	$\Xi_{bc}^{\prime+}\to\Xi_{cc}^{++}a_{1}^{-}$ &$4.44\times 10^{-15}$ &$
	
	1.65\times 10^{-3}$&$\Xi_{bc}^{\prime+}\to\Xi_{cc}^{++}K^{-}$ &$
	
	7.43\times 10^{-17}$ &$2.75\times 10^{-5}$ \\
	$\Xi_{bc}^{\prime+}\to\Xi_{cc}^{++}K^{*-}$ &$1.58\times 10^{-16}$ &$
	
	5.87\times 10^{-5}$&$\Xi_{bc}^{\prime+}\to\Xi_{cc}^{++}D^{-}$ &$
	
	9.77\times 10^{-17}$ &$3.62\times 10^{-5}$ \\
	$\Xi_{bc}^{\prime+}\to\Xi_{cc}^{++}D^{*-}$ &$2.99\times 10^{-16}$ &$
	
	1.11\times 10^{-4}$&$\Xi_{bc}^{\prime+}\to\Xi_{cc}^{++}D_{s}^{-}$ &$
	
	2.49\times 10^{-15}$ &$9.24\times 10^{-4}$ \\
	$\Xi_{bc}^{\prime+}\to\Xi_{cc}^{++}D_{s}^{*-}$ &$7.42\times 10^{-15}$ &$
	
	2.75\times 10^{-3}$& & & \\
	\hline
	$\Xi_{bc}^{\prime0}\to\Lambda_{c}^{+}\pi^{-}$ &$4.51\times 10^{-19}$ &$
	
	6.38\times 10^{-8}$&$\Xi_{bc}^{\prime0}\to\Lambda_{c}^{+}\rho^{-}$ &$
	
	1.41\times 10^{-18}$ &$1.99\times 10^{-7}$ \\
	$\Xi_{bc}^{\prime0}\to\Lambda_{c}^{+}a_{1}^{-}$ &$2.05\times 10^{-18}$ &$
	
	2.90\times 10^{-7}$&$\Xi_{bc}^{\prime0}\to\Lambda_{c}^{+}K^{-}$ &$
	
	3.74\times 10^{-20}$ &$5.28\times 10^{-9}$ \\
	$\Xi_{bc}^{\prime0}\to\Lambda_{c}^{+}K^{*-}$ &$7.41\times 10^{-20}$ &$
	
	1.05\times 10^{-8}$&$\Xi_{bc}^{\prime0}\to\Lambda_{c}^{+}D^{-}$ &$
	
	8.78\times 10^{-20}$ &$1.24\times 10^{-8}$ \\
	$\Xi_{bc}^{\prime0}\to\Lambda_{c}^{+}D^{*-}$ &$1.38\times 10^{-19}$ &$
	
	1.95\times 10^{-8}$&$\Xi_{bc}^{\prime0}\to\Lambda_{c}^{+}D_{s}^{-}$ &$
	
	2.38\times 10^{-18}$ &$3.36\times 10^{-7}$ \\
	$\Xi_{bc}^{\prime0}\to\Lambda_{c}^{+}D_{s}^{*-}$ &$3.43\times 10^{-18}$ &$
	
	4.85\times 10^{-7}$& & & \\
	\hline
	$\Xi_{bc}^{\prime0}\to\Sigma_{c}^{+}\pi^{-}$ &$1.07\times 10^{-18}$ &$
	
	1.51\times 10^{-7}$&$\Xi_{bc}^{\prime0}\to\Sigma_{c}^{+}\rho^{-}$ &$
	
	3.12\times 10^{-18}$ &$4.41\times 10^{-7}$ \\
	$\Xi_{bc}^{\prime0}\to\Sigma_{c}^{+}a_{1}^{-}$ &$4.15\times 10^{-18}$ &$
	
	5.86\times 10^{-7}$&$\Xi_{bc}^{\prime0}\to\Sigma_{c}^{+}K^{-}$ &$
	
	8.84\times 10^{-20}$ &$1.25\times 10^{-8}$ \\
	$\Xi_{bc}^{\prime0}\to\Sigma_{c}^{+}K^{*-}$ &$1.61\times 10^{-19}$ &$
	
	2.27\times 10^{-8}$&$\Xi_{bc}^{\prime0}\to\Sigma_{c}^{+}D^{-}$ &$
	
	2.13\times 10^{-19}$ &$3.02\times 10^{-8}$ \\
	$\Xi_{bc}^{\prime0}\to\Sigma_{c}^{+}D^{*-}$ &$2.24\times 10^{-19}$ &$
	
	3.17\times 10^{-8}$&$\Xi_{bc}^{\prime0}\to\Sigma_{c}^{+}D_{s}^{-}$ &$
	
	5.84\times 10^{-18}$ &$8.26\times 10^{-7}$ \\
	$\Xi_{bc}^{\prime0}\to\Sigma_{c}^{+}D_{s}^{*-}$ &$5.37\times 10^{-18}$ &$
	
	7.60\times 10^{-7}$& & & \\
	\hline
	$\Xi_{bc}^{\prime0}\to\Xi_{cc}^{+}\pi^{-}$ &$9.34\times 10^{-16}$ &$
	
	1.32\times 10^{-4}$&$\Xi_{bc}^{\prime0}\to\Xi_{cc}^{+}\rho^{-}$ &$
	
	3.00\times 10^{-15}$ &$4.23\times 10^{-4}$ \\
	$\Xi_{bc}^{\prime0}\to\Xi_{cc}^{+}a_{1}^{-}$ &$4.44\times 10^{-15}$ &$
	
	6.27\times 10^{-4}$&$\Xi_{bc}^{\prime0}\to\Xi_{cc}^{+}K^{-}$ &$
	
	7.43\times 10^{-17}$ &$1.05\times 10^{-5}$ \\
	$\Xi_{bc}^{\prime0}\to\Xi_{cc}^{+}K^{*-}$ &$1.58\times 10^{-16}$ &$
	
	2.24\times 10^{-5}$&$\Xi_{bc}^{\prime0}\to\Xi_{cc}^{+}D^{-}$ &$
	
	9.77\times 10^{-17}$ &$1.38\times 10^{-5}$ \\
	$\Xi_{bc}^{\prime0}\to\Xi_{cc}^{+}D^{*-}$ &$2.99\times 10^{-16}$ &$
	
	4.23\times 10^{-5}$&$\Xi_{bc}^{\prime0}\to\Xi_{cc}^{+}D_{s}^{-}$ &$
	
	2.49\times 10^{-15}$ &$3.52\times 10^{-4}$ \\
	$\Xi_{bc}^{\prime0}\to\Xi_{cc}^{+}D_{s}^{*-}$ &$7.42\times 10^{-15}$ &$
	
	1.05\times 10^{-3}$& & & \\
	\hline
	$\Omega_{bc}^{\prime0}\to\Xi_{c}^{+}\pi^{-}$ &$3.68\times 10^{-19}$ &$
	
	1.23\times 10^{-7}$&$\Omega_{bc}^{\prime0}\to\Xi_{c}^{+}\rho^{-}$ &$
	
	1.15\times 10^{-18}$ &$3.86\times 10^{-7}$ \\
	$\Omega_{bc}^{\prime0}\to\Xi_{c}^{+}a_{1}^{-}$ &$1.69\times 10^{-18}$ &$
	
	5.64\times 10^{-7}$&$\Omega_{bc}^{\prime0}\to\Xi_{c}^{+}K^{-}$ &$
	
	3.05\times 10^{-20}$ &$1.02\times 10^{-8}$ \\
	$\Omega_{bc}^{\prime0}\to\Xi_{c}^{+}K^{*-}$ &$6.07\times 10^{-20}$ &$
	
	2.03\times 10^{-8}$&$\Omega_{bc}^{\prime0}\to\Xi_{c}^{+}D^{-}$ &$
	
	7.07\times 10^{-20}$ &$2.37\times 10^{-8}$ \\
	$\Omega_{bc}^{\prime0}\to\Xi_{c}^{+}D^{*-}$ &$1.13\times 10^{-19}$ &$
	
	3.77\times 10^{-8}$&$\Omega_{bc}^{\prime0}\to\Xi_{c}^{+}D_{s}^{-}$ &$
	
	1.90\times 10^{-18}$ &$6.37\times 10^{-7}$ \\
	$\Omega_{bc}^{\prime0}\to\Xi_{c}^{+}D_{s}^{*-}$ &$2.79\times 10^{-18}$ &$
	
	9.33\times 10^{-7}$& & & \\
	\hline
	$\Omega_{bc}^{\prime0}\to\Xi_{c}^{\prime+}\pi^{-}$ &$8.76\times 10^{-19}$ &$
	
	2.93\times 10^{-7}$&$\Omega_{bc}^{\prime0}\to\Xi_{c}^{\prime+}\rho^{-}$ &$
	
	2.57\times 10^{-18}$ &$8.58\times 10^{-7}$ \\
	$\Omega_{bc}^{\prime0}\to\Xi_{c}^{\prime+}a_{1}^{-}$ &$3.42\times 10^{-18}$ &$
	
	1.14\times 10^{-6}$&$\Omega_{bc}^{\prime0}\to\Xi_{c}^{\prime+}K^{-}$ &$
	
	7.26\times 10^{-20}$ &$2.43\times 10^{-8}$ \\
	$\Omega_{bc}^{\prime0}\to\Xi_{c}^{\prime+}K^{*-}$ &$1.32\times 10^{-19}$ &$
	
	4.42\times 10^{-8}$&$\Omega_{bc}^{\prime0}\to\Xi_{c}^{\prime+}D^{-}$ &$
	
	1.76\times 10^{-19}$ &$5.89\times 10^{-8}$ \\
	$\Omega_{bc}^{\prime0}\to\Xi_{c}^{\prime+}D^{*-}$ &$1.83\times 10^{-19}$ &$
	
	6.13\times 10^{-8}$&$\Omega_{bc}^{\prime0}\to\Xi_{c}^{\prime+}D_{s}^{-}$ &$
	
	4.81\times 10^{-18}$ &$1.61\times 10^{-6}$ \\
	$\Omega_{bc}^{\prime0}\to\Xi_{c}^{\prime+}D_{s}^{*-}$ &$4.38\times 10^{-18}
	
	$ &$1.46\times 10^{-6}$& & & \\
	\hline
	$\Omega_{bc}^{\prime0}\to\Omega_{cc}^{+}\pi^{-}$ &$8.42\times 10^{-16}$ &$
	
	2.81\times 10^{-4}$&$\Omega_{bc}^{\prime0}\to\Omega_{cc}^{+}\rho^{-}$ &$
	
	2.72\times 10^{-15}$ &$9.08\times 10^{-4}$ \\
	$\Omega_{bc}^{\prime0}\to\Omega_{cc}^{+}a_{1}^{-}$ &$4.05\times 10^{-15}$ &$
	
	1.35\times 10^{-3}$&$\Omega_{bc}^{\prime0}\to\Omega_{cc}^{+}K^{-}$ &$
	
	6.69\times 10^{-17}$ &$2.24\times 10^{-5}$ \\
	$\Omega_{bc}^{\prime0}\to\Omega_{cc}^{+}K^{*-}$ &$1.44\times 10^{-16}$ &$
	
	4.80\times 10^{-5}$&$\Omega_{bc}^{\prime0}\to\Omega_{cc}^{+}D^{-}$ &$
	
	8.72\times 10^{-17}$ &$2.92\times 10^{-5}$ \\
	$\Omega_{bc}^{\prime0}\to\Omega_{cc}^{+}D^{*-}$ &$2.76\times 10^{-16}$ &$
	
	9.24\times 10^{-5}$&$\Omega_{bc}^{\prime0}\to\Omega_{cc}^{+}D_{s}^{-}$ &$
	
	2.22\times 10^{-15}$ &$7.42\times 10^{-4}$ \\
	$\Omega_{bc}^{\prime0}\to\Omega_{cc}^{+}D_{s}^{*-}$ &$6.86\times 10^{-15}$ &$
	
	2.29\times 10^{-3}$& & & \\
	\hline
	\hline
\end{tabular}
\end{table}

The partial decay widths and branching  ratios for the two-body non-leptonic modes of the doubly heavy flavor baryon decays are given in Tables \ref{Tab:NonLep_cc}, \ref{Tab:NonLep_bb_part1}, \ref{Tab:NonLep_bb_part2}, \ref{Tab:NonLep_bc_c}, \ref{Tab:NonLep_bc_c_primed}, \ref{Tab:NonLep_bc_b} and \ref{Tab:NonLep_bc_b_primed}.

There are some remarks  in the non-leptonic modes:
\begin{itemize}
\item As discussed before, the lifetimes of the doubly heavy flavor baryons are of great ambiguity in the theoretical predictions, especially for the baryons with charmed quark, since the significant non-perturbative contributions at the charm scale. Thus we show the decay widths for each decay mode, which is independent on the lifetime of the doubly heavy baryons. The branching fractions are obtained by the decay widths and the lifetimes shown in Table~\ref{Tab:para_doubly_heavy}. 

\item In the charm decays of the doubly charmed baryons and bottom-charm baryons, the branching fractions of the Cabibbo-favored, singly Cabibbo suppressed and doubly Cabibbo-suppressed processes are of the order of $10^{-2}$, $10^{-3}$ and $10^{-4}$, respectively, as expected as the cases in charmed meson and singly charmed baryon decays.

\item In the bottom decays of the doubly bottom baryons and bottom-charm baryons, the branching fractions of $b\to c$ decays are of the order of $10^{-3}\sim10^{-4}$, while those of $b\to u$ decays are suppressed by the CKM matrix element $|V_{ub}|$.

\item For the bottom-charm baryons with the scalar or axial vector $bc$ diquarks, only the lowest-lying states can decay weakly. As it is not clear which state is the lowest-lying one,  we assume that the masses and lifetimes of the two sets of states are the same with each other, $m_{B_{i}^{\prime}}=m_{B_{i}}$ and $\tau_{B_{i}^{\prime}}=\tau_{B_{i}}$. The only difference between $B_{i}^{\prime}\to B_{f}$ and $B_{i}\to B_{f}$ is the mixing coefficients in Tables~\ref{Tab:mixing_bc_c} and \ref{Tab:mixing_bc_b}. Our results would be useful for the studies in the future if the lowest-lying states are determined. 

\item The contributions from the form factors of $f_{3}(q^{2})$ and $g_{3}(q^{2})$ are neglected in the non-leptonic decays. For the modes with a pseudoscalar meson in the
final state, the terms with $f_{3}$ or $g_{3}$ are proportional
to $m_{P}^{2}/M^{2}$ in the heavy quark limit, with $M$ as the mass of the initial-state baryon. No matter for the light pseudoscalar mesons $\pi$ or $K$ in charm or bottom decays, or for the charmed mesons $D$ or $D_{s}$ in bottom decays, these contributions are small and negligible.  As to the processes with a vector or an axial vector meson in the final state, there is no contribution from $f_{3}$ or $g_{3}$ which
is proportional to $q\cdot\epsilon^{*}=0$.

\item Very recently, LHCb has observed the $\Xi_{cc}^{++}$ in the final state of $\Lambda_{c}^{+}K^{-}\pi^{+}\pi^{+}$ with the significance of more than 12$\sigma$~\cite{1707.01621}. The multi-body charmed hadron decays are usually dominated by the resonant contributions, since there are many resonances below the charm scale. 
The external $W$-emission contributions for this four-body process are $\Xi_{cc}^{++}$ decaying into $(csu)$ states and a charged pion, followed by $(csu)$ fragmented into $\Lambda_{c}^{+}K^{-}\pi^{+}$. However, such contributions cannot be large. Only very high excited states of $\Xi_{c}$ can decay into $\Lambda_{c}^{+}K^{-}\pi^{+}$. But the high excited states are more difficult to be produced. The internal $W$-emission amplitudes can contribute to the four-body decay of $\Xi_{cc}^{++} \to \Lambda_{c}^{+}K^{-}\pi^{+}\pi^{+}$~\cite{Yu:2017zst},  as seen in Fig. \ref{fig:4body}.
%

\begin{figure}[!]
\begin{center}
\includegraphics[scale=0.5]{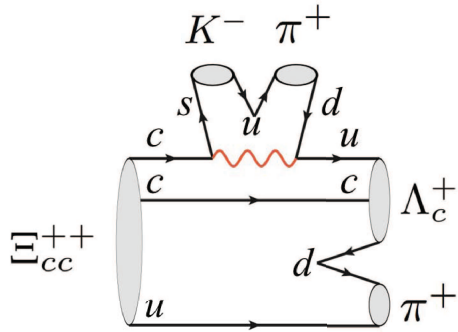}
\end{center}
\caption{Dominant diagram of $\Xi_{cc}^{++}\to\Lambda_{c}^{+}K^{-}\pi^{+}\pi^{+}$.}\label{fig:4body}
\end{figure}

There are many low-lying resonant contributions, for example $\Sigma_{c}^{++}(2455)$ and $\Sigma_{c}^{++}(2520)$ for $\Lambda_{c}^{+}\pi^{+}$, $\overline K^{*0}$ and $(K\pi)_{\rm S-wave}$ for $K^{-}\pi^{+}$. We can naively estimate such contributions. For the internal $W$-emission amplitudes, unlike the case in $B$ meson decays where the non-factorizable contributions are color-suppressed and neglected, the non-factorizable contributions in charm decays are significantly enhanced due to the final-state interacting effects. Empirically, the effective Wilson coefficient in Eq.~(\ref{eq:AB}) is $|a_{2}^{\rm eff}(\mu_{c})|\sim 0.7$ as in $D$ meson decays and in $\Lambda_{c}^{+}$ decays. With this value of $|a_{2}^{\rm eff}|$, we can obtain the branching fraction of $\Xi_{cc}^{++}\to\Sigma_{c}^{++}(2455)\overline K^{*0}=4.1\%$, which is large enough for the experimental measurements. Considering the other resonant contributions, such as $\Sigma_{c}^{++}(2520)\overline K^{*0}$, $\Sigma_{c}^{++}(2455)(K\pi)_{\rm S-wave}$ and $\Sigma_{c}^{++}(2520)(K\pi)_{\rm S-wave}$, the branching fraction of $\Xi_{cc}^{++}\to\Lambda_{c}^{+}K^{-}\pi^{+}\pi^{+}$ could reach   the order of $10\%$. Therefore, the internal $W$-emission contributions are essential to understand the discovery  $\Xi_{cc}^{++}$ in the $\Xi_{cc}^{++}\to\Lambda_{c}^{+}K^{-}\pi^{+}\pi^{+}$  decay mode by  LHCb.

\end{itemize}

\section{Conclusions} 
\label{sec:conclusions}

In the past decades, heavy quark decays have played a very important role in extracting the CKM parameters in the standard model,  understanding the mechanism for the CP violation, and  in  shaping our understanding of dynamics in strong interactions and factorization theorem.  This, however, has only  made use of weak  decays of  the ground state of the heavy mesons/baryons with a single heavy quark.  Weak decays of the doubly-heavy baryons are expected to provide equally important  information.

Very recently, the LHCb collaboration has observed  in the   final state $\Lambda_c K^-\pi^+\pi^+$ a resonant structure  that is identified as the doubly-charmed baryon $\Xi_{cc}^{++}$.  Such an important observation will undoubtedly promote the research on the hadron spectroscopy and also on the weak decays of the doubly  heavy baryons. 
Inspired by this observation,  we have investigated  the decay processes of doubly heavy baryons
$\Xi_{cc}^{++}$, $\Xi_{cc}^{+}$, $\Omega_{cc}^{+}$, $\Xi_{bc}^{(\prime)+}$,
$\Xi_{bc}^{(\prime)0}$, $\Omega_{bc}^{(\prime)0}$, $\Xi_{bb}^{0}$, $\Xi_{bb}^{-}$ and $\Omega_{bb}^{-}$ and focused on the $1/2\to1/2$ transition in this paper.

We have adopted a quark-diquark picture in the calcualtion  the transition form factors. 
At the
quark level these transitions are induced by the weak decays of $c\to d/s$ or $b\to u/c$. We have derived  the form factors of these transitions in the light-front approach and calculated the form factors for both scalar and axial vector diquarks. The obtained form factors  are then applied  to predict the partial widths for the semi-leptonic and non-leptonic decay  of doubly heavy baryons. We find that a number of  decay channels are sizable and can be examined in future measurements at experimental facilities like LHC, Belle II and CEPC.  These results are also useful for a cross-check of  the $\Xi_{cc}^{++}$ and the search for other baryons.

This work can be regarded as a first step towards a comprehensive understanding  of weak decays of doubly-heavy baryons.  The potential  generalizations and improvements    are given as follows. 
\begin{itemize}

\item The $1/2\to 3/2$ transition: 

This work has   focused on the $1/2\to 1/2$ transition,  either the diquark spectator  is a scalar or an axial vector system.  When the diquark spectator is an axial vector, the final baryon may  have spin $3/2$. Such transitions will be calculated in the future.

\item  penguin dominated processes: 

The analysis of nonleptonic decay modes in this work are mainly dominated by  tree-operators. Decay modes induced by penguin operators may have sizable branching fractions as shown in the heavy meson decays.

\item  Non-factorizable contributions: 

To give   more precise predictions on branching ratios and more importantly the CP asymmetries, one has to reliably estimate the non-factorizable contributions. 

\item A comprehensive analysis from QCD: 

As we have discussed at the beginning of Section III, it is a very crude approximation to adopt one heavy and one light quarks as a system to study the heavy to light transition, though this approximation greatly simplifies the calculation. Since the initial and final states contain heavy quark, soft and collinear degrees of freedom, a power counting analysis might be possible in the framework of soft-collinear effective theory.

\end{itemize}

\section*{Acknowledgements} 

The authors are very grateful to   Jibo He, Xiao-Hui Hu,   Run-Hui Li, Ying Li, Cai-Dian L\"u, and Yu-Ming Wang  for useful discussions and valuable comments.
This work is supported  in part   by  National  Natural
Science Foundation of China under Grant
 No.11575110, 11655002, 11735010, 11505083,  Natural  Science Foundation of Shanghai under Grant  No.~15DZ2272100 and No.~15ZR1423100, and  by    Key Laboratory for Particle Physics, Astrophysics and Cosmology, Ministry of Education.



\end{document}